\documentclass[preprint,aps, amsmath,amssymb]{revtex4-1}

\usepackage{MnSymbol}
\usepackage[utf8]{inputenc}
\usepackage{soulutf8}
\usepackage{graphicx}
\usepackage{epstopdf}
\usepackage{amsmath}
\usepackage{slashed}
\usepackage{xcolor}
\usepackage{mathtools}
\usepackage{slashed}
\usepackage{array,multirow}
\usepackage{booktabs}
\usepackage{adjustbox} 
\usepackage{rotating}
\usepackage{blindtext}
\usepackage{physics}
\usepackage{caption}
\usepackage{subcaption}

\begin{document}

\title{Spatial imaging of proton via leading-twist non-skewed GPDs with basis light-front quantization
}

\author{Satvir Kaur}
\email{satvir@impcas.ac.cn}

\affiliation{\small Institute of Modern Physics, Chinese Academy of Sciences, Lanzhou 730000, China}
\affiliation{\small School of Nuclear Science and Technology, University of Chinese Academy of Sciences, Beijing 100049, China}
\affiliation{\small CAS Key Laboratory of High Precision Nuclear Spectroscopy, Institute of Modern Physics, Chinese Academy of Sciences, Lanzhou 730000, China}

\author{Siqi Xu}
\email{xsq234@impcas.ac.cn}

\affiliation{\small Institute of Modern Physics, Chinese Academy of Sciences, Lanzhou 730000, China}
\affiliation{\small School of Nuclear Science and Technology, University of Chinese Academy of Sciences, Beijing 100049, China} 
\affiliation{\small CAS Key Laboratory of High Precision Nuclear Spectroscopy, Institute of Modern Physics, Chinese Academy of Sciences, Lanzhou 730000, China}

\author{Chandan Mondal}
\email{mondal@impcas.ac.cn}

\affiliation{\small Institute of Modern Physics, Chinese Academy of Sciences, Lanzhou 730000, China}
\affiliation{\small School of Nuclear Science and Technology, University of Chinese Academy of Sciences, Beijing 100049, China}
\affiliation{\small CAS Key Laboratory of High Precision Nuclear Spectroscopy, Institute of Modern Physics, Chinese Academy of Sciences, Lanzhou 730000, China}

\author{Xingbo Zhao}
\email{xbzhao@impcas.ac.cn}

\affiliation{\small Institute of Modern Physics, Chinese Academy of Sciences, Lanzhou 730000, China}
\affiliation{\small School of Nuclear Science and Technology, University of Chinese Academy of Sciences, Beijing 100049, China} 
\affiliation{\small CAS Key Laboratory of High Precision Nuclear Spectroscopy, Institute of Modern Physics, Chinese Academy of Sciences, Lanzhou 730000, China}

\author{James P. Vary}
\email{jvary@iastate.edu}

\affiliation{\small Department of Physics and Astronomy, Iowa State University, Ames, Iowa 50011, USA}

\collaboration{BLFQ Collaboration}

\begin{abstract} 
The internal image of the proton is unveiled by examining the generalized parton distributions (GPDs) at zero skewness, within the basis light-front quantized environment. Several distributions emerge when a quark is sampled with different currents depending upon the helicity arrangements of the active quark and the proton target. We investigate six of the eight leading-twist proton GPDs of the valence quarks, the helicity conserving distributions $(H, E, \tilde{H})$ and the helicity non-conserving $(H_T,E_T,\tilde{H}_T)$ distributions at skewness set to zero ($\zeta=0$). We consider purely transverse momentum transfer and, hence, obtain results describe only the proton's two-dimensional structure in the transverse plane.
We present the Mellin moments of these distribution functions, where the first moment produces a form factor and the second Mellin moments help extract the information on partonic contributions to the hadronic angular momentum. We compare our results for the Mellin moments with those from lattice QCD and other approaches where available.
We also present the GPDs in transverse position space.

\end{abstract}

\maketitle
\section{Introduction}

Probing the hadron's complex internal structure provides knowledge of the non-perturbative aspects of Quantum Chromodynamics (QCD) and insights into fundamental questions such as the nature of confinement. Generalized parton distributions (GPDs) are three-dimensional functions that convey structural details of the hadron. For example, from these functions, one can obtain information about the distribution of partons in the plane transverse to the direction in which the hadron is moving. Alternatively, one can deduce the distribution of the longitudinal momentum carried by the partons. These distribution functions have the potential to address one of the major issues in hadron physics -- the proton spin problem. This is simply because the GPDs provide information on the orbital motion of the partons in conjunction with their spatial flavor distributions. 

Multi-variable GPDs are functions of $(x,\zeta,t)$ where $x$ is the longitudinal momentum fraction held by the parton, $\zeta(= -\Delta^+/2 P^+)$ and $t(=\Delta^2)$ define the longitudinal momentum transfer from the initial to the final state of a hadron and the square of the total momentum transferred respectively.
While they are not probabilistic functions, their two-dimensional (2-d) Fourier transforms from transverse momentum transfer to the impact-parameter plane in the absence of the longitudinal momentum transfer provide a probabilistic interpretation of the GPDs~\cite{Burkardt:2000za, Burkardt:2002hr}. 

Further value can be derived from GPDs by implementing certain limits that provide notable 1-d distributions.
For instance, the first Mellin moments of different GPDs reproduce different form factors depending upon the helicity configurations of both quark and proton. Also, one can retrieve the parton distribution functions (PDFs) at the forward limit of the GPDs, i.e. when there is no momentum transfer from the initial to the final state of the proton. Further, at the $t \rightarrow 0$ limit, one can find the connection of GPDs via second Mellin moments to the quark's and gluon's angular momentum distribution inside the hadron. An indirect connection has been stated between the basic mechanical properties of the proton, like pressure, shear distributions etc. and the GPDs~\cite{Burkert:2018bqq}. It is worth noticing that the different helicity configurations of the active parton and the hadron give rise to different GPDs and they, in turn, provide a bounty of information on the hadron structure and its spin.   

The GPDs are generally classified into two categories: the chiral-even GPDs and the chiral-odd GPDs based on whether the quark helicity is preserved or not. At leading-twist, the chiral-even GPDs are further divided into unpolarized $(H(x,\zeta,t), E(x,\zeta,t))$ and helicity-dependent GPDs $(\tilde{H}(x,\zeta,t), \tilde{E}(x,\zeta,t))$, where $H$ and $\tilde{H}$ appear when the helicity of the proton is conserved in the initial and final states, which is not the case for $E$ and $\tilde{E}$. The GPD $\tilde{E}$ can be evaluated by considering the momentum transfer in the longitudinal direction. Since we focus on extracting the GPDs at $\zeta=0$, $\tilde{E}$ is beyond the scope of this work. To clarify, one can potentially determine $\tilde{E}$ by conducting a calculation at a nonzero $\zeta$ value in order to isolate this GPD from the parametrization. Then, substituting $\zeta=0$ may provide the nonzero distribution. Hence, the approach used in this study initializes $\zeta$ as zero at the outset of the calculation, making $\tilde{E}$ unattainable using this approach.
These GPDs are convoluted with other quantities when forming representations of amplitudes for hard exclusive processes, such as deeply virtual Compton scattering (DVCS)~\cite{Ji:1996nm, Radyushkin:1997ki, Belitsky:2001ns} and deeply virtual meson production (DVMP)~\cite{Collins:1996fb}. 

Extensive experimental efforts have been undertaken to investigate GPDs. One can cite, for example, H1~\cite{H1:2001nez, H1:2005gdw}, ZEUS~\cite{ZEUS:1998xpo, ZEUS:2003pwh}, HERMES at DESY~\cite{HERMES:2001bob, HERMES:2006pre, HERMES:2008abz}, Hall A~\cite{JeffersonLabHallA:2006prd, JeffersonLabHallA:2007jdm}, CLAS at Jefferson Lab (JLab)~\cite{CLAS:2001wjj, CLAS:2006krx, CLAS:2007clm}, COMPASS at CERN~\cite{dHose:2004usi}. Recently, these distributions have been determined by analysing the world electron scattering data~\cite{Hashamipour:2021kes, Hashamipour:2020kip}. 

There are a total of four chiral-odd GPDs, also known as transversity GPDs: $H_T(x,\zeta,t)$, $E_T(x,\zeta,t)$, $\tilde{H}_T(x,\zeta,t)$ and $\tilde{E}_T(x,\zeta,t)$. The distribution $\tilde{E}_T$ vanishes for $\zeta=0$, since it is an odd function under the transformation $\zeta \rightarrow -\zeta$. Such GPDs are quite challenging to measure through hard exclusive processes. Nevertheless, it has been proposed that these GPDs could be probed in diffractive double meson electroproduction~\cite{Ivanov:2002jj, Enberg:2006he, ElBeiyad:2010pji}. Theoretical efforts have shown the possibility of describing the hard exclusive electroproduction of pseudoscalar mesons by a hard scattering mechanism involving the leading-twist chiral-odd GPDs of the nucleon~\cite{Ahmad:2008hp, Goloskokov:2009ia, Goloskokov:2011rd, Goldstein:2012az, Goldstein:2013gra, Kroll:2016aop}. The first evidence of the existence of these GPDs was given by the COMPASS collaboration where the exclusive production of $\rho^0$ mesons was studied by scattering muons off transversely polarized protons~\cite{COMPASS:2013fsk}. Results from GPDs-based model calculations were found to be in agreement with the data. Further, results of exclusive $\pi^0$ and $\eta$ electroproduction by CLAS collaboration confirm the direct experimental accessibility of transversity GPDs~\cite{CLAS:2012cna, Kubarovsky:2016yaa}. It is noteworthy that experiments are planned to extract GPDs at upcoming facilities such as the Electron-Ion-Collider (EIC)~\cite{AbdulKhalek:2021gbh}, the EIC in China (EIcC)~\cite{Anderle:2021wcy, Cao:2023wyz} and the $12$ GeV upgrade program at JLab~\cite{Dudek:2012vr, Burkert:2018nvj}.

Theoretically, the proton GPDs have drawn immense attention. Several QCD inspired models have been developed to understand the proton structure~(see, for example, Refs.~\cite{Ji:1997gm, Boffi:2002yy, Boffi:2003yj, Pasquini:2005dk, Scopetta:2004wt, Goeke:2001tz, Penttinen:1999th, Petrov:1998kf, Mondal:2015uha, Mondal:2017wbf, Maji:2017ill, Chakrabarti:2015ama, Freese:2020mcx}), but deriving the theoretical links with QCD remains a challenge. For instance, numerous studies on moments which are related to the GPDs in certain limits are presented in the lattice QCD approach~\cite{Bhattacharya:2022aob, Alexandrou:2021bbo, Alexandrou:2022dtc, Guo:2022upw, Alexandrou:2020zbe, Gockeler:2005cj, QCDSF:2006tkx, Alexandrou:2019ali} and Dyson-Schwinger equations (DSE) approach~\cite{Chen:2022odn, Xu:2015kta, Chen:2020wuq, Cui:2020rmu}. 
Unlike the Euclidean methods, the QCD observables are directly obtainable in Minkowski space-time. However, no such method has been developed so far. The basis light-front quantization (BLFQ) approach~\cite{Vary:2009gt, Zhao:2014xaa, Wiecki:2014ola, Li:2015zda, Li:2017mlw, Mondal:2019jdg, Xu:2021wwj, Nair:2022evk} has potential to achieve this goal of solving the QCD from first principles when QCD interactions alone are included. Since we consider only the valence Fock sector, this work utilizes an effective Hamiltonian, in essence a QCD-inspired model, in order to solve for observables. 

We adopt the BLFQ approach which is a convenient framework defined on the light-front to get the hadron spectra and its structure while obtaining the light-front wave functions (LFWFs) through diagonalization of the Hamiltonian. One can access the distributions of sea quarks and gluons in this approach by including the higher Fock sector representations of a hadron. This approach has successfully described the QCD bound states of mesons~\cite{Li:2015zda, Li:2017mlw, Lan:2019vui, Jia:2018ary, Tang:2018myz, Lan:2019rba, Lan:2019img, Adhikari:2021jrh, Lan:2021wok} and baryons~\cite{Mondal:2019jdg, Xu:2021wwj, Liu:2022fvl, Peng:2022lte, Hu:2022ctr}. Recently, these states have been expanded in the Fock space including one dynamical gluon component for the pion $\ket{q\bar{q}g}$~\cite{Lan:2021wok} and the nucleon $\ket{qqqg}$~\cite{Xu:2022abw}. Within this approach, one now has access to the gluon distributions based on coupling defined by QCD. 

In this work, we employ the LFWFs to study the proton GPDs by taking into account the valence Fock sector $\ket{qqq}$, where the chiral-even GPDs along with the other applications have already been studied~\cite{Mondal:2019jdg, Xu:2021wwj, Liu:2022fvl}. Note that the proton form factors (FFs), such as electromagnetic FFs and axial-vector FFs have been evaluated in this approach and have been found consistent with the available experimental data~\cite{Mondal:2019jdg, Xu:2021wwj}. Further, the 1-d proton PDFs, particularly the unpolarized, helicity-dependent and transversity PDFs have been examined by comparing them with the available global fits and measured data~\cite{Mondal:2019jdg, Xu:2021wwj}. Overall, the results have been found in agreement with the data. In Ref.~\cite{Liu:2022fvl}, the angular momentum distributions have been explored, which have been evaluated using the unpolarized and helicity-dependent GPDs. The previous results are encouraging and motivate us to extend this approach to study the chiral-odd GPDs, when the quark helicity flips unlike for the case of chiral-even GPDs. 
The 3-d chiral-odd GPDs are the extended version of the transversity PDF and tensor form factors, which provide the spatial tomography of the proton when the valence quarks are transversely polarized. In this way, the chiral-odd GPDs provide important details on the correlation between the angular momentum and spin of quarks inside the proton.

Our aim is to investigate the structure of the proton through its GPDs and other observables in greater detail in order to provide a better understanding using the BLFQ approach. Our selected observables include different Mellin moments (so-called generalized form factors) and impact-parameter dependent GPDs.
\section{Basis light-front quantization approach}
In the BLFQ approach, an eigenvalue problem of the Hamiltonian, $H_{\rm eff}\ket{\Psi}=M_{H}^2\ket{\Psi}$, is solved on the light-front (LF). The eigensolutions provide LFWFs, and the eigenvalues are recognized as the hadronic mass spectra ($M_H$). The former play crucial roles in understanding the detailed structure of QCD bound state systems.


The baryonic state on which the Hamiltonian operator would act, is expanded at fixed LF time as
\begin{equation}
\ket{\Psi}= \psi_{(q q q)} \ket{q q q} + \psi_{(q q q q \bar{q})} \ket{q q q q \bar{q}} + \psi_{(q q q g)} \ket{q q q g} + .\;.\;. \;,
\label{Fock-space}
\end{equation}
where $q, \bar{q}$ and $g$ represent quark, antiquark and gluon Fock particles respectively. The significance of LFWFs $\psi_{(q q q)},\psi_{(q q q q \bar{q})},\psi_{(q q q g)},...$ is to provide the probability amplitudes for the Fock states defined by $\ket{q q q}, \ket{q q q q \bar{q}}, \ket{q q q g}$ and so on. In this work, we consider only the valence Fock state, i.e., the first term in Eq.~\eqref{Fock-space}.

The effective Hamiltonian of the baryonic systems in our chosen Fock space is defined as~\cite{Mondal:2019jdg, Xu:2021wwj, Liu:2022fvl, Hu:2022ctr, Peng:2022lte}
\begin{eqnarray}
H_{\rm eff}&=&\sum_i \frac{{\bf k}^2_{\perp i}+m_i^2}{x_i} + \frac{1}{2}\sum_{i\neq j} \kappa^4 \left(x_i x_j ({\bf r}_{\perp i} -{\bf r}_{\perp j})^2 - \frac{\partial_{x_i}(x_i x_j \partial_{x_j})}{(m_i+m_j)^2} \right) \nonumber\\
&+& \frac{1}{2}\sum_{i \neq j} \frac{4 \pi C_F \alpha_s}{Q^2_{ij}} \bar{u}_{s_i^\prime}(k^\prime_i) \gamma^\mu u_{s_i} (k_i) \bar{u}_{s_j^\prime}(k^\prime_j) \gamma^\nu u_{s_j} (k_j) g_{\mu \nu} \;.
\label{Hamiltonian}
\end{eqnarray}

The first term in Eq.~\eqref{Hamiltonian} expresses the kinetic energy with $m_{i}$ being mass of the valence quark; $x_i$ and ${\bf k}_{\perp i}$ symbolize the longitudinal momentum fraction and transverse momentum carried by $i$th constituent of the system with $\sum_i x_i = 1$ and $\sum_i {\bf k}_{\perp i} = 0$. The second term in Eq.~\eqref{Hamiltonian} expresses the confining potential, which is separately defined in transverse and longitudinal directions through the soft-wall LF holographic QCD~\cite{Brodsky:2014yha} and the phenomenological modelling~\cite{Li:2015zda} respectively. With regard to the compatibility of our harmonic oscillator (HO) basis function in the transverse direction (details about the basis are given in the ensuing paragraphs), the holographic QCD provides us with the HO potential. In the nonrelativistic limit, the total confinement potential emerges as a 3-d HO potential~\cite{Xu:2021wwj, Mondal:2019jdg, Li:2015zda, Hu:2022ctr}.
The third term in Eq.~\eqref{Hamiltonian} refers to the one gluon exchange interactions with coupling constant $\alpha_s$, which underlies the dynamical spin structure in the LFWFs.
Here, $u_{s_i}(k_i)$ represents the Dirac spinor with $s_i$ and $k_i$ being the spin and momentum carried by the $i$th valence quark.
$C_F$ and $g_{\mu\nu}$ define the color factor and the metric tensor respectively. Further, the square of the average four-momentum transfer is expressed as $Q^2_{ij}=-q^2= - \frac{1}{2}\left((k^\prime_i - k_i)^2 + (k^\prime_j - k_j)^2\right)$. 


Our goal is to follow BLFQ and evaluate the Hamiltonian, defined in Eq.~\eqref{Hamiltonian}, in a suitably-truncated basis and diagonalize it to produce the baryon mass spectra and corresponding wave functions. For the BLFQ basis, we choose the 2-d HO basis and the discretized plane-wave basis in transverse and longitudinal directions respectively to expand $\ket{\Psi}$~\cite{Vary:2009gt, Zhao:2014xaa}. The ortho-normalized 2-d HO basis function in the transverse direction is given by
\begin{equation}
\phi_{n,m}({\bf k}_\perp;b)= \frac{\sqrt{2}}{b (2\pi)^{3/2}}\sqrt{\frac{n !}{(n+\vert m \vert)!}} e^{-k_\perp^2/2 b^2} \left(\frac{|k_\perp|}{b}\right)^{|m|} L_n^{\vert m \vert}\left(\frac{k^2_\perp}{b^2}\right) e^{i m \theta}\;,
\end{equation}
where $b$ is the HO scale parameter. The quantum numbers $n$ and $m$ represent the radial excitation and angular momentum projection respectively of a particle in a 2-d HO. $L_n^{\vert m \vert}$ represents the associated Laguerre polynomial.

In the discretized plane-wave basis, the longitudinal momentum fraction of $i$th particle is represented by $x_i=\frac{p^+_i}{P^+} = \frac{k_i}{K}$ with the dimensionless quantity being $k=\frac{1}{2},\frac{3}{2},\frac{5}{2},.\;.\;.$ The values of $k$ are chosen to signify the choice of anti-periodic boundary conditions. Note that $K=\sum_{i} k_i$. In addition, the total angular momentum projection is defined for many-body basis states as $M_J=\sum_i (m_i + \lambda_i)$ with $\lambda$ being the quark helicity. The effective Hamiltonian in Eq.~\eqref{Hamiltonian} conserves $M_J$ which leads to efficiencies in numerical calculations. We select $M_J = 1/2$ to solve for the proton spectroscopy.

Apart from restricting the Fock space, a further truncation is necessary to limit the basis size within each Fock sector. With our chosen basis, further truncation can be achieved by specifying two basis parameters $(K)$ and $(N_{\rm max})$. The former is conserved by the effective Hamiltonian, which is held fixed and controls the basis in the longitudinal direction. Meanwhile, $N_{\rm max}$ limits a total transverse quantum number $N_\alpha = \sum_l (2 n_l + \vert m_l \vert + 1)$ for multi-particle basis state $\ket{\alpha}$ such that $N_\alpha \leq N_{\rm max}$. This parameter acts as an ultaviolet (UV) and an infrared (IR) regulator for the LFWFs with $\Lambda_{\rm UV} \approx b \sqrt{N_{\rm max}}$ and $\Lambda_{\rm IR} \approx b/\sqrt{N_{\rm max}}$ respectively~\cite{Zhao:2014xaa}.


The resulting LFWFs in momentum space are expressed as
\begin{equation}
\Psi^\Lambda_{\lbrace x_i, {\bf k}_{\perp i}, \lambda_i \rbrace} = \bra{P,\Lambda} \lbrace x_i, {\bf k}_{\perp i}, \lambda_i \rbrace \rangle = \sum_{\lbrace n_i, m_i \rbrace} \left( \psi^{\Lambda}_{\lbrace x_i,n_i,m_i,\lambda_i \rbrace} \prod_i \phi_{n_i,m_i} ({\bf k}_{\perp i };b) \right) \;,
\end{equation}
with $\psi^{\Lambda}_{\lbrace x_i,n_i,m_i,\lambda_i \rbrace}= \bra{P,\Lambda} \lbrace x_i, n_i, m_i, \lambda_i \rbrace \rangle$ being the LFWF in BLFQ's chosen basis and with $P$ and $\Lambda$ being the momentum and helicity of the target spin-1/2 composite system respectively. 


To produce our LFWFs, the basis truncation parameters in the transverse and longitudinal directions are taken as $N_{\rm max}=10$ and $K=16.5$ respectively~\cite{Mondal:2019jdg, Xu:2021wwj}. Besides this, other model parameters are fixed in a way that they provide the known nucleon mass and the electromagnetic form factors~\cite{Mondal:2019jdg, Xu:2021wwj}, leading us to adopt
$\lbrace m_{q/{\rm K.E.}},m_{q/{\rm OGE}}, \kappa, \alpha_s \rbrace = \lbrace 0.3~{\rm GeV}, 0.2~{\rm GeV}, 0.34~{\rm GeV}, 1.1 \pm 0.1 \rbrace $ where $m_{q/{\rm K.E.}}$ and $m_{q/{\rm OGE}}$ represent the quark masses in kinetic energy and OGE interaction terms with the HO scale parameter being $b=0.6$ GeV. The calculated LFWFs for the valence Fock sector using these parameters, which imply a model scale $\mu^2_0=0.195 \pm 0.020$ GeV$^2$~\cite{Mondal:2019jdg, Xu:2021wwj}, are employed to provide physical observables and distribution functions of the valence quarks inside the proton. In this work, we specifically study the proton GPDs using these LFWFs. 
\section{Generalized Parton Distributions (GPDs)}
The 3-d spatial distributions are categorized as chiral-even and chiral-odd GPDs and are defined through the non-forward matrix elements of the bilocal operators between hadronic states. The connection between the correlator functions and various GPDs strictly depends upon the bilocal operator. 
Note that, for the present work, we restrict ourselves to the Dokshitzer-Gribov-Lipatov-Altarelli-Parisi (DGLAP) region, $\zeta<x<1$, where the number of partons in the initial and the final states remains conserved. Also, the momentum transfer in the longitudinal direction is taken to be zero. Accordingly, these distributions are classified and parameterized as we now show.

The distributions where the quark does not transfer helicity are parameterized as~\cite{Diehl:2003ny}
\begin{align}
\int \frac{{\rm d}z^-}{8\pi} e^{\iota x P^+ z^-/2} \bra{P^\prime, \Lambda^\prime} \bar{\varphi}(0) \gamma^+ \varphi(z^-) \ket{P,\Lambda} \vert_{z^+ = {\bf z}_\perp=0} &= \frac{1}{2\bar{P}^+} \bar{u}(P^\prime,\Lambda^\prime) \bigg(H^q(x,\zeta,t) \gamma^+  \nonumber\\
& \left. + E^q(x,\zeta,t) \frac{\iota \sigma^{+ \alpha}\Delta_{\alpha}}{2M_P} \right) u(P, \Lambda)\;, \\
\int \frac{{\rm d}z^-}{8\pi} e^{\iota x P^+ z^-/2} \bra{P^\prime, \Lambda^\prime} \bar{\varphi}(0) \gamma^+ \gamma_5 \varphi(z^-) \ket{P,\Lambda} \vert_{z^+ = {\bf z}_\perp=0} &= \frac{1}{2\bar{P}^+} \bar{u}(P^\prime,\Lambda^\prime)\bigg(\tilde{H}^q(x,\zeta,t) \gamma^+ \gamma_5 \nonumber\\
& \left. + \tilde{E}^q(x,\zeta,t) \frac{\gamma_5 \Delta^+}{2M_P} \right)u(P, \Lambda)\;.
\end{align}
On the other hand, the transversity distributions where the quark transfers helicity are parameterized as~\cite{Diehl:2003ny} 
\begin{align}
\int \frac{{\rm d}z^-}{8\pi} & e^{\iota x P^+ z^-/2} \bra{P^\prime, \Lambda^\prime} \bar{\varphi}(0)  \sigma^{+j} \gamma_5 \varphi(z^-) \ket{P,\Lambda} \vert_{z^+ = {\bf z}_\perp=0} = \frac{1}{2\bar{P}^+} \bar{u}(P^\prime,\Lambda^\prime) \bigg(H_T^q(x,\zeta,t) \sigma^{+j} \gamma_5 \nonumber\\
& \left. +\tilde{H}_T^q (x,\zeta,t) \frac{\epsilon^{+j\alpha \beta} \Delta_\alpha \bar{P}_\beta}{M_P^2} + E_T^q (x,\zeta,t) \frac{\epsilon^{+j\alpha \beta} \Delta_\alpha \gamma_\beta}{2M_P} + \tilde{E}_T^q (x,\zeta,t) \frac{\epsilon^{+j\alpha \beta} \bar{P}_\alpha \gamma_\beta}{M_P}  \right) u(P,\Lambda)\;,
\end{align}
with $M_P$ being the mass of the target spin-$1/2$ composite system which is the proton in our case.  
We choose a frame such that the momenta of the target proton at the initial and final state, at $\zeta=0$, become 
\begin{align}
P &= \left(P^+, \frac{M_P^2}{P^+} ,{\bf 0}_\perp \right) \;,\\
P^{\prime} &= \left(P^+,\frac{M_P^2+\mathbf{\Delta}_\perp^2}{P^+} ,- \mathbf{\Delta}_\perp \right)\;.
\end{align}
The overlap representation of the GPDs in terms of the LFWFs for $\zeta=0$ are expressed as 
\begin{align}
H^q(x,0,t) &= \sum_{\lbrace \lambda_i \rbrace} \int \left[{\rm d}\mathcal{X}{\rm d}\mathcal{K}_\perp \right] \Psi^{\uparrow *}_{\lbrace x^\prime_i, {\bf k}_{\perp i}^\prime , \lambda_i \rbrace} \Psi^{\uparrow}_{\lbrace x_i, {\bf k}_{\perp i} , \lambda_i \rbrace} \delta(x-x_1)\;, \\
E^q(x,0,t) &= -\frac{2M}{(\Delta^1 - \iota \Delta^2)} \sum_{\lbrace \lambda_i \rbrace} \int \left[{\rm d}\mathcal{X}{\rm d}\mathcal{K}_\perp \right] \Psi^{\uparrow *}_{\lbrace x^\prime_i, {\bf k}_{\perp i}^\prime , \lambda_i \rbrace} \Psi^{\downarrow}_{\lbrace x_i, {\bf k}_{\perp i} , \lambda_i \rbrace} \delta(x-x_1)\;,\\
\tilde{H}^q(x,0,t) &= \sum_{\lbrace \lambda_i \rbrace} \int \left[{\rm d}\mathcal{X}{\rm d}\mathcal{K}_\perp \right] \lambda_1 \Psi^{\uparrow *}_{\lbrace x^\prime_i, {\bf k}_{\perp i}^\prime , \lambda_i \rbrace} \Psi^{\uparrow}_{\lbrace x_i, {\bf k}_{\perp i} , \lambda_i \rbrace} \delta(x-x_1)\;,\\
 E_T^q(x,0,t) &+ 2 \tilde{H}_T^q(x,0,t) = \sum_{\lbrace \lambda^\prime_i, \lambda_i \rbrace} \int \left[{\rm d}\mathcal{X}{\rm d}\mathcal{K}_\perp \right] \Psi^{\uparrow *}_{\lbrace x^\prime_i, {\bf k}_{\perp i}^\prime , \lambda^\prime_i \rbrace} \Psi^{\uparrow}_{\lbrace x_i, {\bf k}_{\perp i} , \lambda_i \rbrace} \delta(x-x_1)\;, \\
H_T^q(x,0,t) &= \sum_{\lbrace \lambda^\prime_i, \lambda_i \rbrace} \int \left[{\rm d}\mathcal{X}{\rm d}\mathcal{K}_\perp \right] \Psi^{\uparrow *}_{\lbrace x^\prime_i, {\bf k}_{\perp i}^\prime , \lambda^\prime_i \rbrace} \Psi^{\downarrow}_{\lbrace x_i, {\bf k}_{\perp i} , \lambda_i \rbrace} \delta(x-x_1)\;, \\
\tilde{H}^q_T(x,0,t) & = - \sum_{\lbrace \lambda_i^\prime, \lambda_i \rbrace} \int \left[{\rm d}\mathcal{X}{\rm d}\mathcal{K}_\perp \right] \Psi^{\downarrow *}_{\lbrace x^\prime_i, {\bf k}_{\perp i}^\prime , \lambda^\prime_i \rbrace} \Psi^{\uparrow}_{\lbrace x_i, {\bf k}_{\perp i} , \lambda_i \rbrace} \, \delta_{\lambda_i, -\lambda^\prime_i} \, \delta(x-x_1) \;,
\end{align}
where
\begin{align}
\left[{\rm d}\mathcal{X}{\rm d}\mathcal{K}_\perp \right] = \prod_{i=1}^3 \frac{{\rm d}x_i \, {\rm d}^2{\bf k}_{\perp i}}{16 \pi^3} \,16\pi^3 \, \delta\left( 1-\sum_{i=1}^3 x_i \right) \, \delta^2\left( \sum_{i=1}^3 {\bf k}_{\perp i}\right)\;,
\end{align}
 with the longitudinal momentum fraction and the transverse momentum for the active quark being $x_1^\prime=x_1$ and ${\bf k}_{\perp 1}^\prime = {\bf k}_{\perp 1} + (1-x_1) \mathbf{\Delta}_\perp$. For the spectators these momenta become $x_i^\prime = x_i$ and ${\bf k}_{\perp i}^\prime = {\bf k}_{\perp i} - x_i \mathbf{\Delta}_\perp$. Here, $t=-\mathbf{\Delta}_\perp^2$ when the skewness $\zeta=0$.
 
\begin{figure}[hbt!]
     \centering
     \begin{subfigure}[b]{0.43\textwidth}
         \centering
         \includegraphics[width=\textwidth]{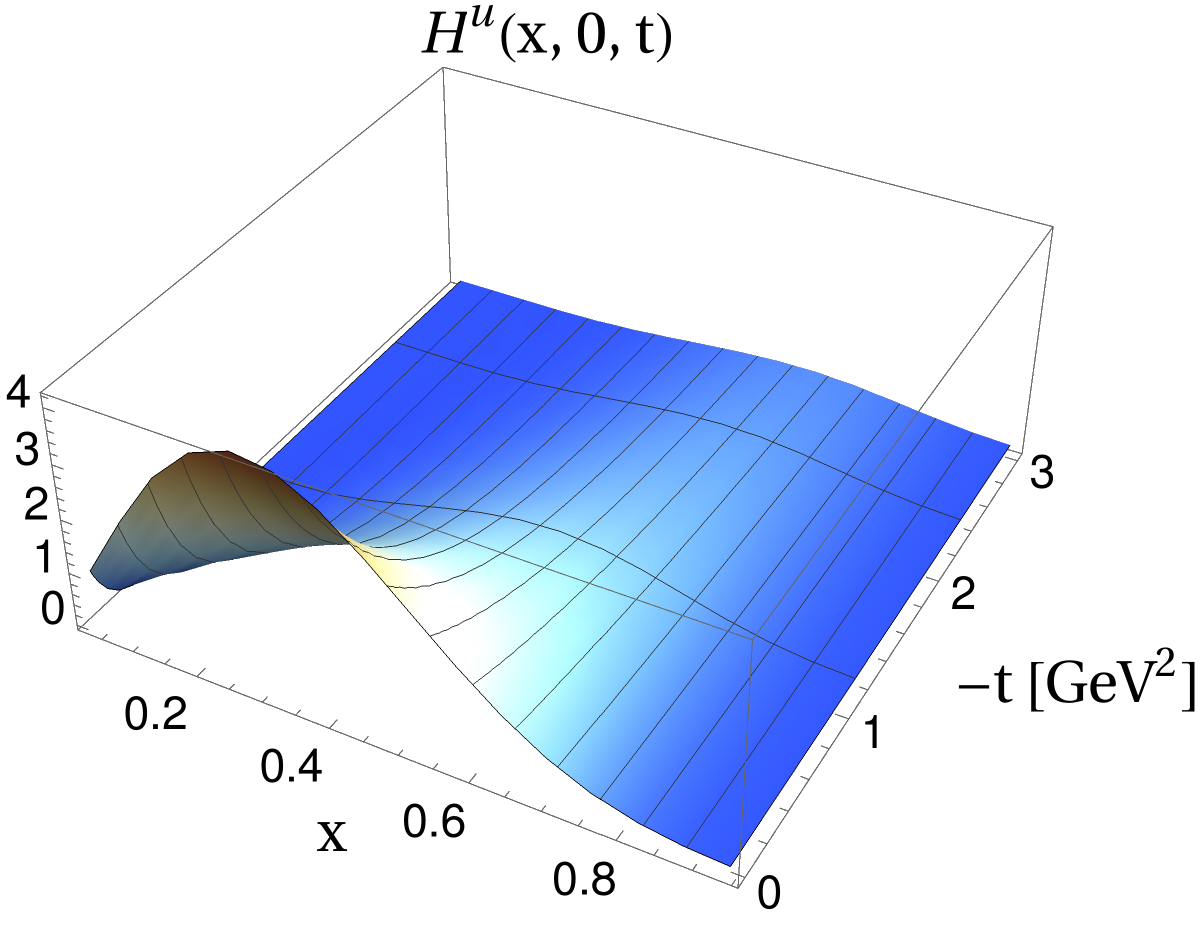}
         \caption{}
         \label{fig:H_u}
     \end{subfigure}
     \hfill
     \begin{subfigure}[b]{0.43\textwidth}
         \centering
         \includegraphics[width=\textwidth]{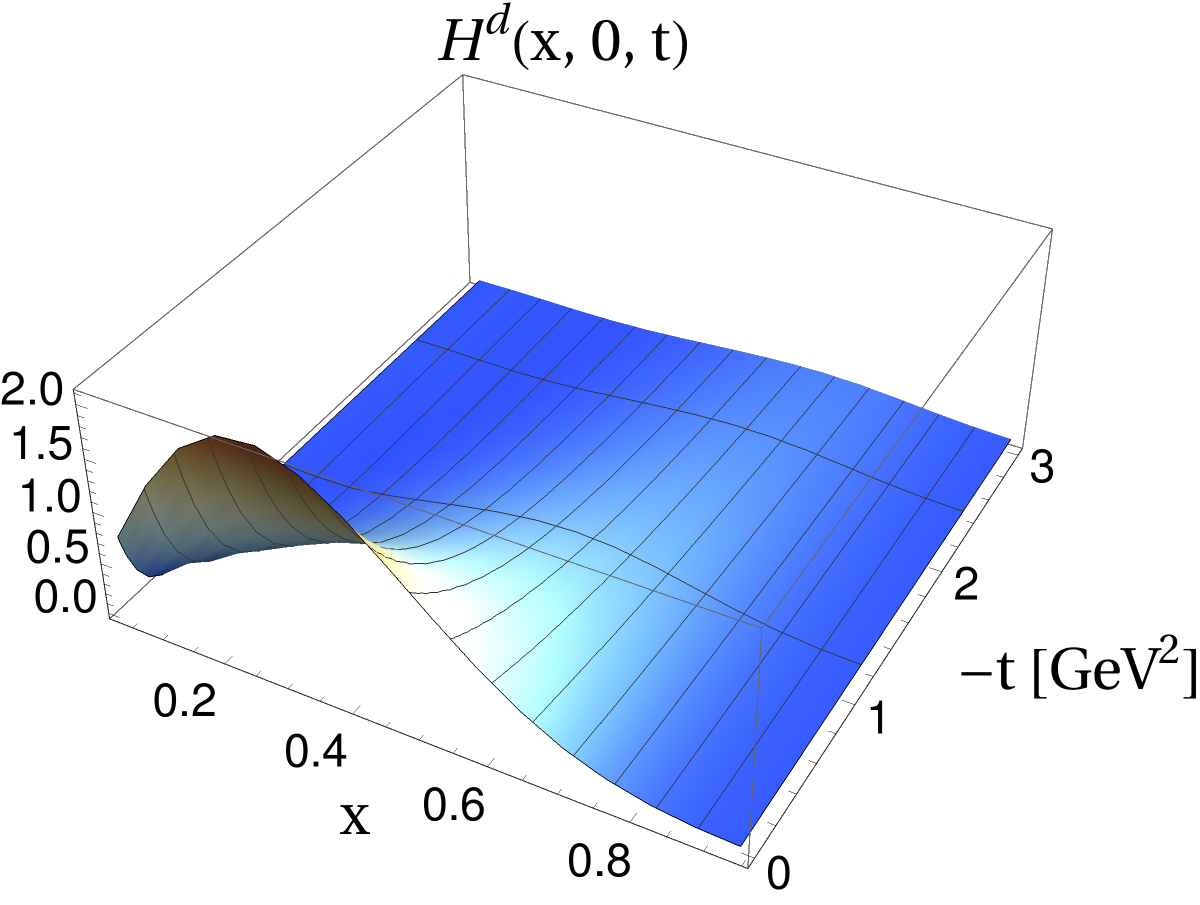}
        \caption{}
         \label{fig:H_d}
     \end{subfigure}
     \hfill
     \begin{subfigure}[b]{0.43\textwidth}
         \centering
         \includegraphics[width=\textwidth]{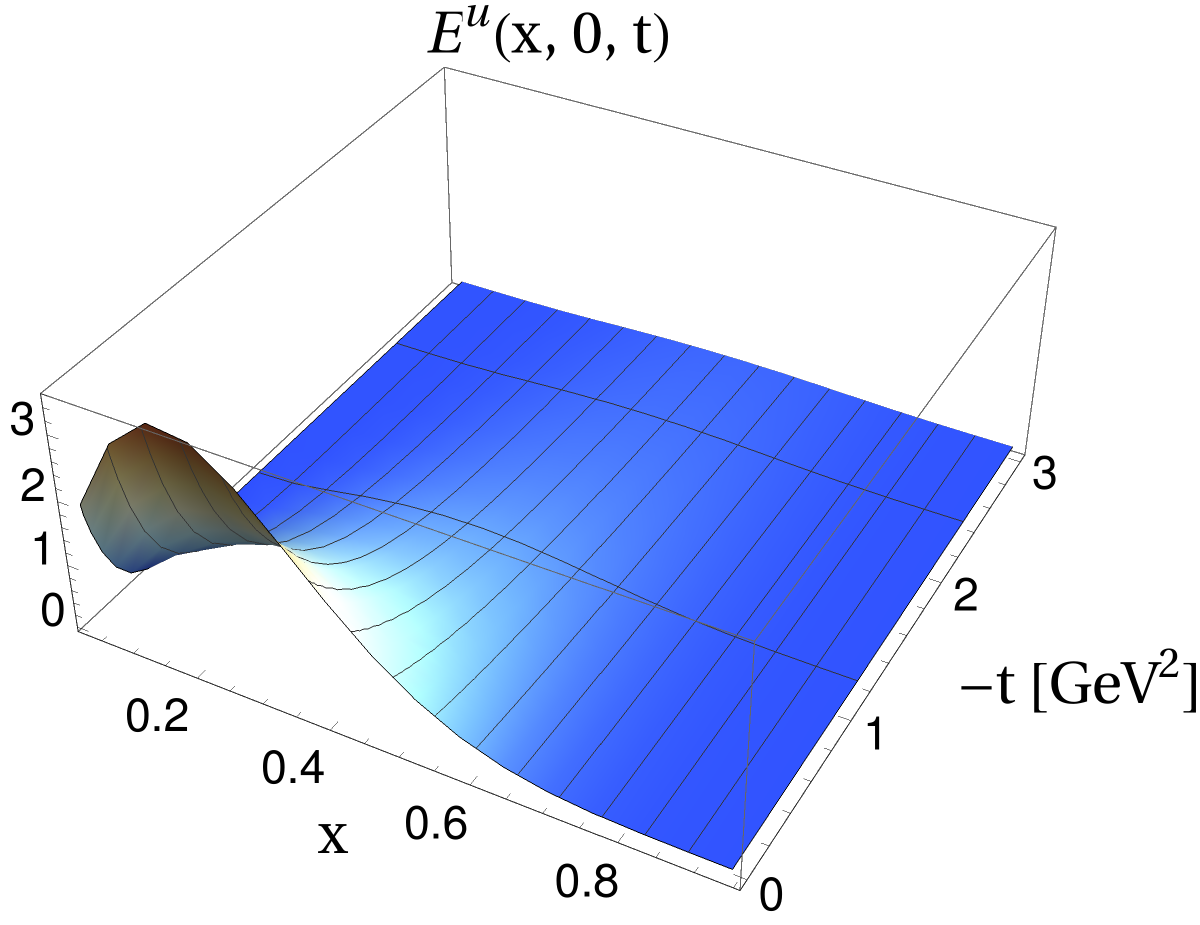}
       \caption{}
         \label{fig:E_u}
     \end{subfigure}
       \hfill
          \begin{subfigure}[b]{0.43\textwidth}
         \centering
         \includegraphics[width=\textwidth]{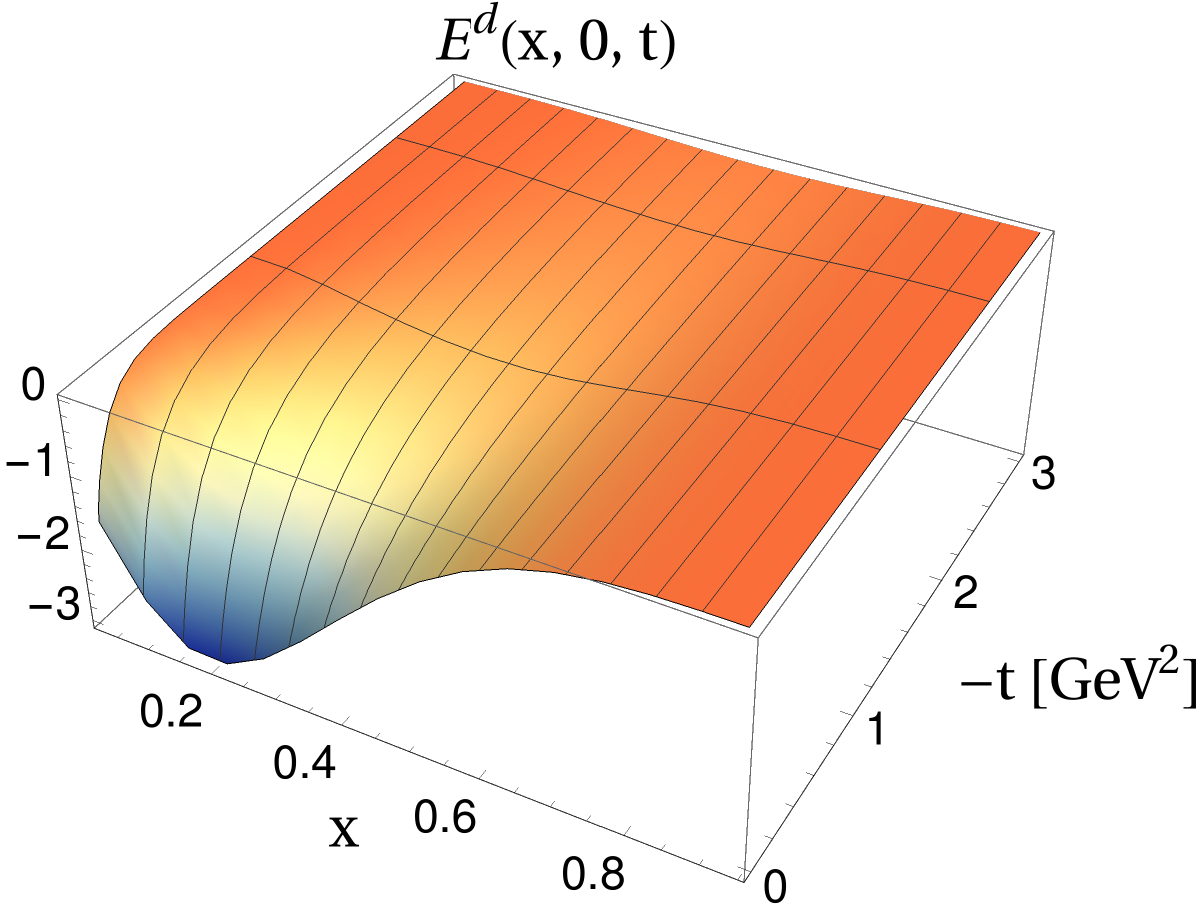}
         \caption{}
         \label{fig:E_d}
     \end{subfigure}
     \hfill
     \begin{subfigure}[b]{0.43\textwidth}
         \centering
         \includegraphics[width=\textwidth]{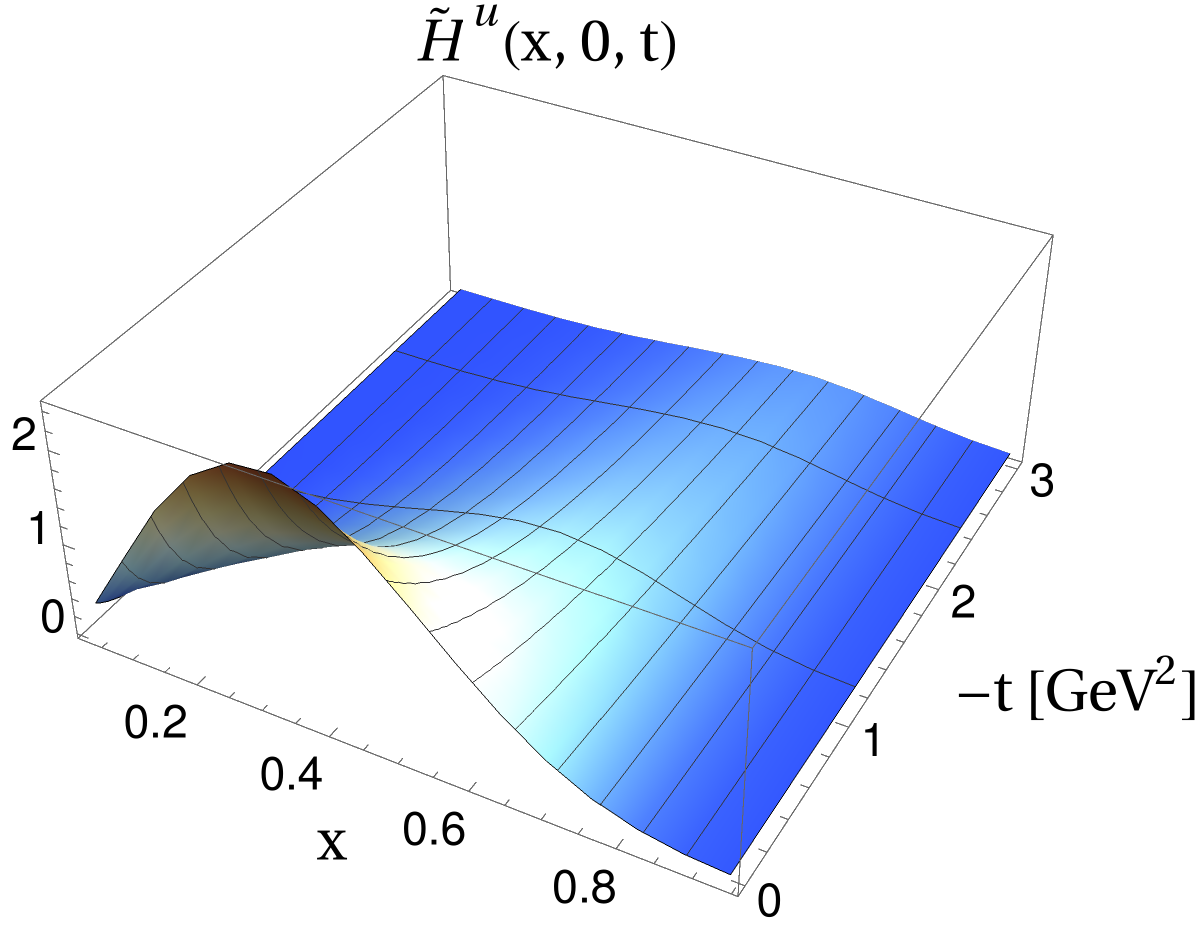}
      \caption{}
         \label{fig:Htilde_u}
     \end{subfigure}
     \hfill
     \begin{subfigure}[b]{0.43\textwidth}
         \centering
         \includegraphics[width=\textwidth]{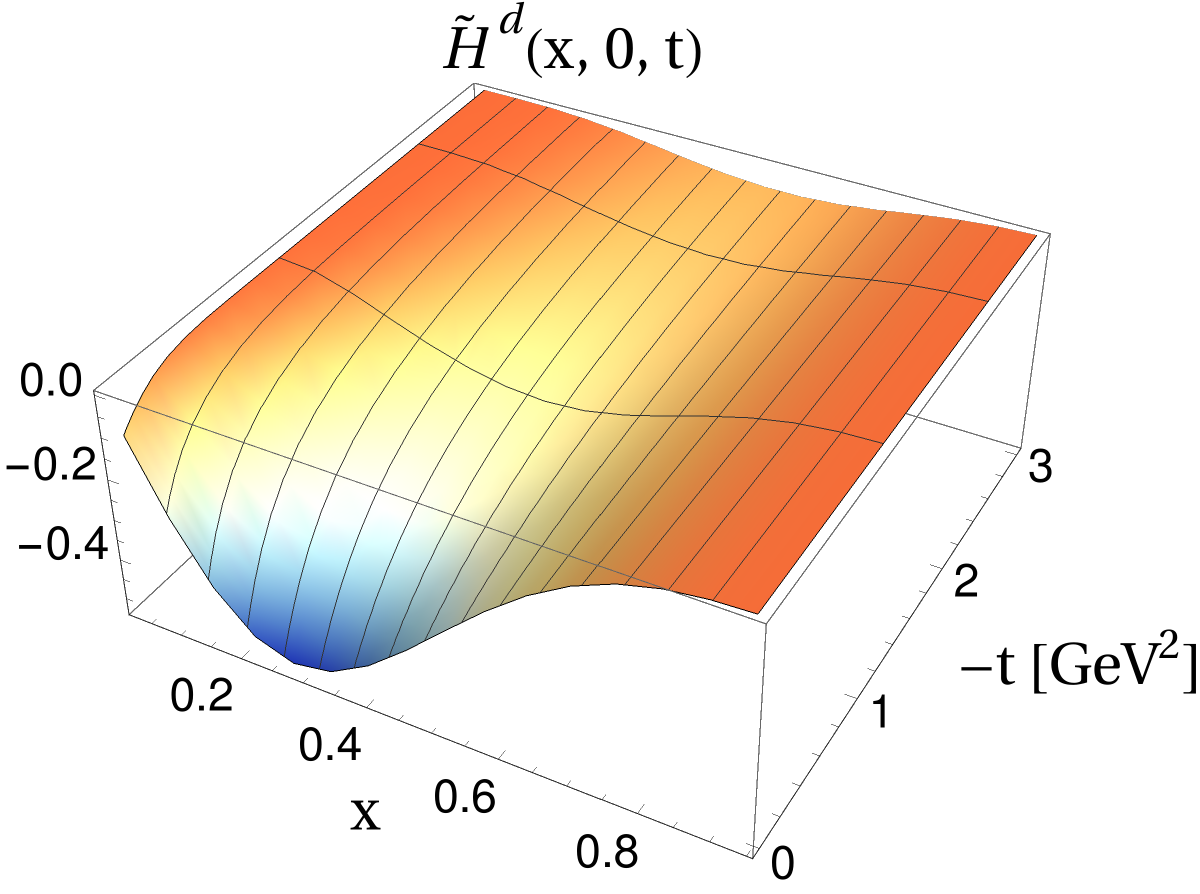}
        \caption{}
         \label{fig:Htilde_d}
     \end{subfigure}
        \caption{The chiral-even GPDs: (a) $H(x,0,t)$, (c) $E(x,0,t)$ and (e) $\tilde{H}(x,0,t)$ for the $u$-quark, where the respective GPDs for the $d$-quark are shown in (b), (d) and (f). The GPDs are presented with respect to $x$ and $-t$ (in GeV$^2$).}
        \label{fig:chiral_even_GPDs}
\end{figure}

We show the results of the chiral-even GPDs $(H,E,\tilde{H})$ for the valence quarks in Fig.~\ref{fig:chiral_even_GPDs}, where the distribution functions are plotted with respect to the light-cone momentum $(x)$ and the square of the total momentum transferred to the final proton state $(-t)$. The GPDs for $u$ and $d$ quarks are presented in the left and the right panels of Fig.~\ref{fig:chiral_even_GPDs} respectively. We find that the distributions have their maxima when the proton does not transfer transverse momentum to its final state and the struck quark inside the proton carries less than $50\%$ of the proton's longitudinal momentum.
As expected, by increasing the momentum transfer in the transverse direction, the distribution peak shifts gradually towards the higher values of $x$ accompanied by a continuous drop in the magnitude. At the large $x$-region, all the distributions eventually decay and become independent of $t$. However, this decay is observed to be faster for $E^q$ than the other GPDs ($H^q$ and $\tilde{H}^q$). As the anomalous magnetic moment and the axial charge are measured to be negative for the $d$-quark, the connected GPDs, $E^d$ and $\tilde{H}^d$, are correspondingly negative. All the mentioned features appear to be model-independent as they have also been observed in other QCD inspired models~\cite{Ji:1997gm, Boffi:2002yy, Boffi:2003yj, Mondal:2015uha, Mondal:2017wbf, Freese:2020mcx}.

When there is no momentum transfer $(t=0)$, these distributions reproduce the valence quark distribution functions, particularly, the unpolarized and helicity-dependent functions, i.e., $H^q(x,0,0)=f^q(x)$ and $\tilde{H}^q(x,0,0)=g^q(x)$. These 1-d functions have been studied previously in the BLFQ approach by considering  both the Fock sector containing valence quarks ($\ket{qqq}$)~\cite{Mondal:2019jdg, Xu:2021wwj} and one beyond this sector ($\ket{qqqg}$)~\cite{Xu:2022abw}.
Additionally, the detailed interpretation of the moments, which are functions of $t$, is given below in Section~\ref{Mellin-moments}.

\begin{figure}[hbt!]
     \centering
     \begin{subfigure}[b]{0.43\textwidth}
         \centering
         \includegraphics[width=\textwidth]{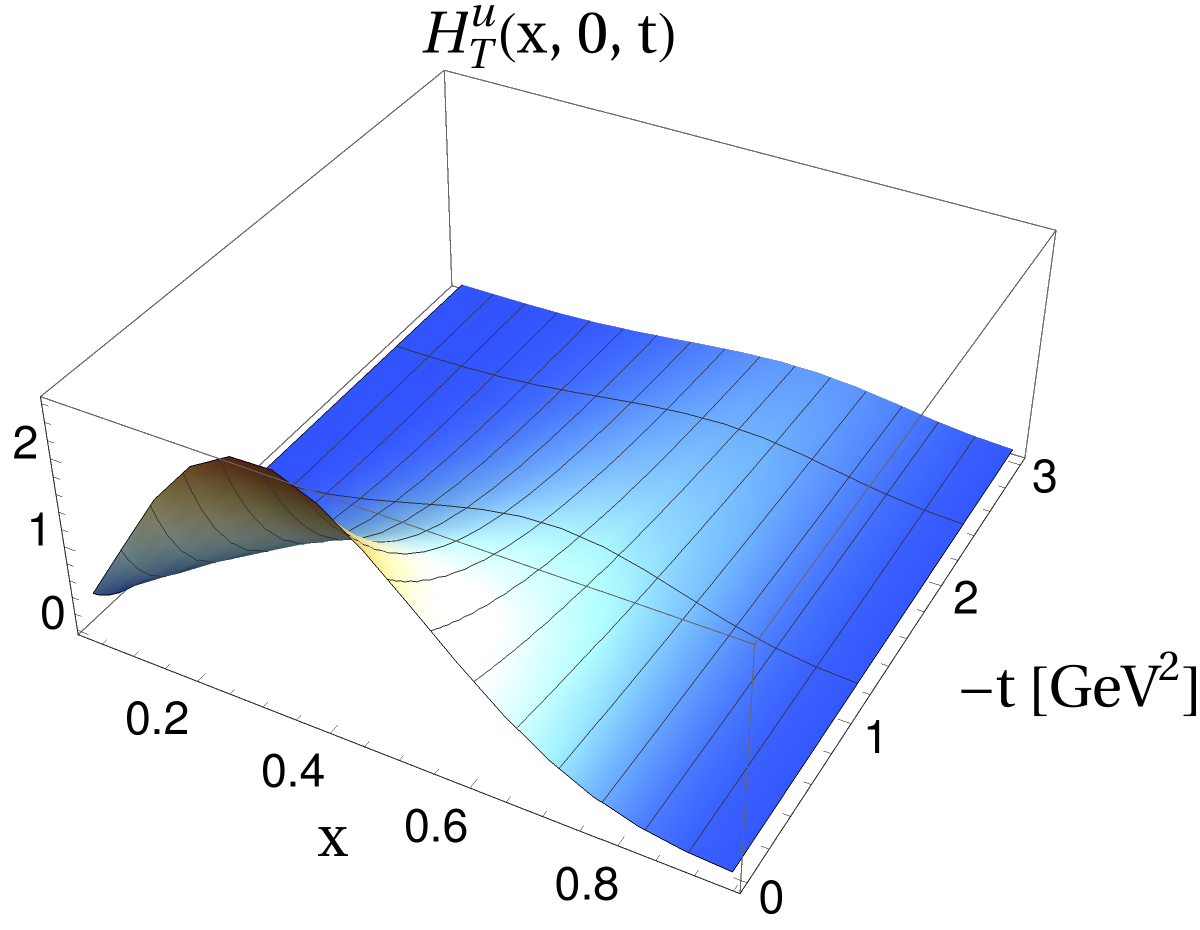}
         \caption{}
         \label{fig:HT_u}
     \end{subfigure}
     \begin{subfigure}[b]{0.43\textwidth}
         \centering
         \includegraphics[width=\textwidth]{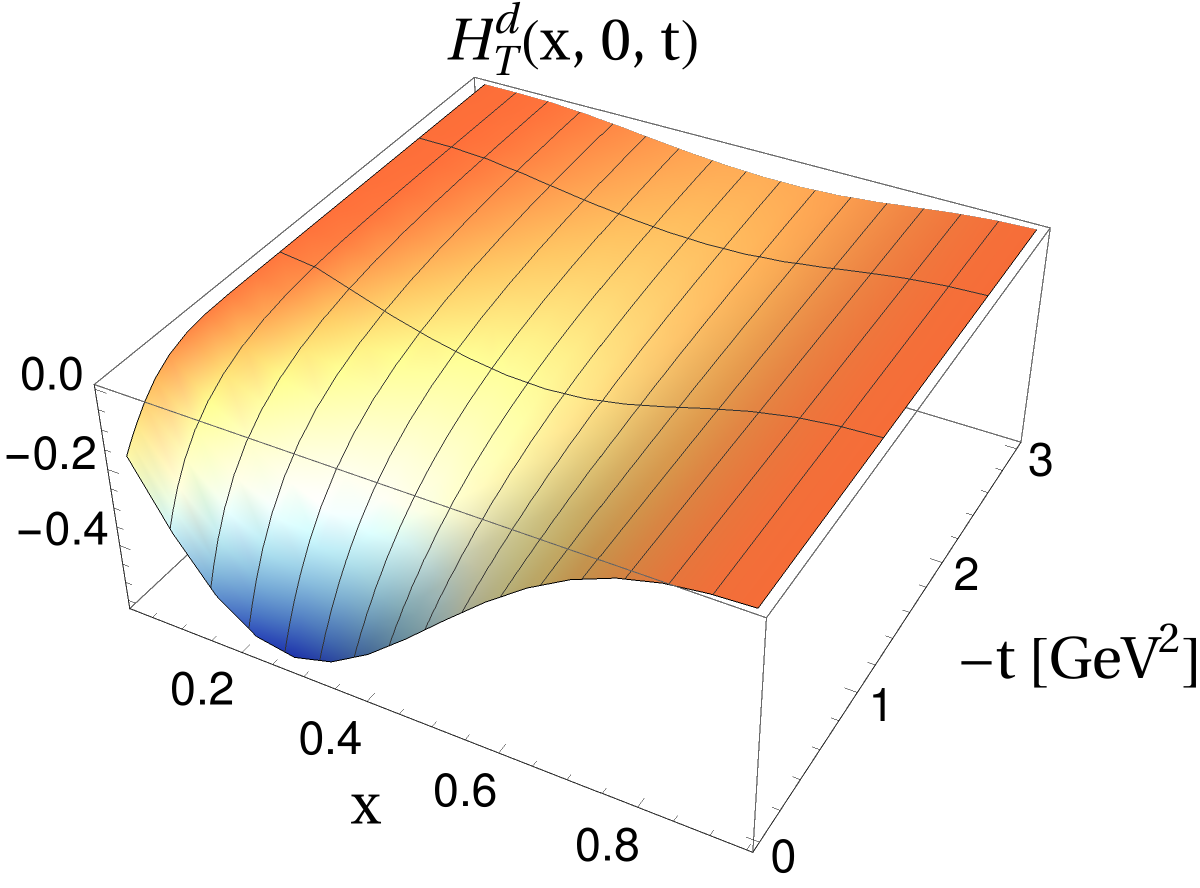}
        \caption{}
         \label{fig:HT_d}
     \end{subfigure}
     \begin{subfigure}[b]{0.43\textwidth}
         \centering
         \includegraphics[width=\textwidth]{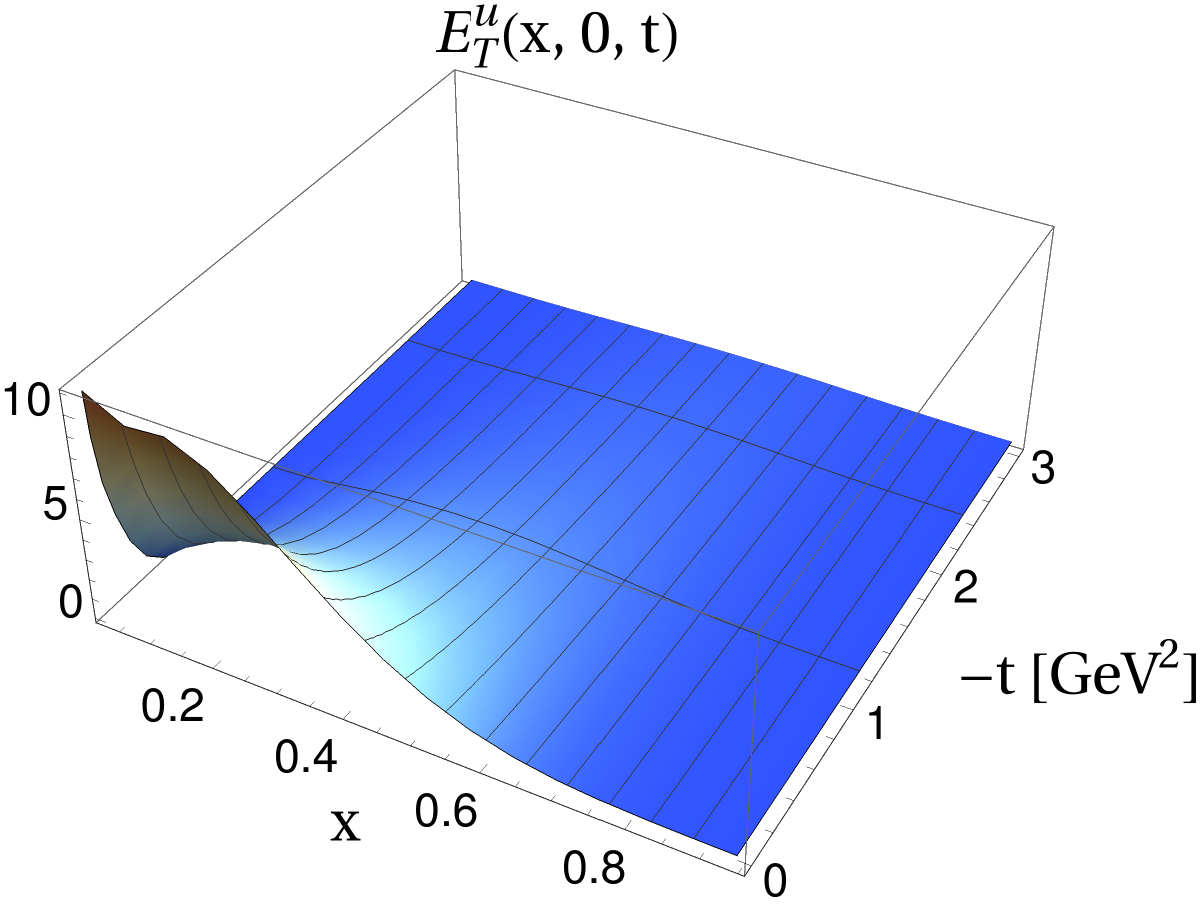}
       \caption{}
         \label{fig:ET_u}
     \end{subfigure}
          \begin{subfigure}[b]{0.43\textwidth}
         \centering
         \includegraphics[width=\textwidth]{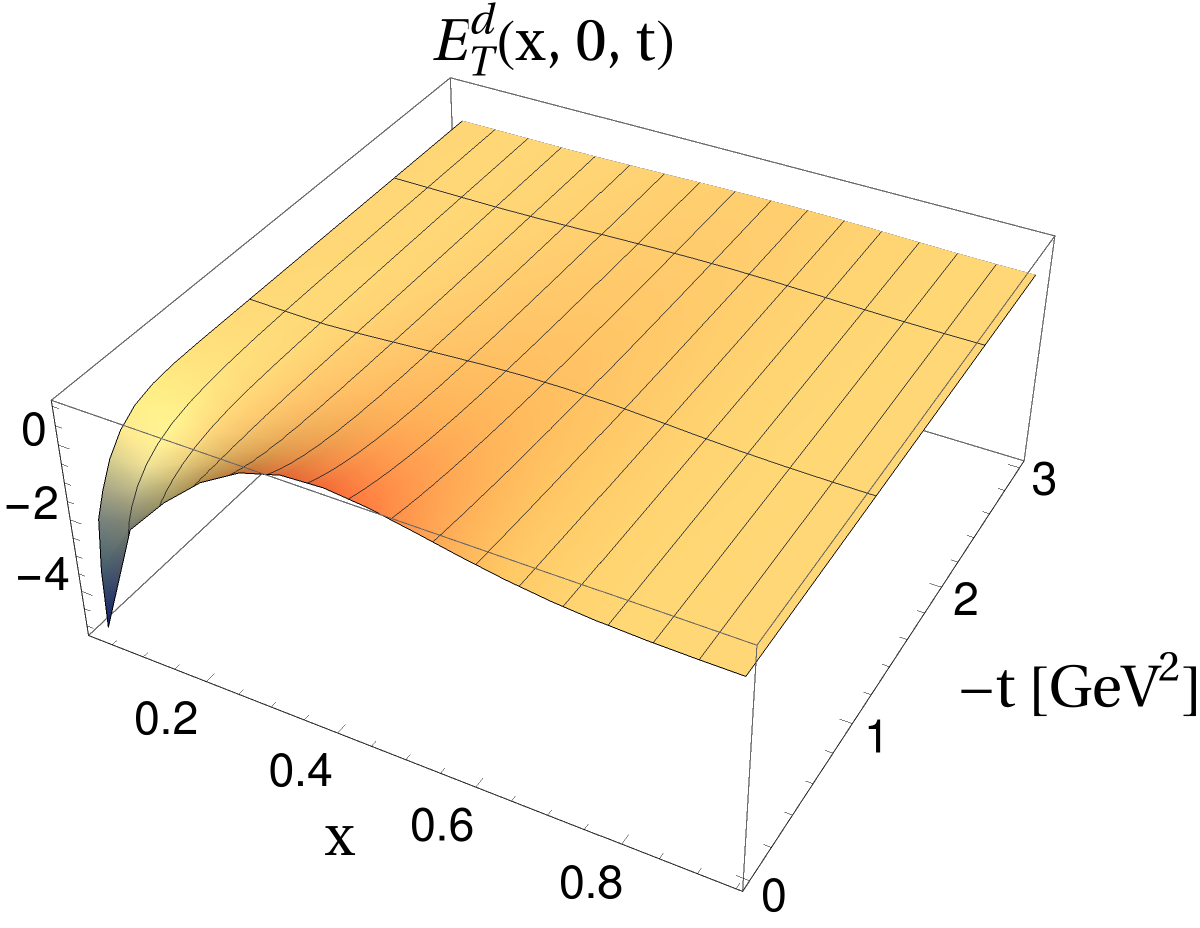}
         \caption{}
         \label{fig:ET_d}
     \end{subfigure}
     \begin{subfigure}[b]{0.43\textwidth}
         \centering
         \includegraphics[width=\textwidth]{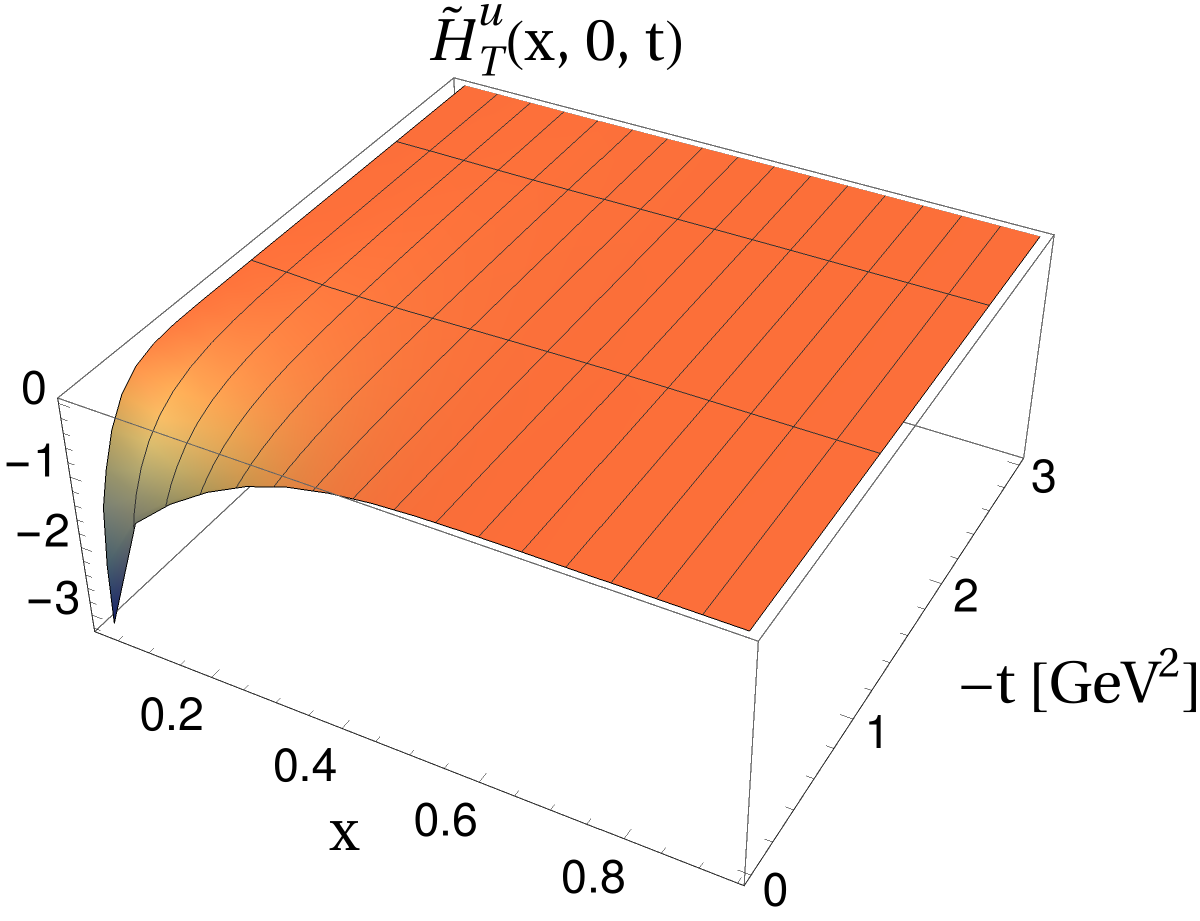}
      \caption{}
         \label{fig:HTtilde_u}
     \end{subfigure}
     \begin{subfigure}[b]{0.43\textwidth}
         \centering
         \includegraphics[width=\textwidth]{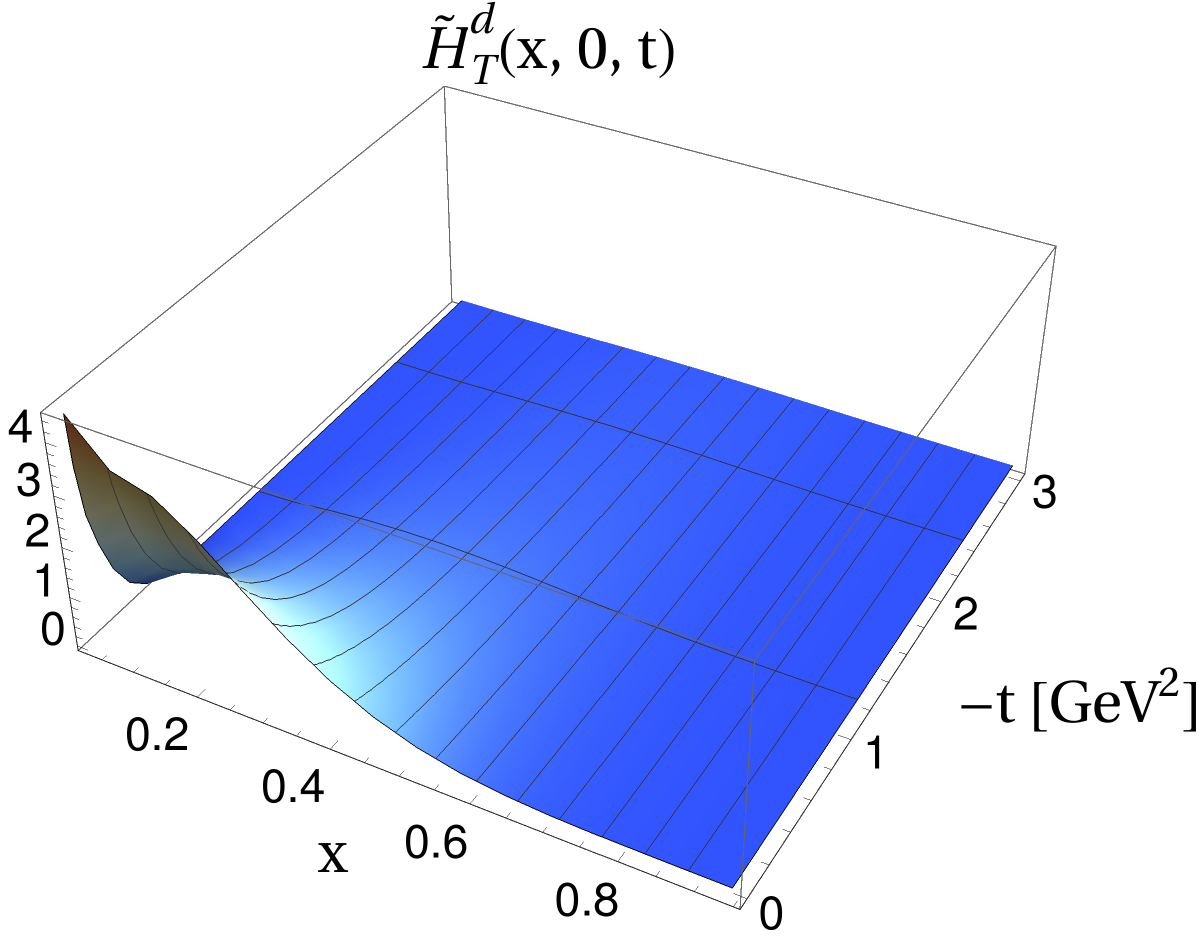}
        \caption{}
         \label{fig:HTtilde_d}
     \end{subfigure}
        \caption{The chiral-odd GPDs (a) $H_T(x,0,t)$, (c) $E_T(x,0,t)$ and (e) $\tilde{H}_T(x,0,t)$ for the $u$-quark, where the respective GPDs for the $d$-quark are shown in (b), (d) and (f). The GPDs are presented with respect to $x$ and $-t$ (in GeV$^2$).}
        \label{fig:chiral_odd_GPDs}
\end{figure}

The 3-d graphical representations of the chiral-odd proton GPDs $(H_T, E_T, \tilde{H}_T)$ for the $u$-quark and the $d$-quark are shown in the left and right panels of Fig.~\ref{fig:chiral_odd_GPDs} respectively. Similar to the helicity conserving GPDs, these helicity non-conserving distribution functons are illustrated as functions of $x$ and $-t$. 
All the helicity flip distributions show similar behavior as for the case of helicity non-flip GPDs, except the behavior of $E^q_T$ and $\tilde{H}^q_T$ in the small $x$ region. In that case, the peaks observed near $x \rightarrow 0$ are model-dependent~\cite{Pasquini:2005dk, Chakrabarti:2015ama}. All the flavor distribution peaks move along $x$, when the momentum transfer from the initial proton is increased gradually. A noteworthy distribution is the combination of two chiral-odd GPDs, $2\tilde{H}^q_T+E^q_T$ which provides the details on the angular momentum contribution at certain limits and is reducible to the tensor form factor. We observe zero crossing points in $E^d_T$ along $x$, which has also been observed in other models~\cite{Pasquini:2005dk, Chakrabarti:2015ama}. 
Our GPDs' results, for the case of $H_T^u$ and $\tilde{H}_T^u$, are observed to be opposite to that of the $d$-quark distributions.
Similar to $H^q$ and $\tilde{H}^q$, the helicity non-conserving GPD $H^q_T$ is reducible to transversity PDF, $H_T^q(x,0,0)=h^q(x)$ and has been previously studied in this approach~\cite{Mondal:2019jdg, Xu:2021wwj, Xu:2022abw}.

\begin{figure}
\begin{minipage}{1.0\textwidth}
\centering
\includegraphics[width=0.48\linewidth, height=0.178\textheight]{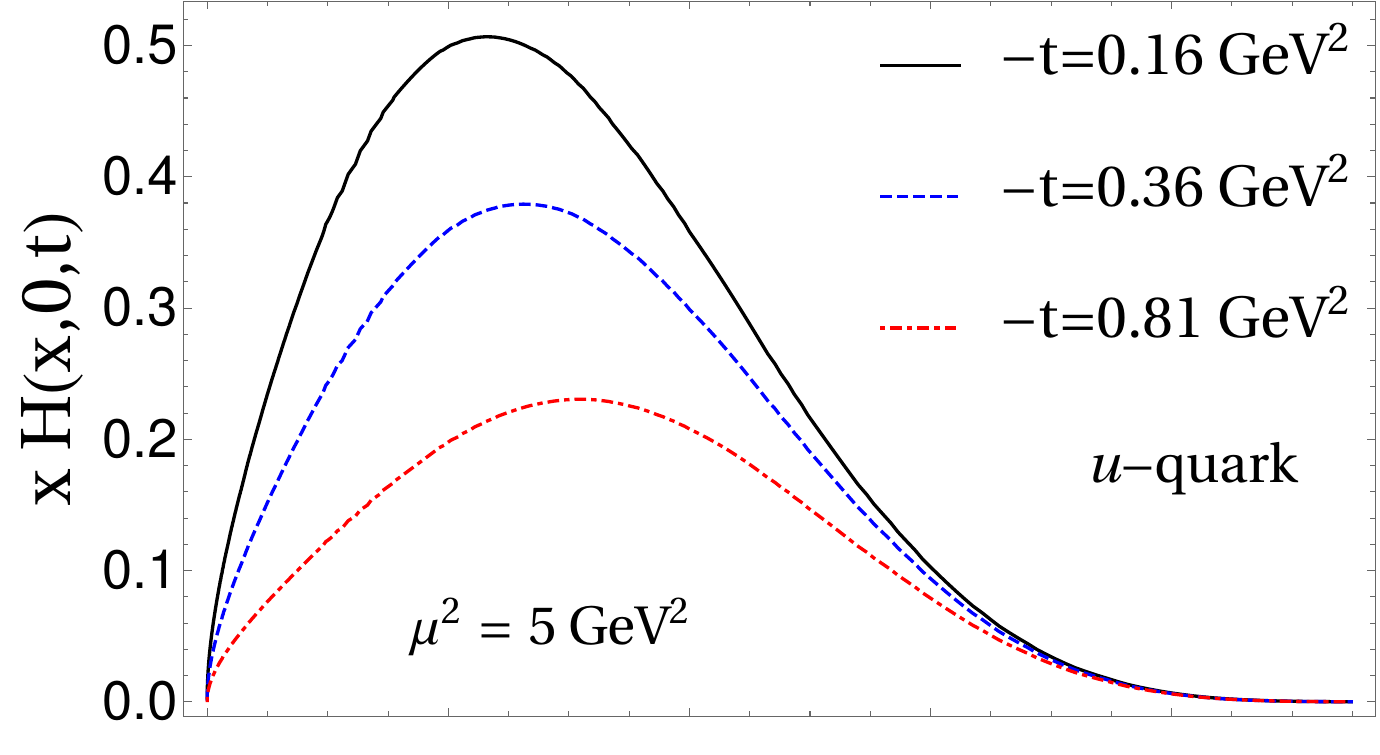}
\includegraphics[width=0.46\linewidth, height=0.178\textheight]{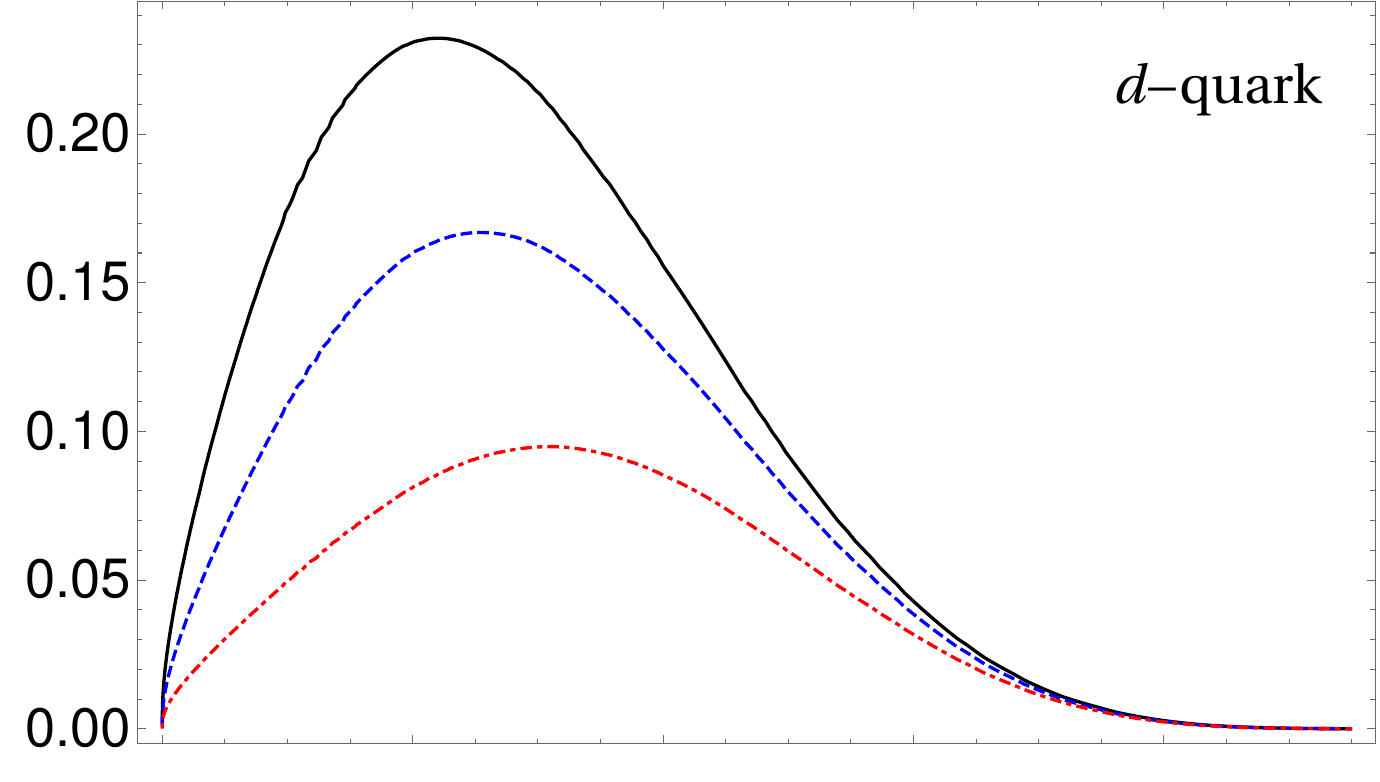}
\end{minipage}

\begin{minipage}{1.0\textwidth}
\centering
\includegraphics[width=0.482\linewidth, height=0.215\textheight]{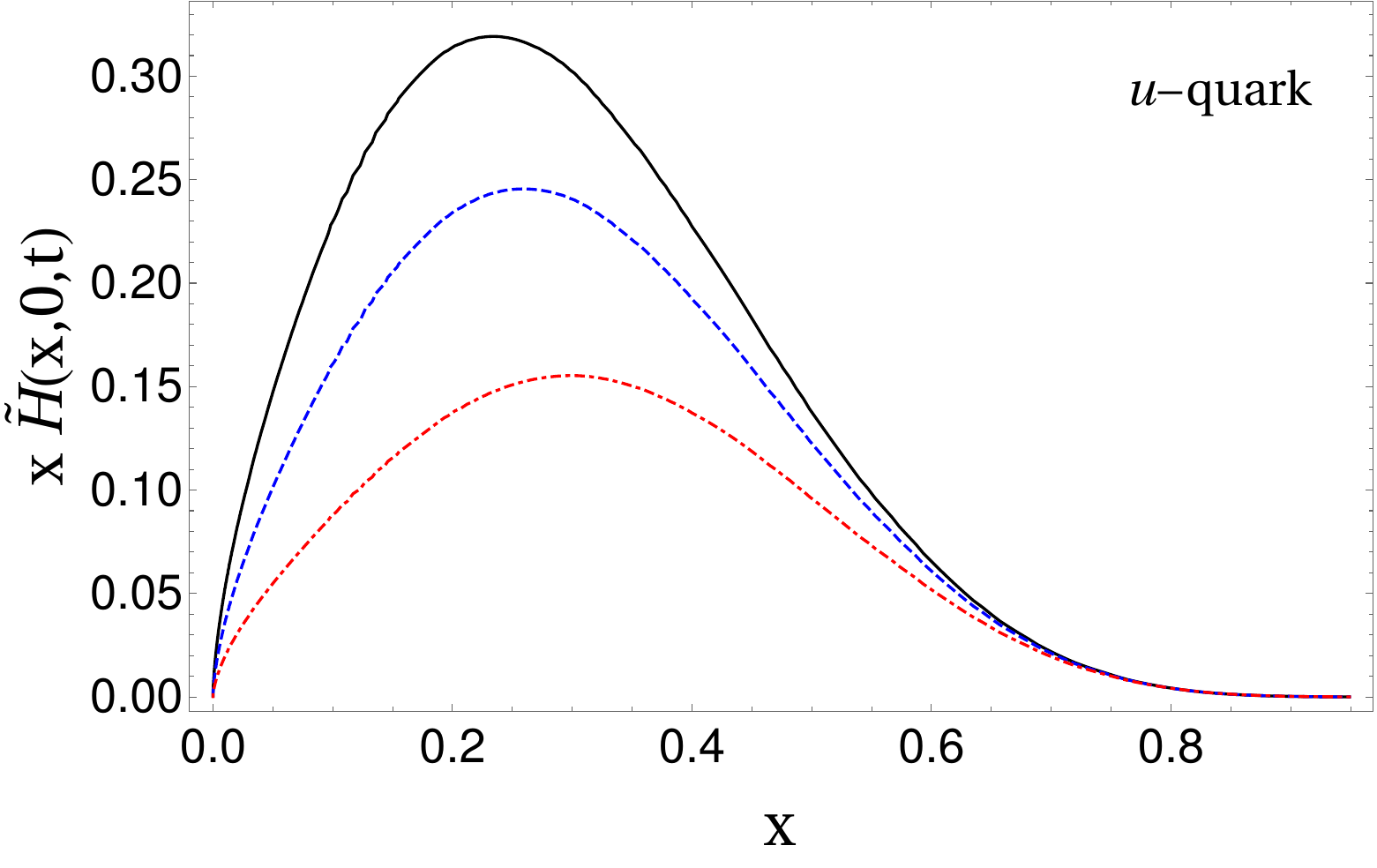}  \hspace*{-0.29cm}
\includegraphics[width=0.47\linewidth, height=0.21\textheight]{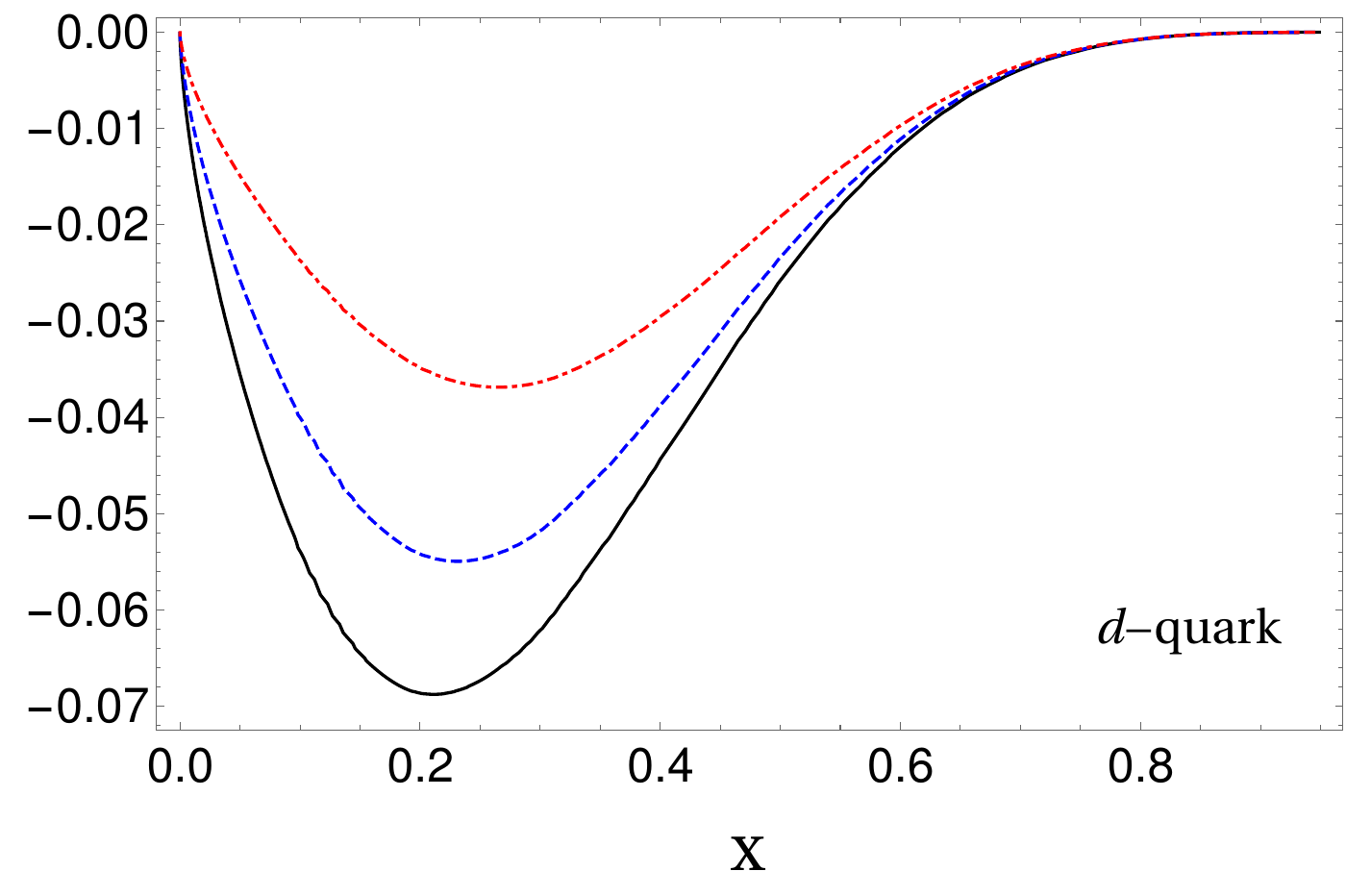}
\end{minipage}
\caption{The GPDs $H^q$ and $\tilde{H}^q$ multiplied by $x$ evolved to the scale $\mu^2=5$ GeV$^2$ with respect to $x$.}
\label{fig:evolved-xGPDs}
\end{figure}

We perform the QCD evolution to obtain the GPDs at higher scale $\mu^2$ using the next-to-next-to-leading order (NNLO) Dokshitzer-Gribov-Lipatov-Altarelli-Parisi (DGLAP) equations of QCD~\cite{Dokshitzer:1977sg, Gribov:1972ri, Altarelli:1977zs, Belitsky:1997pc, Freund:2001bf,Scopetta:2003et, Diehl:2004cx, Diehl:2007zu, Maji:2017ill}. To solve these equations numerically, we use the Higher Order Perturbative Parton Evolution (HOPPET) toolkit~\cite{Salam:2008qg}. We show the evolved chiral-even GPDs, unpolarized GPD $H^q$ and helicity GPD $\tilde{H}^q$, for both $u$ and $d$ quarks at different values of momentum transferred ($t$) in Fig.~\ref{fig:evolved-xGPDs}. The evolution is performed from the initial scale $\mu^2_0=0.195$ GeV$^2$ to $\mu^2=5$ GeV$^2$. The method is not well-established for the evolution of transversity GPD, hence we refrain from evaluating this at the higher scale.

\subsection{Mellin moments of GPDs}
\label{Mellin-moments}
For zero skewness $(\zeta=0)$, the moments of the valence quark GPDs are defined as
\begin{equation}
{\rm [Mellin-moment]}_{n0}^q(t) = \int_0^1 {\rm d}x \, x^{n-1} \, {\rm [GPD]}^q(x,0,t) \;,
\label{Eq::Mellin-moment}
\end{equation}
where $n=1,2,3,...$ represent first, second, third moments and so on. The first moments of GPDs provide form factors depending upon the helicity configurations of the active quark and the proton. Specifically, the first moments of the unpolarized GPDs, $H^q(x,0,t)$ and $E^q(x,0,t)$, provide the Dirac and Pauli form factors, $F_1^q(t)$ and $F_2^q(t)$, respectively. The helicity-dependent GPDs, $\tilde{H}^q(x,0,t)$ and $\tilde{E}^q(x,0,t)$, give the axial-vector form factor $G_A^q(t)$ and the pseudo-scalar form factor $G_P^q(t)$ respectively. Lastly, the tensor form factors $g_T^q(t)$ and $\kappa_T^q(t)$ are provided by the chiral-odd GPDs.
Mathematically,
\begin{align}
F^q_1(t) &= A^q_{10}(t)=\int {\rm d}x H^q(x,0,t) \hspace{0.5cm},\hspace{0.5cm} F^q_2(t) = B^q_{10}(t)= \int {\rm d}x E^q(x,0,t)\;, \label{Eq:Dirac_and_Pauli_FF} \\
G^q_A(t) &= \tilde{A}^q_{10}(t)=\int {\rm d}x \tilde{H}^q(x,0,t) \hspace{0.5cm},\hspace{0.5cm} G^q_P(t) = \tilde{B}^q_{10}(t)=\int {\rm d}x \tilde{E}^q(x,0,t)\;,\label{Eq:Axial_FF} \\
g^q_T(t)&=A^q_{T10}(t) = \int {\rm d}x H^q_T(x,0,t) \hspace{0.5cm},\hspace{0.5cm} \kappa^q_T(t)=\bar{B}^q_{T10}(t) = \int {\rm d}x [E^q_T(x,0,t)+2\tilde{H}^q_T(x,0,t)] \;, \label{Eq:tensor_FF}
\end{align}

%
\begin{figure}[hbt!]
        \centering
     \begin{subfigure}[b]{0.49\textwidth}
         \centering
         \includegraphics[width=\textwidth]{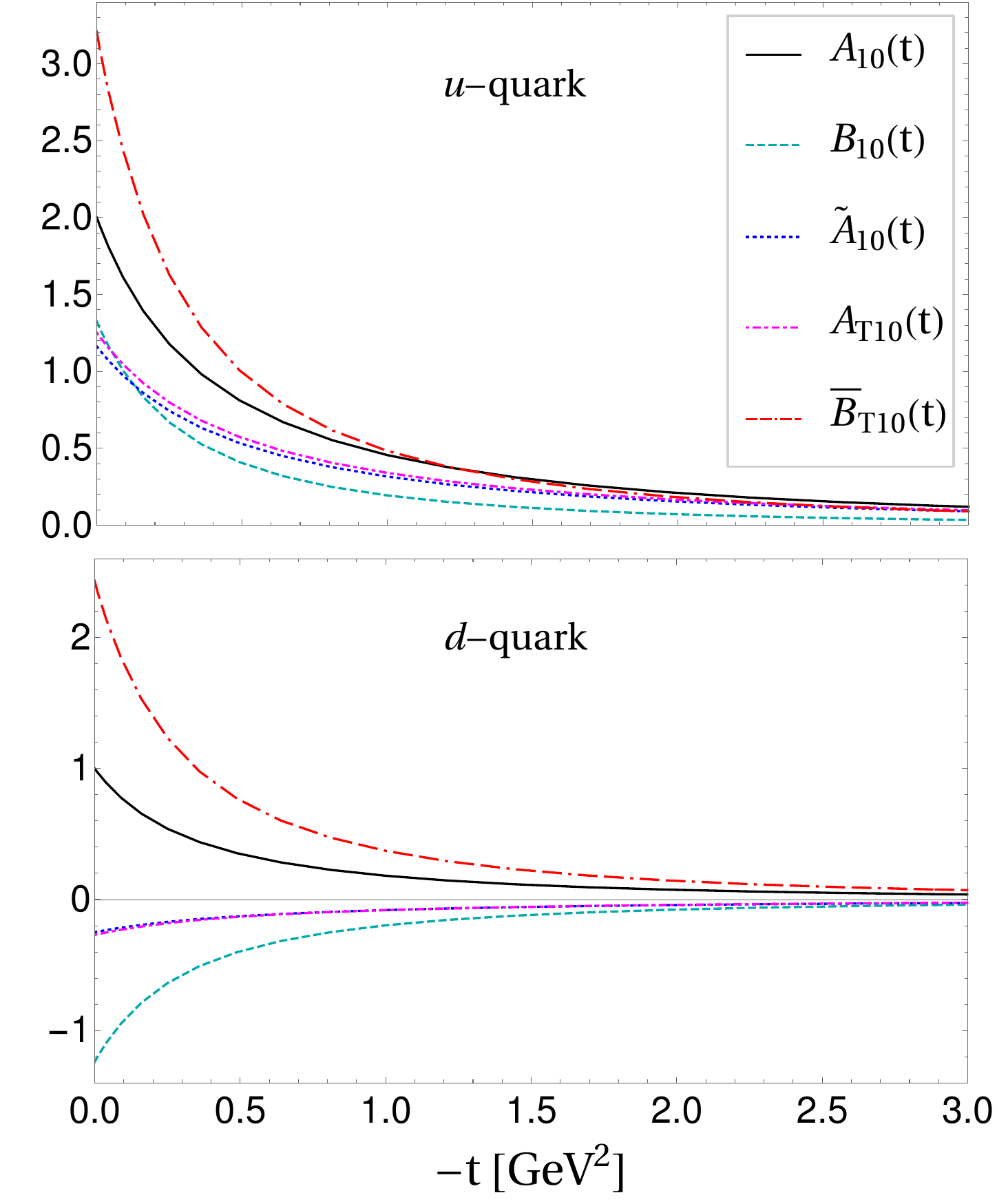}
         \caption{}
         \label{fig:first-moment}
     \end{subfigure}
     \begin{subfigure}[b]{0.49\textwidth}
         \centering
         \includegraphics[width=\textwidth]{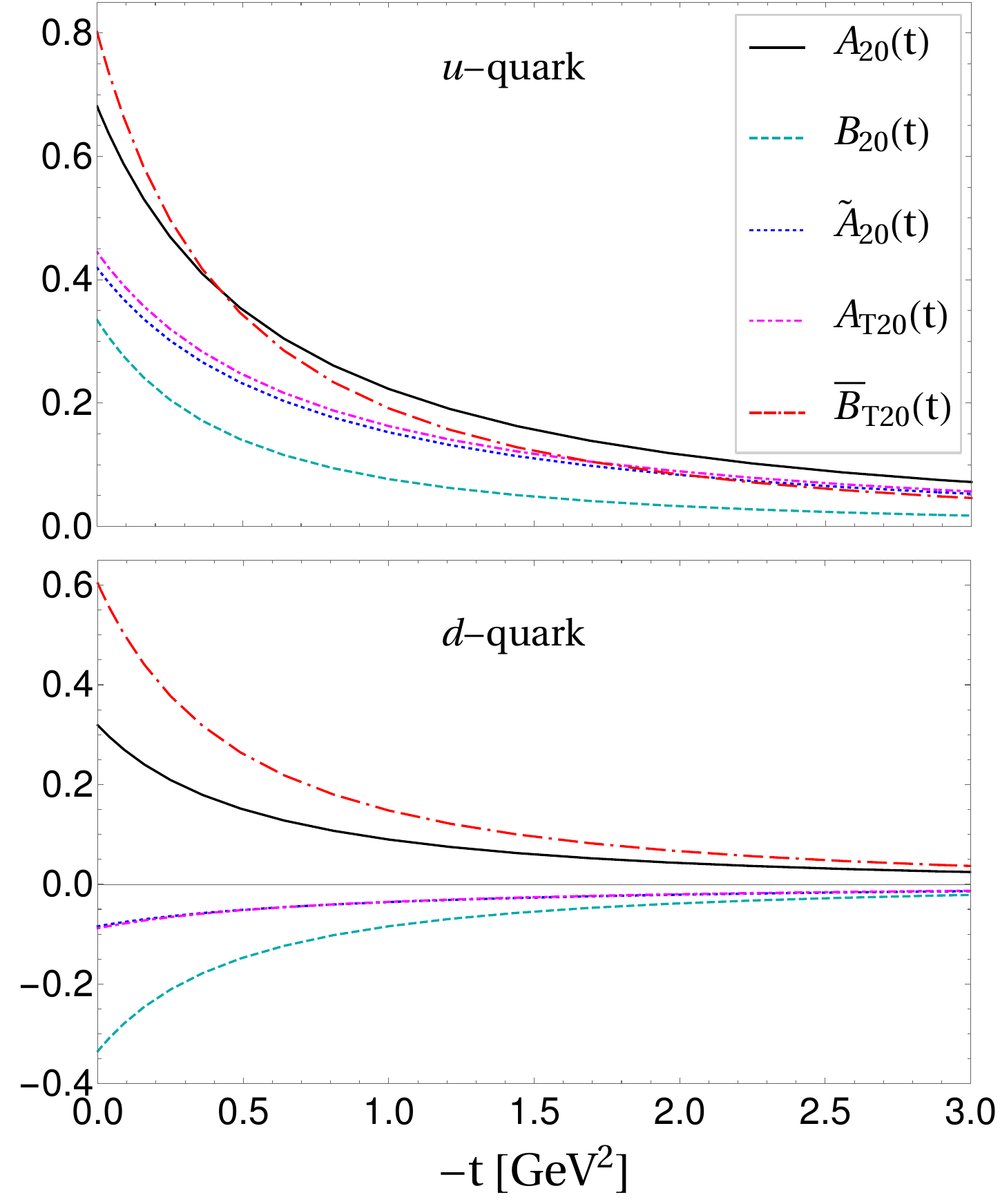}
        \caption{}
         \label{fig:second-moment}
     \end{subfigure}
        \caption{(a) The first and (b) second Mellin moments of GPDs, also known as generalized form factors, for the $u$-quark (upper) and the $d$-quark (lower) w.r.t. the square of the momentum transfer $-t$ (in GeV$^2$).}
        \label{fig:Mellin_moment}
\end{figure}

and
\begin{align}
A^q_{20}(t)&=\int {\rm d}x x H^q(x,0,t) \hspace{0.5cm},\hspace{0.5cm} B^q_{20}(t)= \int {\rm d}x x E^q(x,0,t)\;, \\
\tilde{A}^q_{20}(t)&=\int {\rm d}x x \tilde{H}^q(x,0,t) \hspace{0.5cm},\hspace{0.5cm} \tilde{B}^q_{20}(t)=\int {\rm d}x x \tilde{E}^q(x,0,t)\;, \\
A^q_{T20}(t) &= \int {\rm d}x x H^q_T(x,0,t) \hspace{0.5cm},\hspace{0.5cm}  \bar{B}^q_{T20}(t) = \int {\rm d}x x [E^q_T(x,0,t)+2\tilde{H}^q_T(x,0,t)] \;.
\label{Eq:second_moment}
\end{align}

Now, in the forward limit $(t=0)$, the Dirac form factors exhibit the normalization as
\begin{align}
F_1^{u}(0) =2 \hspace{1cm}, \hspace{1cm} F_1^{d}(0) = 1 \;,
\end{align} 
the Pauli form factors are regarded as the anomalous magnetic moments,
\begin{align}
F_2^{q}(0) = \kappa^{q}  \;,
\end{align}
the axial-vector form factors and pseudoscalar form factors are regarded as the axial-vector coupling constant (axial charge) and pseudo-scalar coupling constant,
\begin{align}
G_A^{q}(0) = g_A^q  \hspace{1cm} {\rm and} \hspace{1cm} G_P^q(0) = g_P^q\;,
\end{align}
respectively, and the tensor form factors identify with the tensor charge $g_T^q$ and tensor magnetic moment $\kappa^q_T$.

In Fig.~\ref{fig:first-moment}, we show the first Mellin moment of both chiral-even and chiral-odd GPDs, defined in Eq.~\eqref{Eq:Dirac_and_Pauli_FF}-\eqref{Eq:tensor_FF}. As discussed earlier, the first moment of unpolarized and helicity GPDs represent the Dirac, Pauli and axial form factors. The detailed study of these form factors can be found in Ref.~\cite{Mondal:2019jdg, Xu:2021wwj}, where reasonable agreement with the experimental data has been observed. Along with these form factors, we evaluate tensor form factors which are connected with $A^q_{T10}$ and $\bar{B}^q_{T10}$.

\begin{figure}
\begin{minipage}{1.0\textwidth}
\centering
\includegraphics[width=0.48\linewidth, height=0.20\textheight]{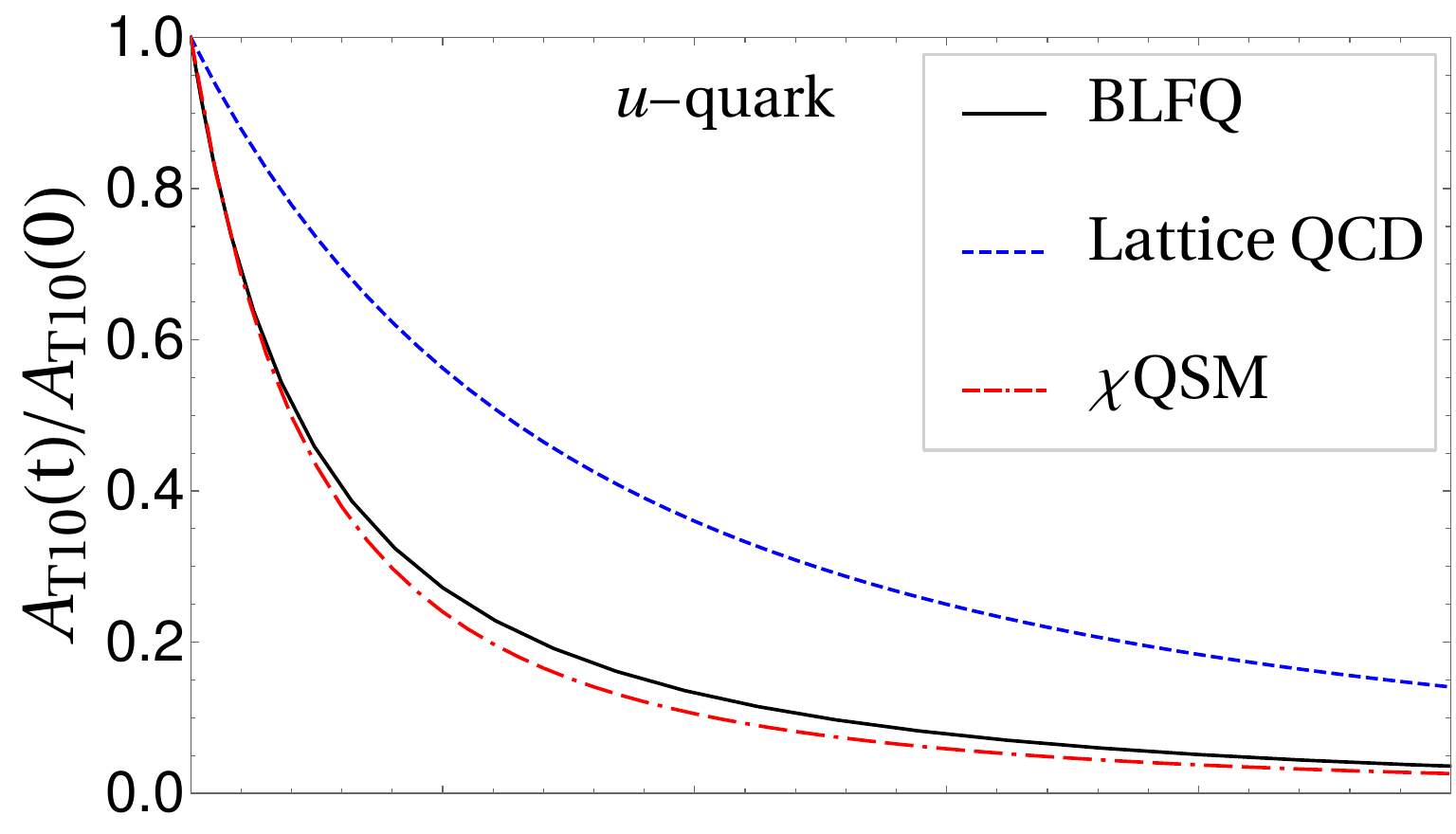}
\includegraphics[width=0.46\linewidth, height=0.193\textheight]{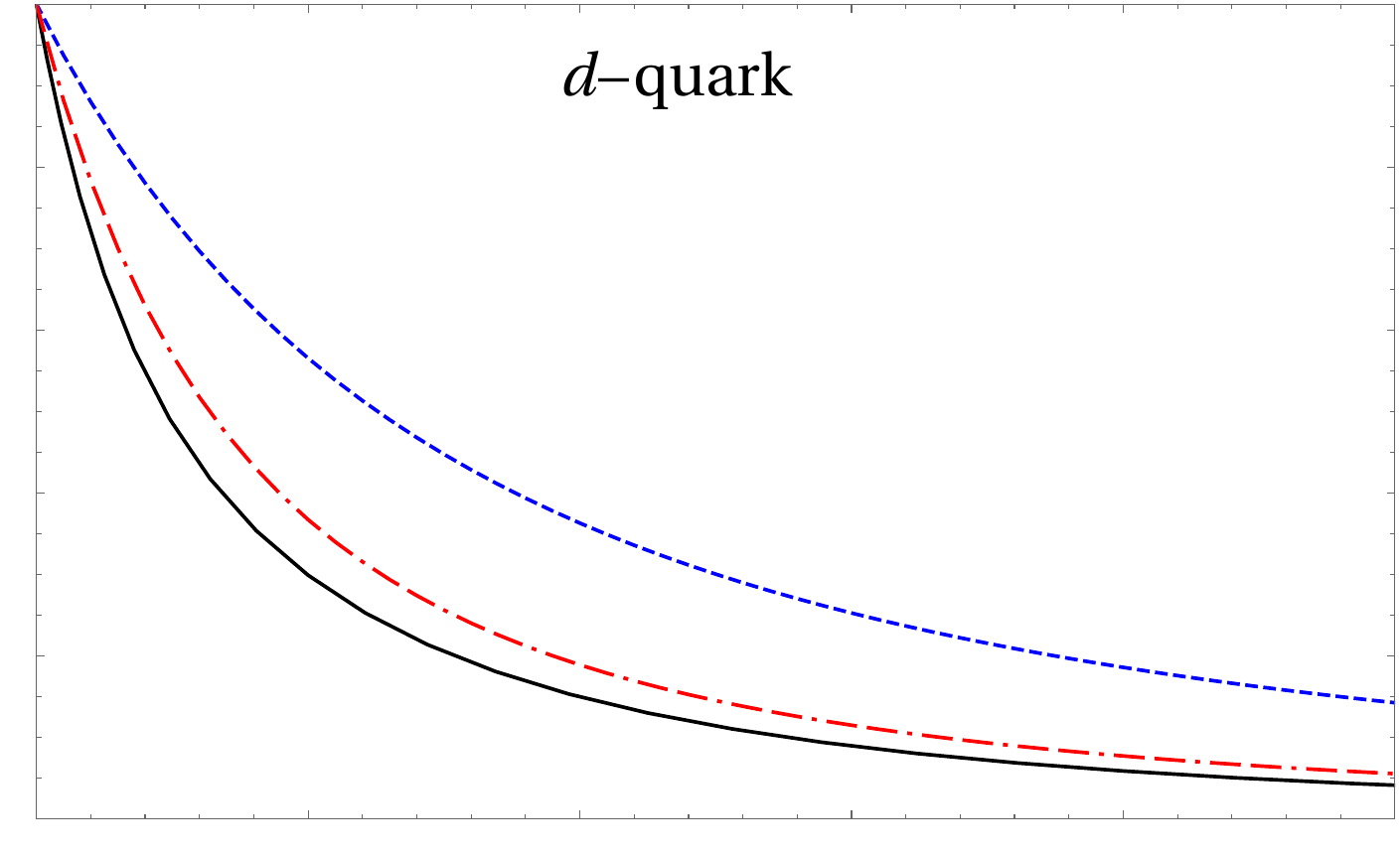}
\end{minipage}

\begin{minipage}{1.0\textwidth}
\centering
\includegraphics[width=0.48\linewidth, height=0.23\textheight]{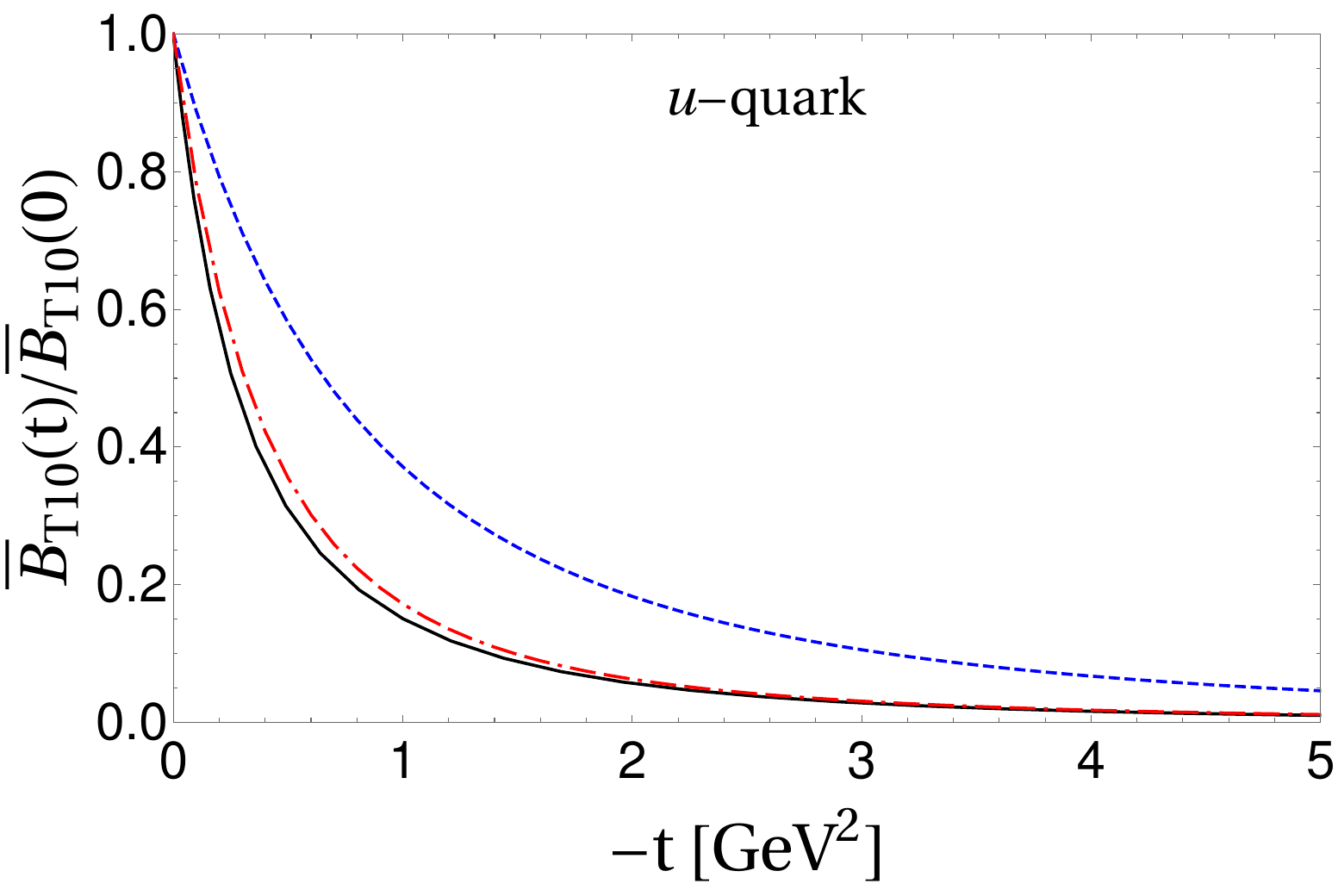} 
\includegraphics[width=0.462\linewidth, height=0.223\textheight]{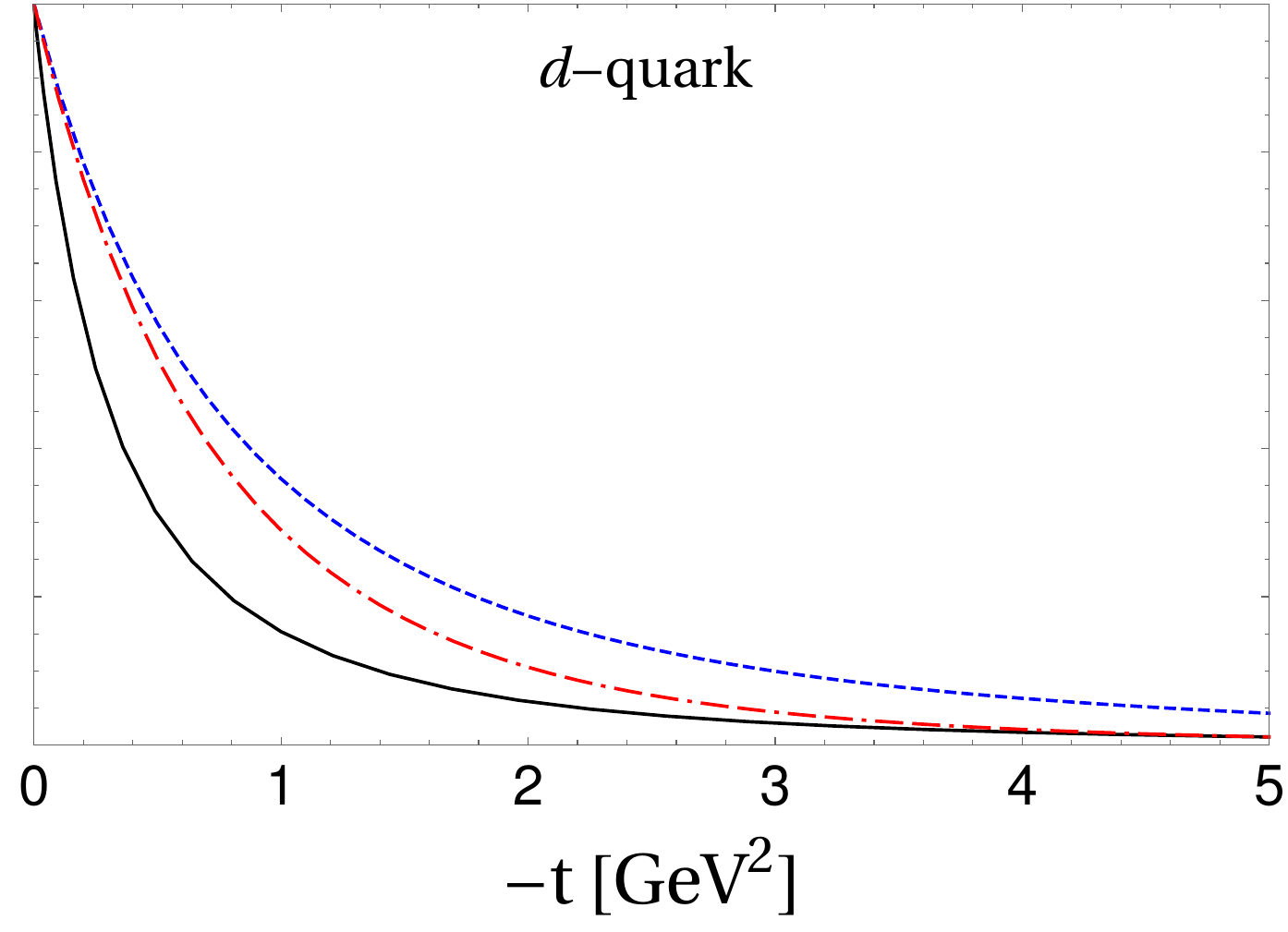}
\end{minipage}
\caption{The normalized first Mellin moments with respect to $-t$ (in GeV$^2$) compared with the predictions of the lattice QCD approach~\cite{Gockeler:2005cj,QCDSF:2006tkx} and the chiral quark soliton model ($\chi$QSM)~\cite{Ledwig:2011qw, Ledwig:2010tu}.}
\label{fig:comparison_First_Mellin_moment}
\end{figure}

The transversity GPDs are connected with the tensor FFs $A^q_{T10}(=g^q_T(t))$ and $\bar{B}^q_{T10}(=\kappa^q_T(t))$, and are illustrated in Fig.~\ref{fig:first-moment}. When comparing with other published studies, we find that our tensor FFs qualitatively agree with the lattice QCD predictions for both $u$ and $d$ quarks~\cite{Gockeler:2005cj,QCDSF:2006tkx} and with other model predictions~\cite{Ledwig:2010tu, Pasquini:2005dk, Ledwig:2011qw, Chakrabarti:2015ama}. We illustrate the $t$-dependence of the first Mellin moments of chiral-odd GPDs, and compare our predictions with that of the lattice QCD approach~\cite{Gockeler:2005cj,QCDSF:2006tkx} and the chiral quark soliton model ($\chi$QSM)~\cite{Ledwig:2011qw, Ledwig:2010tu} in Fig.~\ref{fig:comparison_First_Mellin_moment}. Since the lattice QCD method and the $\chi$QSM have predicted their results at $4$ GeV$^2$ and $0.36$ GeV$^2$ respectively which are considerably different from our model scale $\mu^2_0=0.195 \pm 0.020$ GeV$^2$~\cite{Mondal:2019jdg, Xu:2021wwj}, the comparison between them is only qualitative though some similarities are apparent. Our model scale could be significantly increased when we increase our BLFQ basis spaces to include Fock components beyond the valence sector. We anticipate that, with these planned model improvements, our results may then become more comparable with the lattice QCD and other model predictions. 

In Table~\ref{table:first_moment}, we summarize our predictions for the first Mellin moments or the generalized form factors of the valence quark GPDs, when there is no momentum transfer from the initial to the final state of proton $(t=0)$. Similar to the other model results, we also observe that the tensor charge $g^q_T(0)(=A^q_{T10})$ is larger than the axial charge $g_A^q(0)(=\tilde{A}^q_{10}(0))$, regardless of the sign. However, the difference is observed to be small in our BLFQ computation.

\begin{figure}
\begin{minipage}{1.0\textwidth}
\centering
\includegraphics[width=0.48\linewidth, height=0.20\textheight]{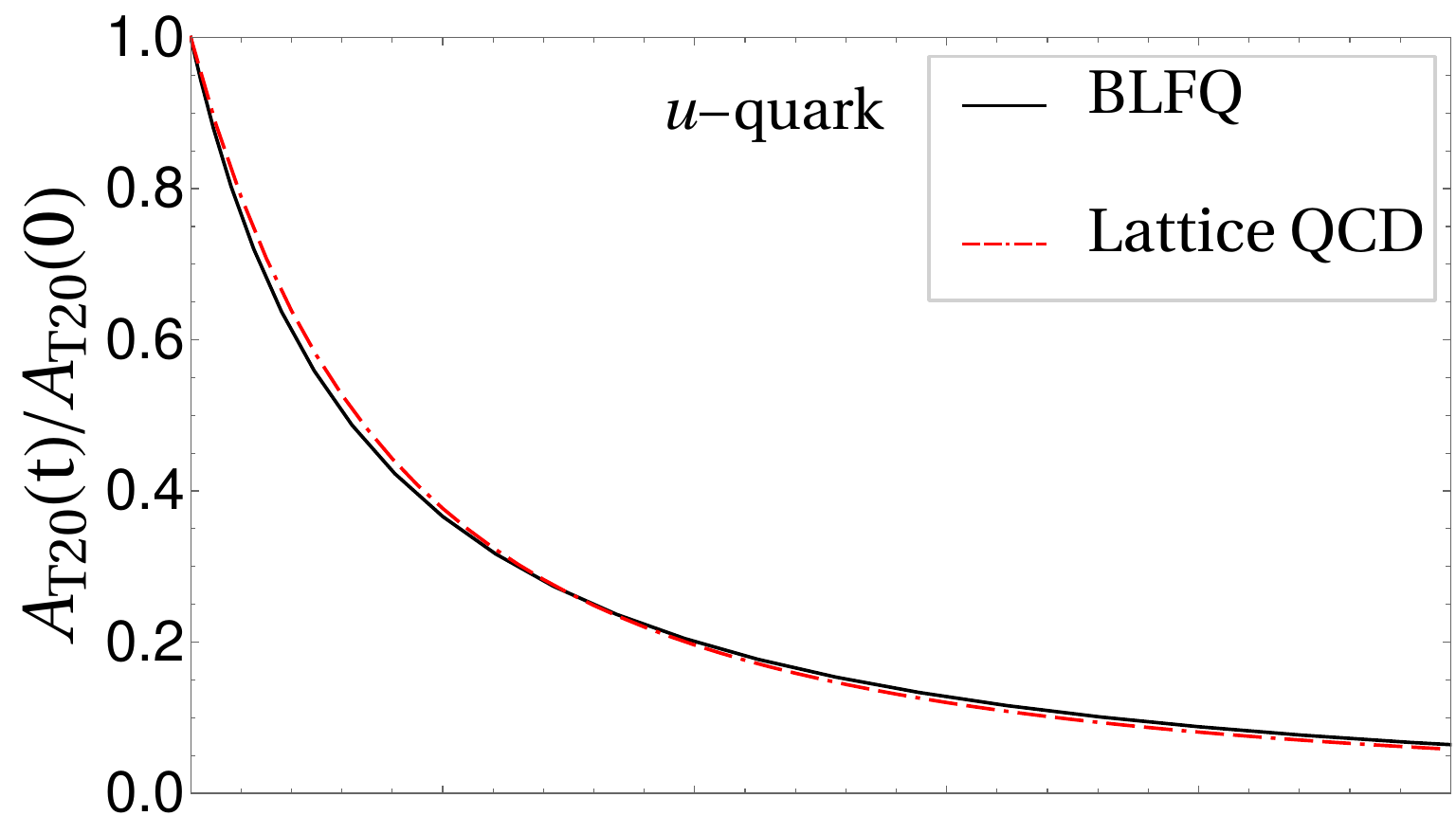}
\includegraphics[width=0.46\linewidth, height=0.193\textheight]{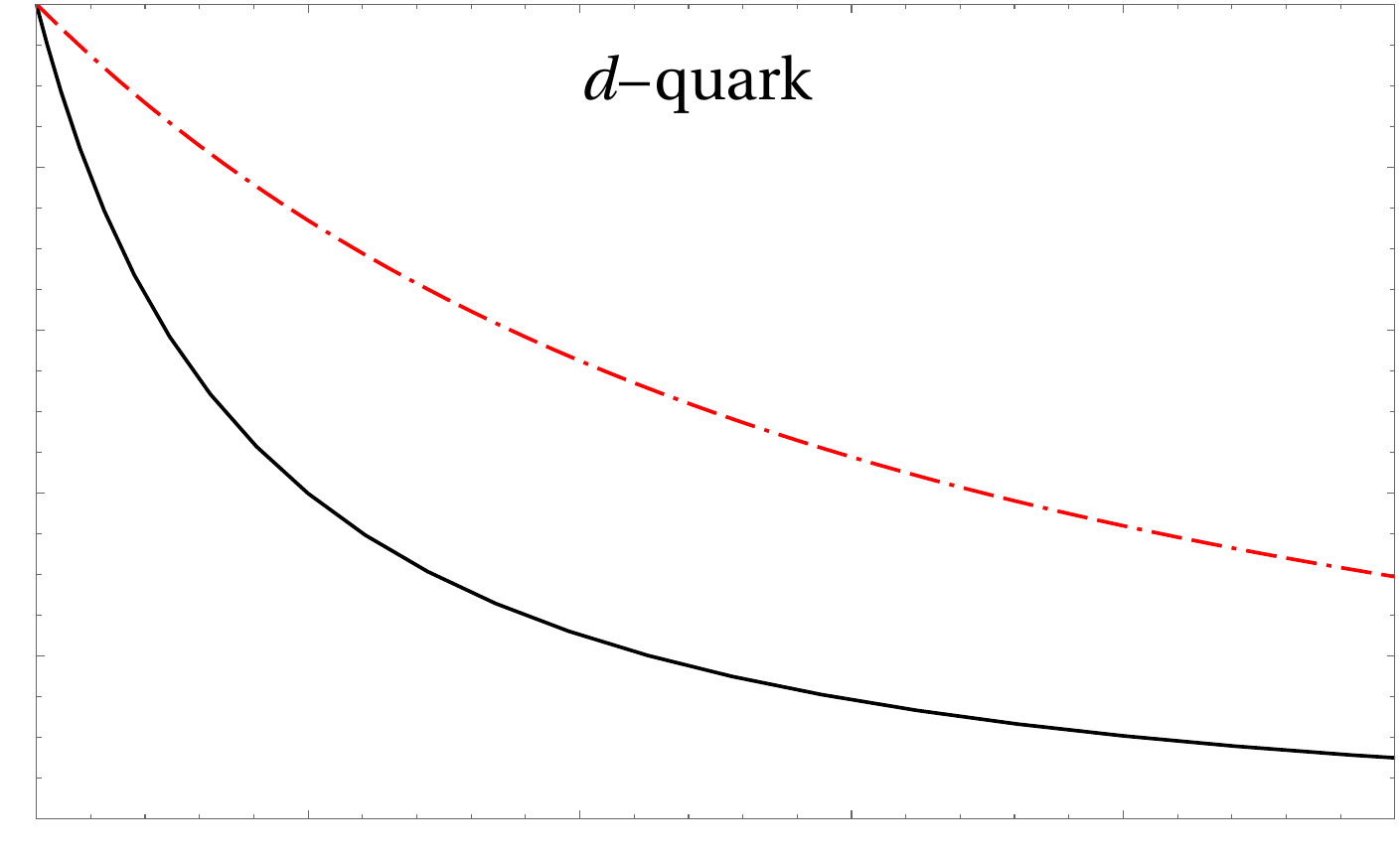}
\end{minipage}

\begin{minipage}{1.0\textwidth}
\centering
\includegraphics[width=0.48\linewidth, height=0.23\textheight]{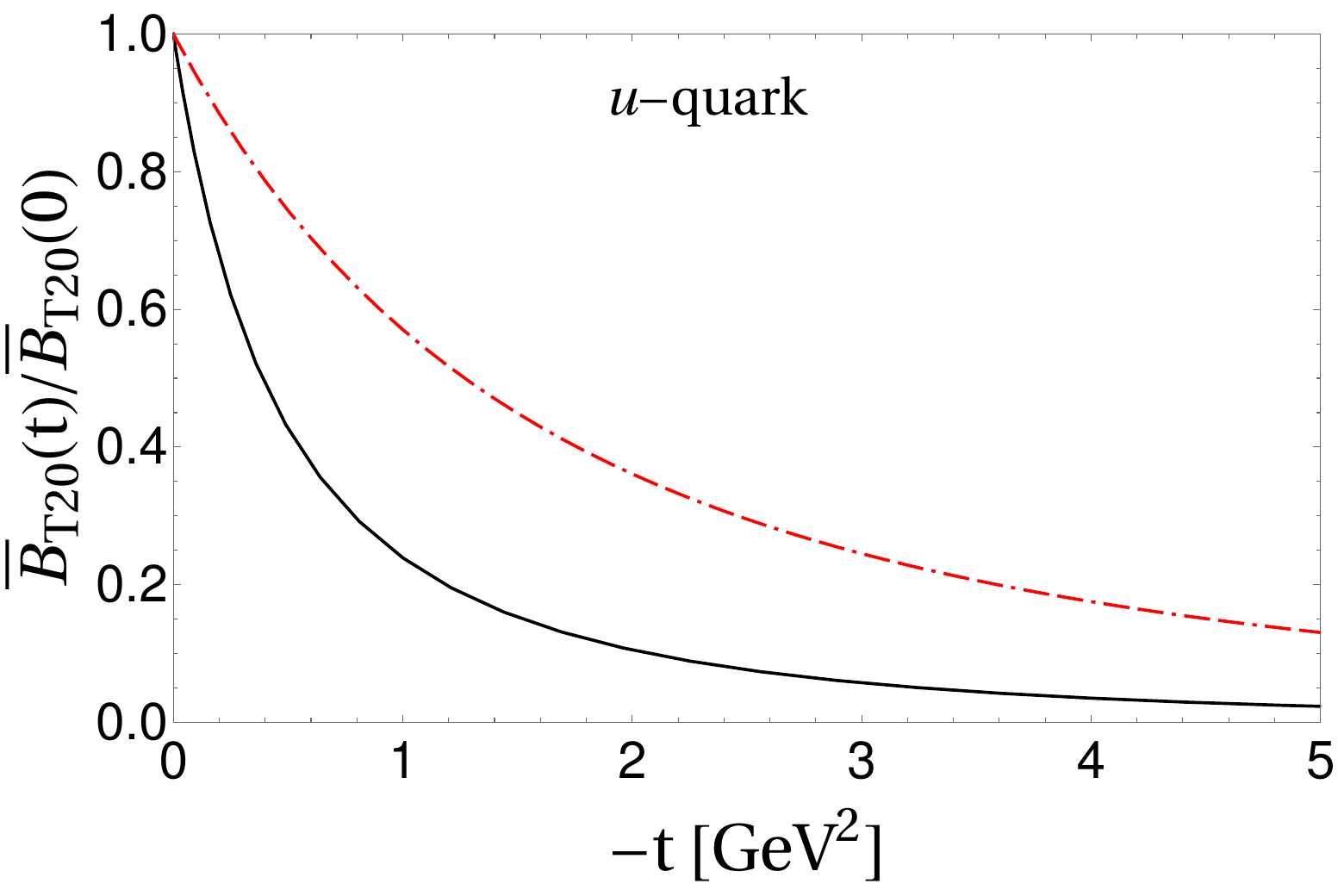} 
\includegraphics[width=0.462\linewidth, height=0.223\textheight]{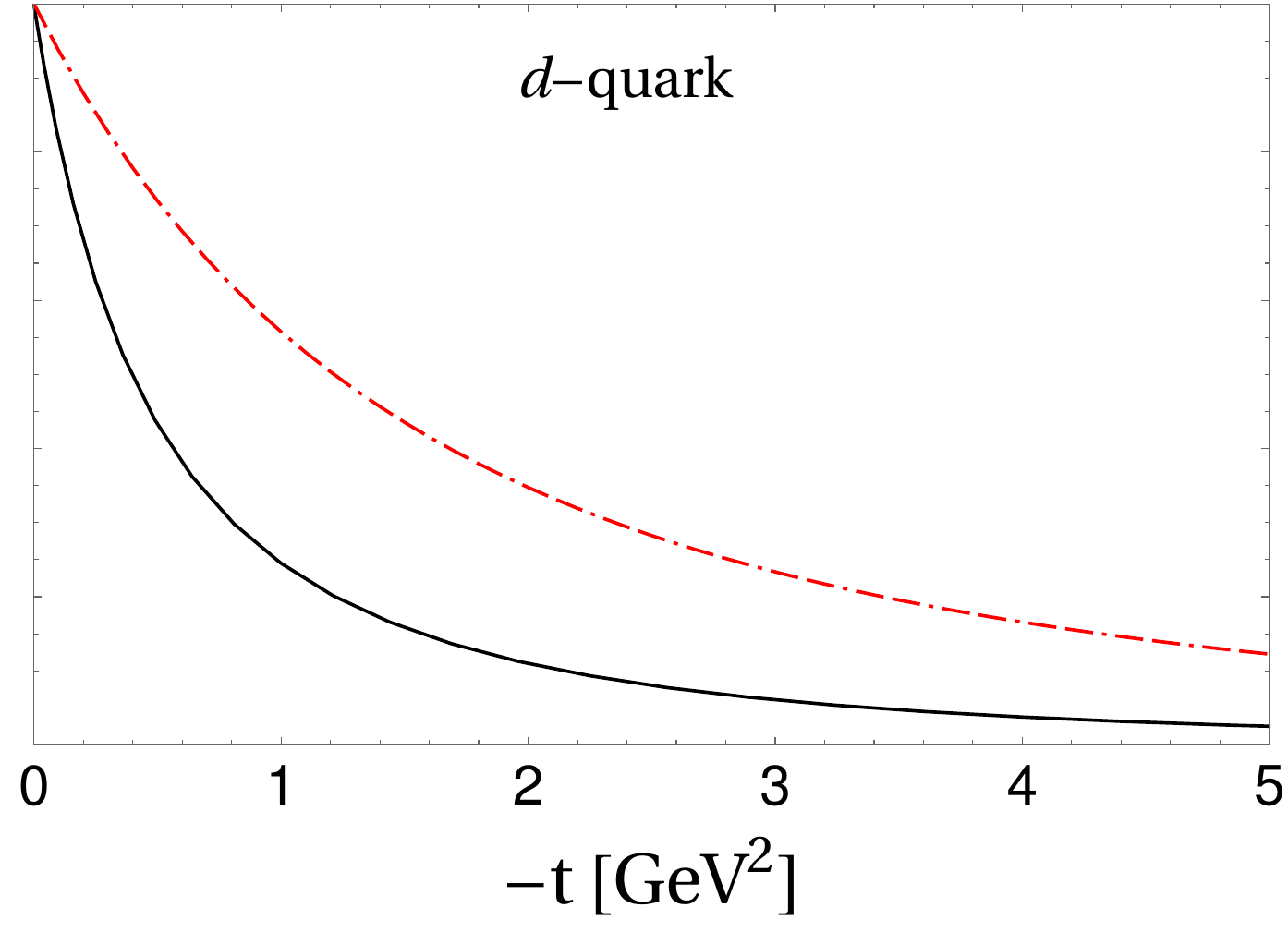}
\end{minipage}
\caption{The normalized second Mellin moments with respect to $-t$ (in GeV$^2$) compared with the predictions of the lattice QCD approach~\cite{Gockeler:2005cj,QCDSF:2006tkx}.}
\label{fig:comparison_second_Mellin_moment}
\end{figure}



 \begin{table}[hbt!]
 \begin{tabular}{|  l |  c    c    c   c   c |}
 \cline{1-6}
 Quantity  & $A_{10}(0)$ & $B_{10}(0)$ & $\tilde{A}_{10}(0)$ & $A_{T10}(0)$ & $\bar{B}_{T10}(0)$  \\
 \cline{1-6} 
 $u$-quark & 2 & 1.367 & 1.162 & 1.251 & 3.208 \\
 $d$-quark & 1 & -1.279 & -0.249 & -0.270 & 2.432 \\
 \cline{1-6} 
\end{tabular}
\caption{The values of first moment of the GPDs at $t=0$ for $u$ and $d$ quarks in BLFQ.}
\label{table:first_moment}
\end{table}

\begin{figure}[hbt!]
     \centering
         \includegraphics[width=0.6\textwidth]{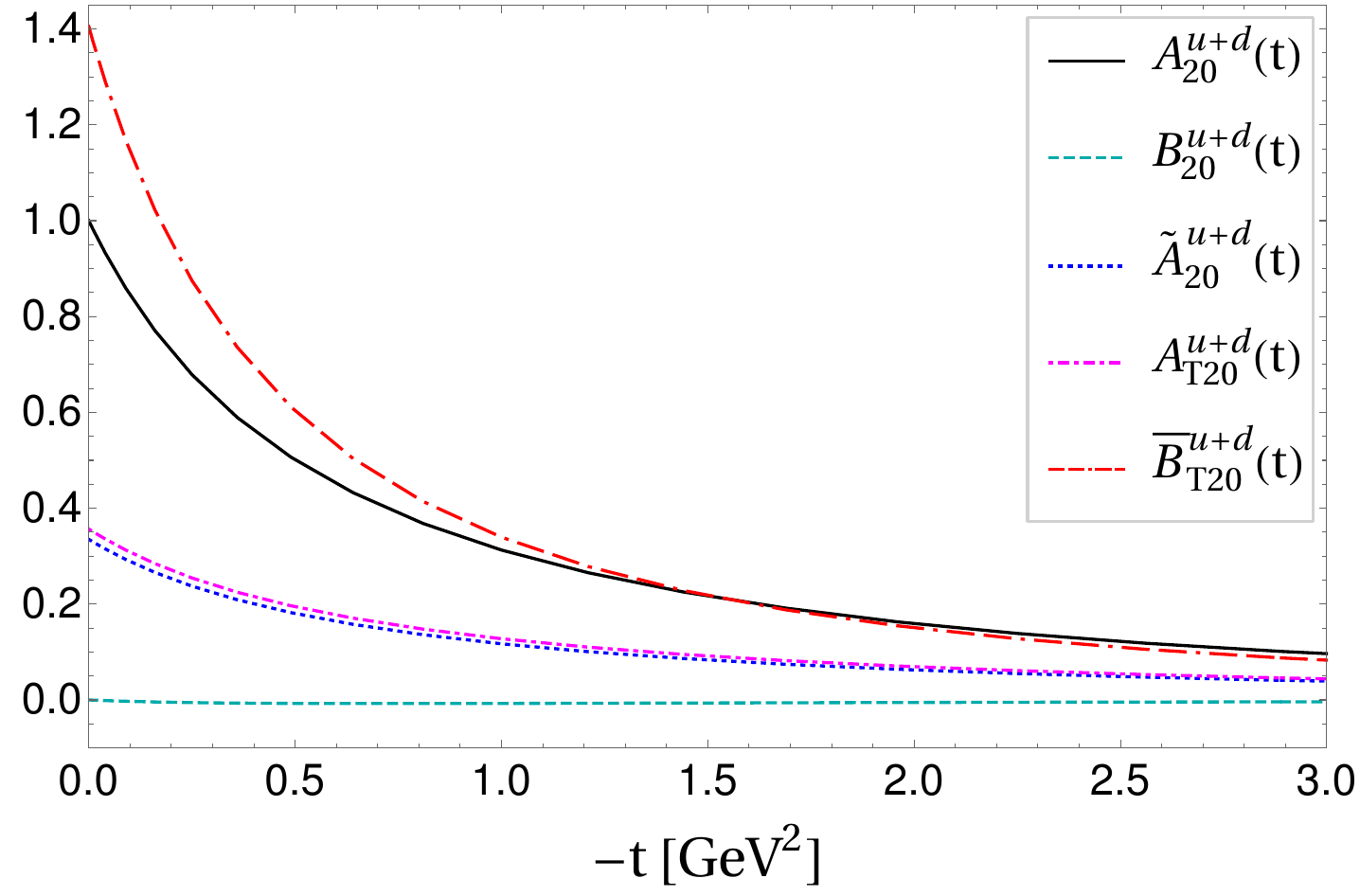}
        \caption{The isoscalar generalized form factors w.r.t. the square of the momentum transfer $-t$ (in GeV$^2$).}
        \label{fig:second_Mellin_moment}
\end{figure}

\begin{figure}[hbt!]
     \centering
         \includegraphics[width=0.65\textwidth]{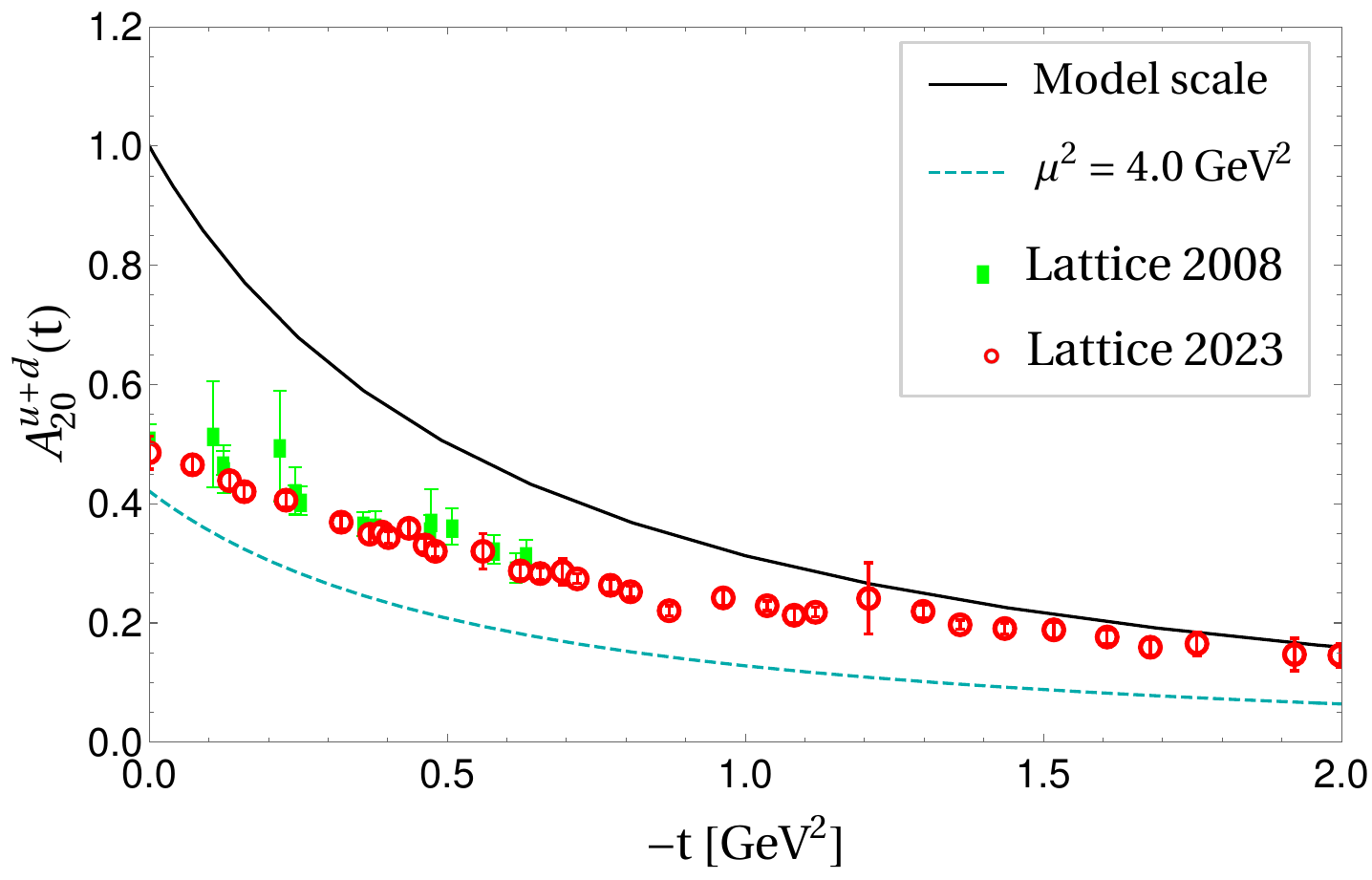}
        \caption{The isoscalar generalized form factor $A_{20}$ w.r.t. the square of the momentum transfer $-t$ (in GeV$^2$) in comparison with lattice QCD predictions~\cite{LHPC:2007blg, Hackett:2023rif}.}
        \label{fig:A20-comparison}
\end{figure}

\begin{figure}
\begin{minipage}{1.0\textwidth}
\centering
\includegraphics[width=0.45\linewidth, height=0.2\textheight]{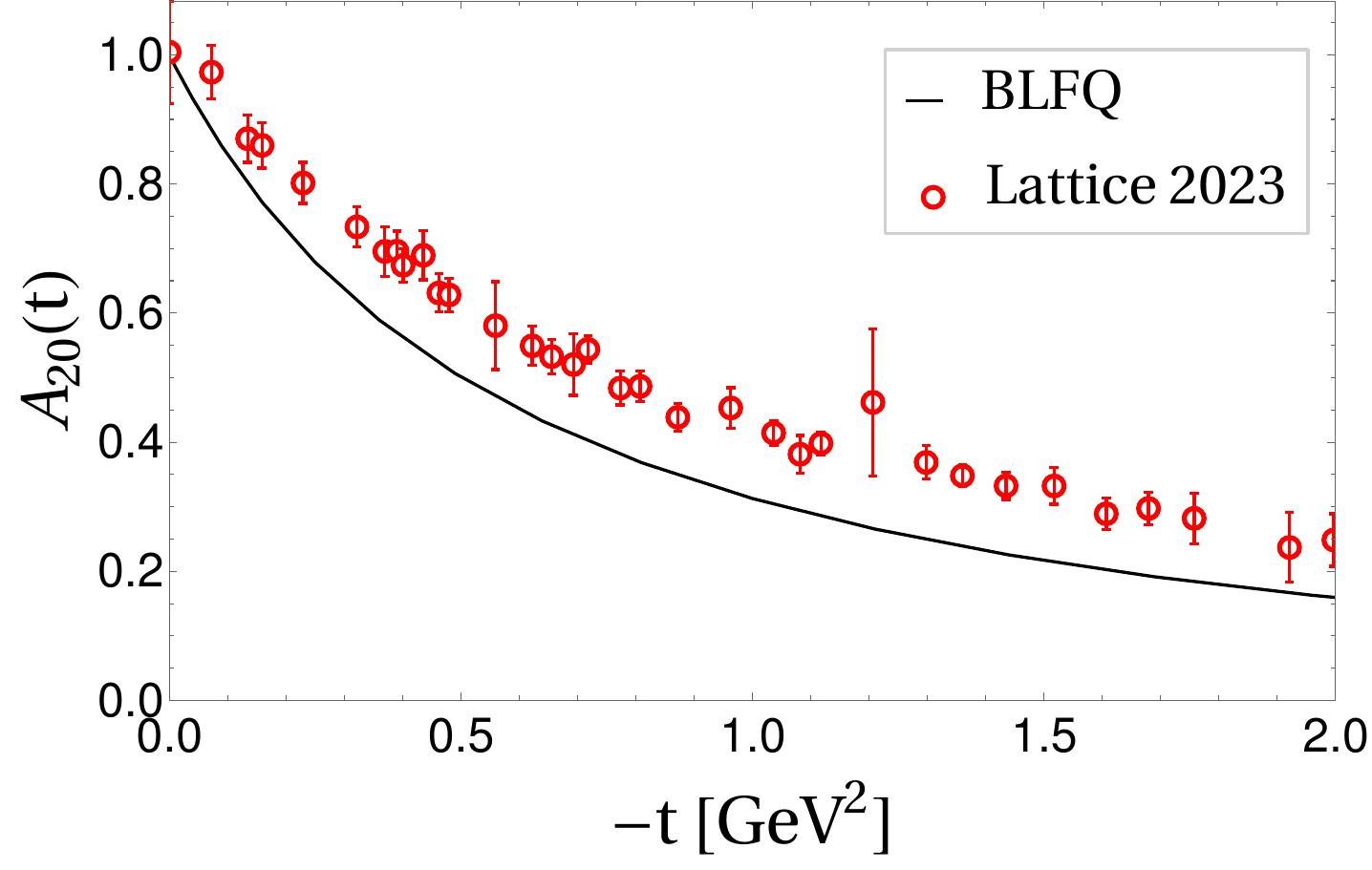}
\includegraphics[width=0.45\linewidth, height=0.206\textheight]{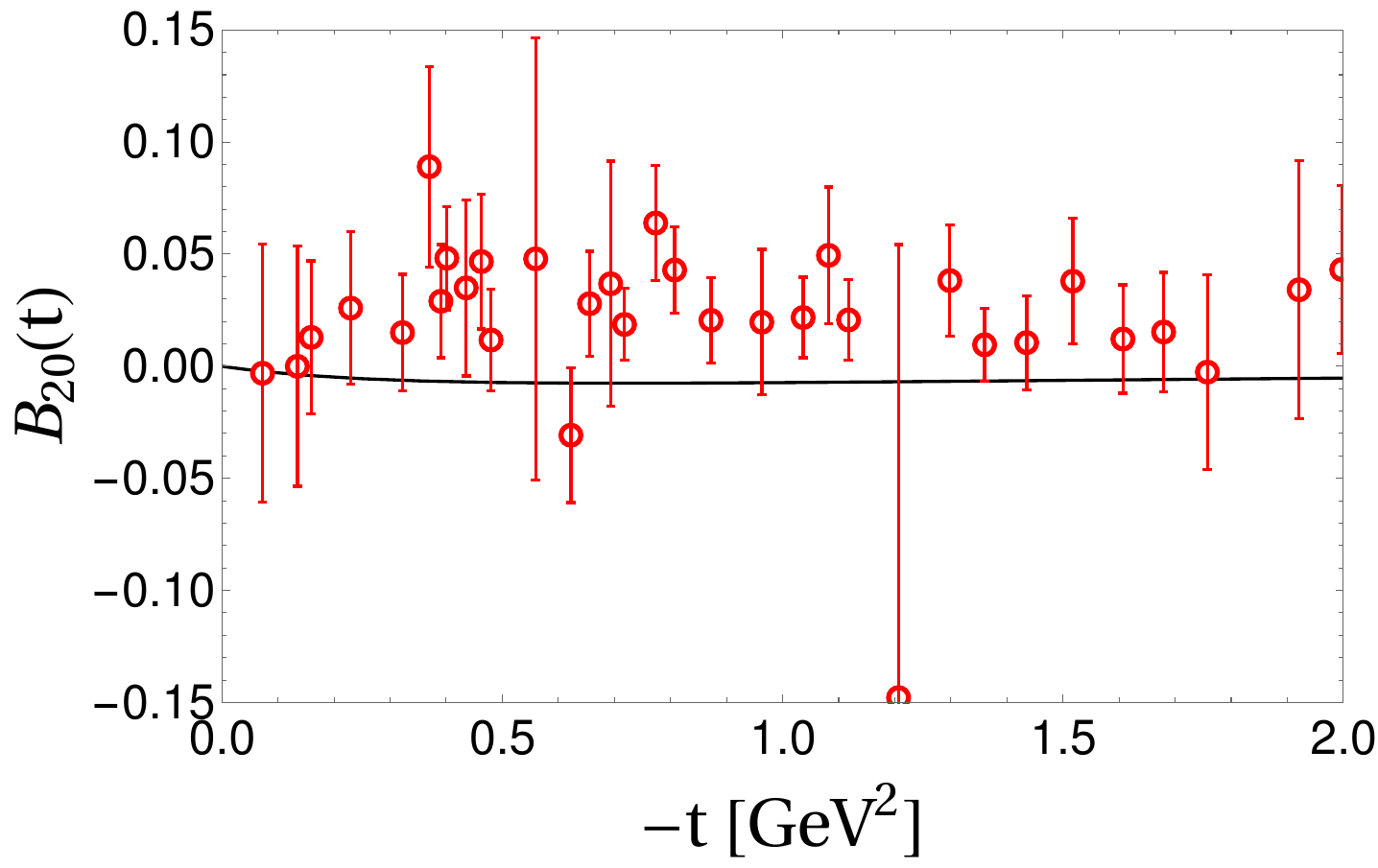}
\end{minipage}
\caption{The form factors $A_{20}(t)$ and $B_{20}(t)$ w.r.t. the square of the momentum transfer $-t$ in comparison with lattice QCD predictions~\cite{Hackett:2023rif}.}
\label{fig:At-Jt-comparison}
\end{figure}


In Fig.~\ref{fig:second-moment}, we present the second moment of the proton GPDs where, through a visual comparison to published lattice QCD results~\cite{Gockeler:2005cj,QCDSF:2006tkx}, we again find similarities in the qualitative behavior. To elucidate the comparison with lattice QCD results, we show the $t$-dependence of the second Mellin moments of the chiral-odd GPDs in Fig.~\ref{fig:comparison_second_Mellin_moment}. Again, the predictions are made at different scales so that the comparison is only qualitative. The second moments give information about the gravitational form factors and our results for $u$ and $d$ quark contributions at $t=0$ are shown in Table~\ref{table:second_moment}. Further, by simply adding the second Mellin moments for $u$ and $d$ quarks, we obtain the isoscalar generalized FFs that we present in Fig.~\ref{fig:second_Mellin_moment}. Unlike other generalized FFs, $B^{u+d}_{20}$ is observed to be independent of $-t$, and it appears to come out close to zero in our calculations.
Furthermore, in Fig.~\ref{fig:A20-comparison}, we have shown the comparison of our predictions of $A^{u+d}_{20}$ by evolving it to the scale of the lattice QCD predictions~\cite{LHPC:2007blg, Hackett:2023rif}, i.e., to $\mu^2=4$ GeV$^2$. A clear discrepancy has been observed which can be attributed to the fact that our calculations are based on only the leading Fock sector.

Since the GFFs $A_{20}(t)$ and $B_{20}(t)$ summing over all partons are scale invariant, we compare these GFFs in our BLFQ approach with those in the lattice QCD~\cite{Hackett:2023rif} in Fig.~\ref{fig:At-Jt-comparison}. Within our current treatment of the BLFQ approach, only $u$ and $d$ quarks contribute to these quantities, while the lattice QCD considers contribution from $u$, $d$, $s$ quarks, and gluons. We find that our results are somewhat underestimated compared to the lattice QCD predictions. 
This quantitative difference is expected as we consider only the valence Fock sector. On the other hand, the lattice QCD predictions involve first-principle calculations considering the quarks as well as the gluons. Our predictions may become closer to those of the lattice QCD when we explicitly include gluons and the sea quarks in our approach.

 \begin{table}[hbt!]
 \begin{tabular}{|  l |  c    c    c   c   c |}
 \cline{1-6}
 Quantity  & $A_{20}(0)$ & $B_{20}(0)$ & $\tilde{A}_{20}(0)$ & $A_{T20}(0)$ & $\bar{B}_{T20}(0)$  \\
 \cline{1-6} 
 $u$-quark & 0.681 & 0.335 & 0.419 & 0.445 & 0.802 \\
 $d$-quark & 0.319 & -0.335 & -0.084 & -0.088 & 0.604 \\
 \cline{1-6} 
\end{tabular}
\caption{The values of second moment of the GPDs at $t=0$ for $u$ and $d$ quarks in BLFQ.}
\label{table:second_moment}
\end{table}

According to Ji sum rule, the second moment of the chiral-even GPDs give the partonic contribution to the total angular momentum of the proton~\cite{Ji:1996ek}, we have
\begin{equation}
J^z_q = \frac{1}{2} \int dx x \left[H^q(x,0,0) +E^q(x,0,0)\right] = \frac{1}{2}\left[A^q_{20}(0)+B^q_{20}(0)\right] \;.
\label{eq:angular-momentum}
\end{equation}
From Table~\ref{table:second_moment}, we observe that Ji sum rule is satisfied in our model as we get $J_q^z=1/2$ following Eq.~\eqref{eq:angular-momentum}. The detailed study on the total angular momentum contribution of the partons, when they do not flip their helicities, can be found in Ref.~\cite{Liu:2022fvl}.

On the other hand, the second moments of chiral-odd GPDs are connected with the angular momentum carried by the partons with the transverse spin along the $\widehat{x}$ direction in an unpolarized proton, $J_q^x$. According to Burkardt~\cite{Burkardt:2005hp}, this quantity is one half of the expectation value of the transversity asymmetry
\begin{equation}
\langle \delta^x J^x_q \rangle = \frac{1}{2} \int dx x \left[H^q_T(x,0,0) + 2 \tilde{H}^q_T(x,0,0) + E^q_T(x,0,0) \right] = \frac{1}{2}\left[A^q_{T20}(0)+ \bar{B}^q_{T20}(0)\right] \;.
\end{equation}
The obtained values of transversity asymmetry for $u$-quark and $d$-quark in our model as well as a comparison with other predictions from the harmonic oscillator (HO) model~\cite{Pasquini:2005dk}, hypercentral constituent quark model (CQM)~\cite{Pasquini:2005dk}, and chiral quark soliton model (CQSM)~\cite{Wakamatsu:2008ki} are shown in Table~\ref{table:transverse-asymmetry}. We observe that our predictions are close to those of the HO model. Note that the methods chosen for comparison may have different initial scales, which could be a significant source of the differences in the results among the models.

 \begin{table}[hbt!]
 \begin{tabular}{|  c   c    c    c   c |}
 \cline{1-5}
 ~Transversity asymmetry~~  & ~~BLFQ~~ & ~~HO~~ & ~~Hypercentral CQM~~ & ~~CQSM~ \\
 \cline{1-5} 
 $\langle \delta^x J^x_u \rangle$ & 0.62 & 0.68 & 0.39 & 0.49 \\
 $\langle \delta^x J^x_d \rangle$ & 0.26 & 0.28 & 0.10 & 0.22 \\
 \cline{1-5} 
\end{tabular}
\caption{Our transversity asymmetry values for $u$-quark and $d$-quark in the proton compared with the predictions of harmonic oscillator (HO) model~\cite{Pasquini:2005dk}, hypercentral constituent quark model (CQM)~\cite{Pasquini:2005dk}, and chiral quark soliton model (CQSM)~\cite{Wakamatsu:2008ki}.}
\label{table:transverse-asymmetry}
\end{table}
 

\subsection{Impact-parameter dependent GPDs}
\begin{figure}[hbt!]
     \centering
     \begin{subfigure}[b]{0.41\textwidth}
         \centering
         \includegraphics[width=\textwidth]{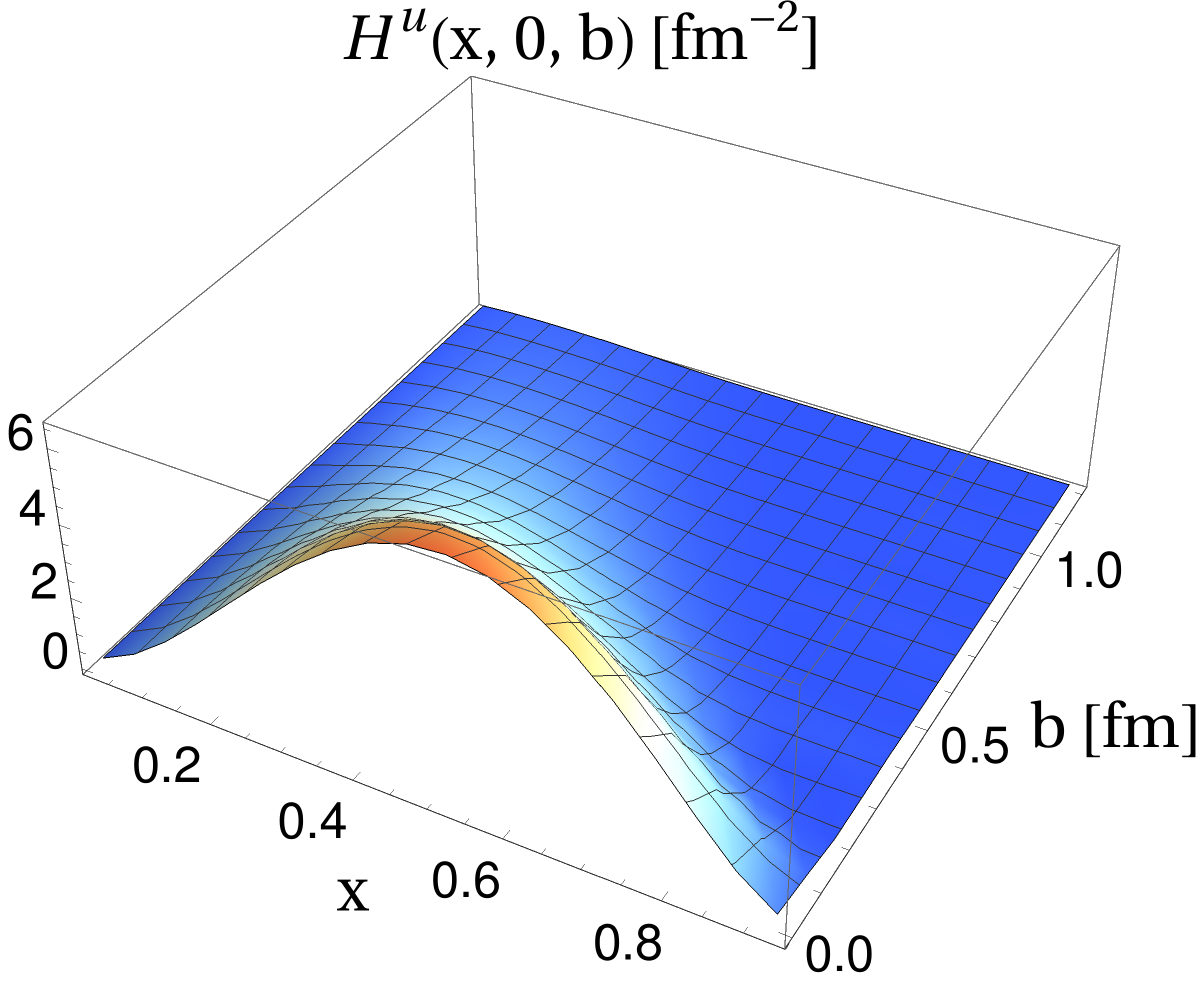}
         \caption{}
         \label{fig:H_u_b}
     \end{subfigure}
     \begin{subfigure}[b]{0.41\textwidth}
         \centering
         \includegraphics[width=\textwidth]{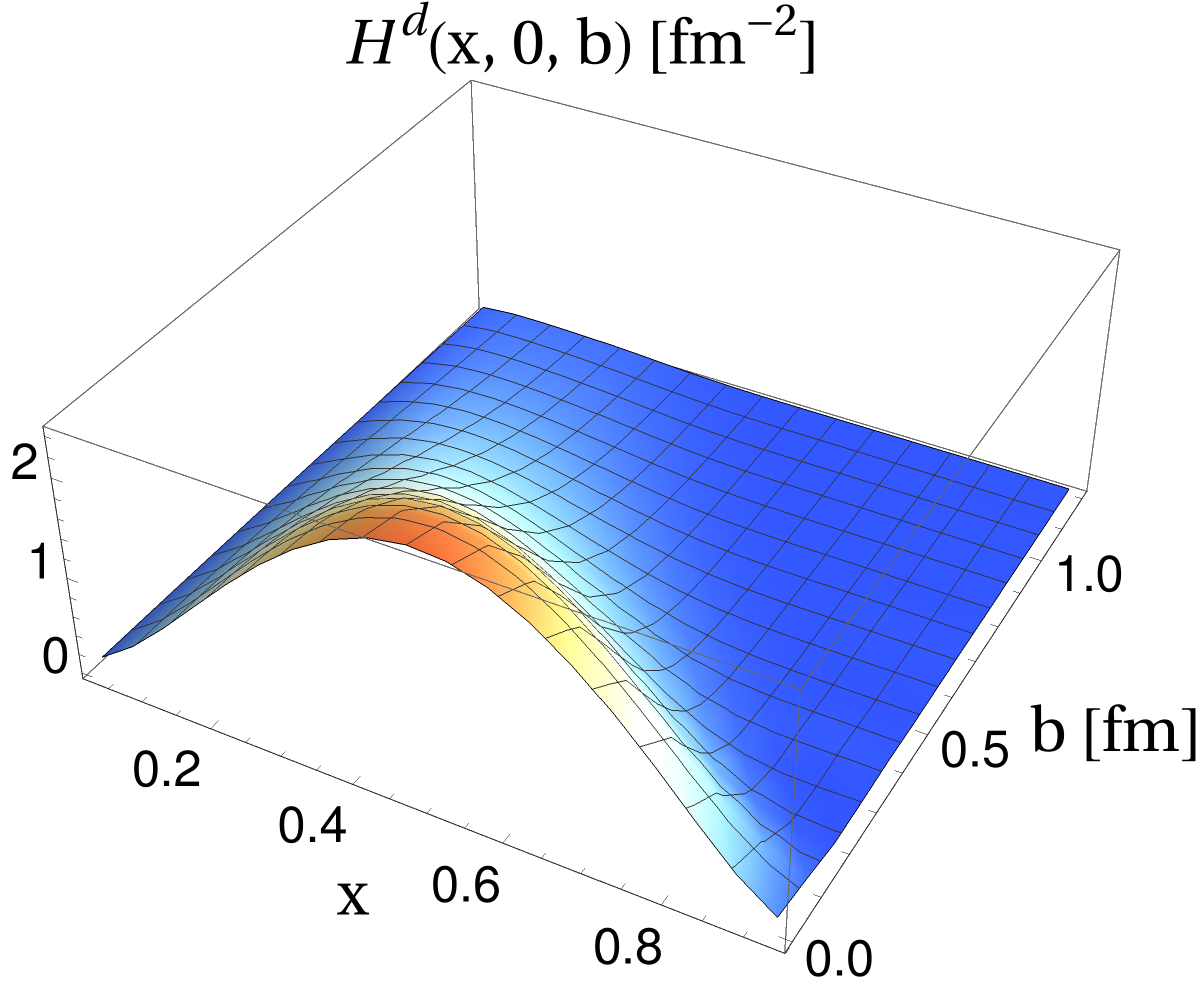}
        \caption{}
         \label{fig:H_d_b}
     \end{subfigure}
     \begin{subfigure}[b]{0.41\textwidth}
         \centering
         \includegraphics[width=\textwidth]{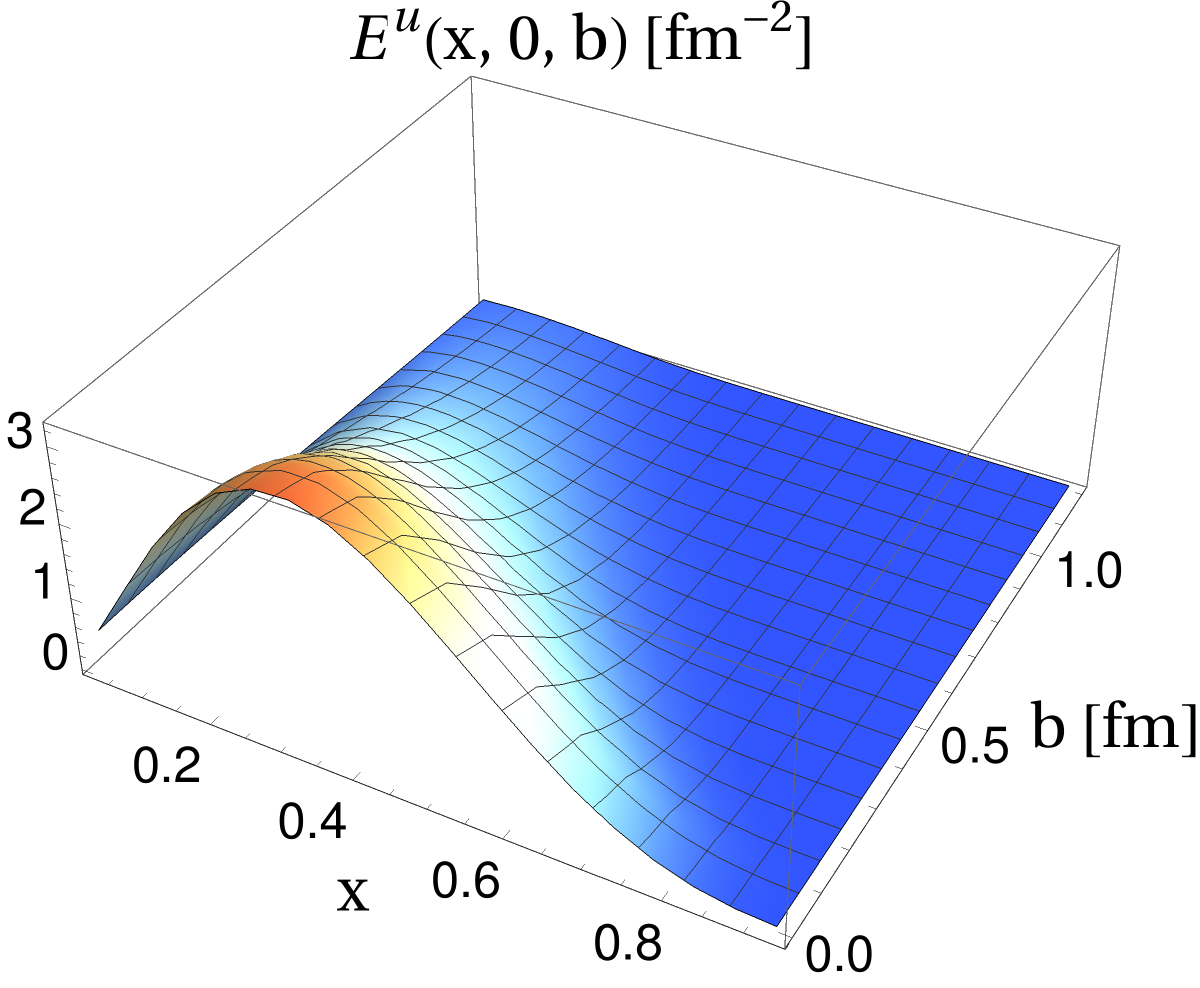}
       \caption{}
         \label{fig:E_u_b}
     \end{subfigure}
          \begin{subfigure}[b]{0.41\textwidth}
         \centering
         \includegraphics[width=\textwidth]{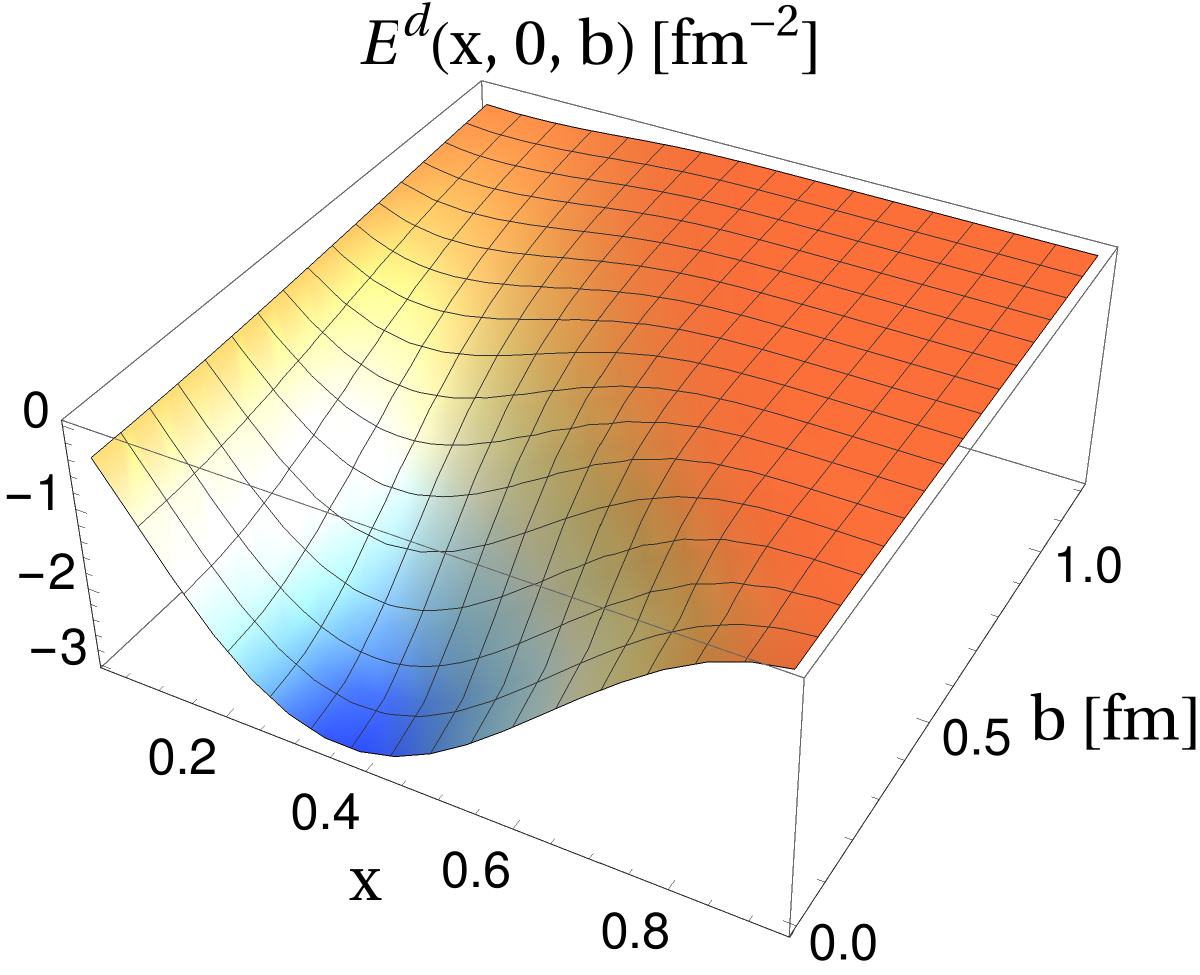}
         \caption{}
         \label{fig:E_d_b}
     \end{subfigure}
     \begin{subfigure}[b]{0.41\textwidth}
         \centering
         \includegraphics[width=\textwidth]{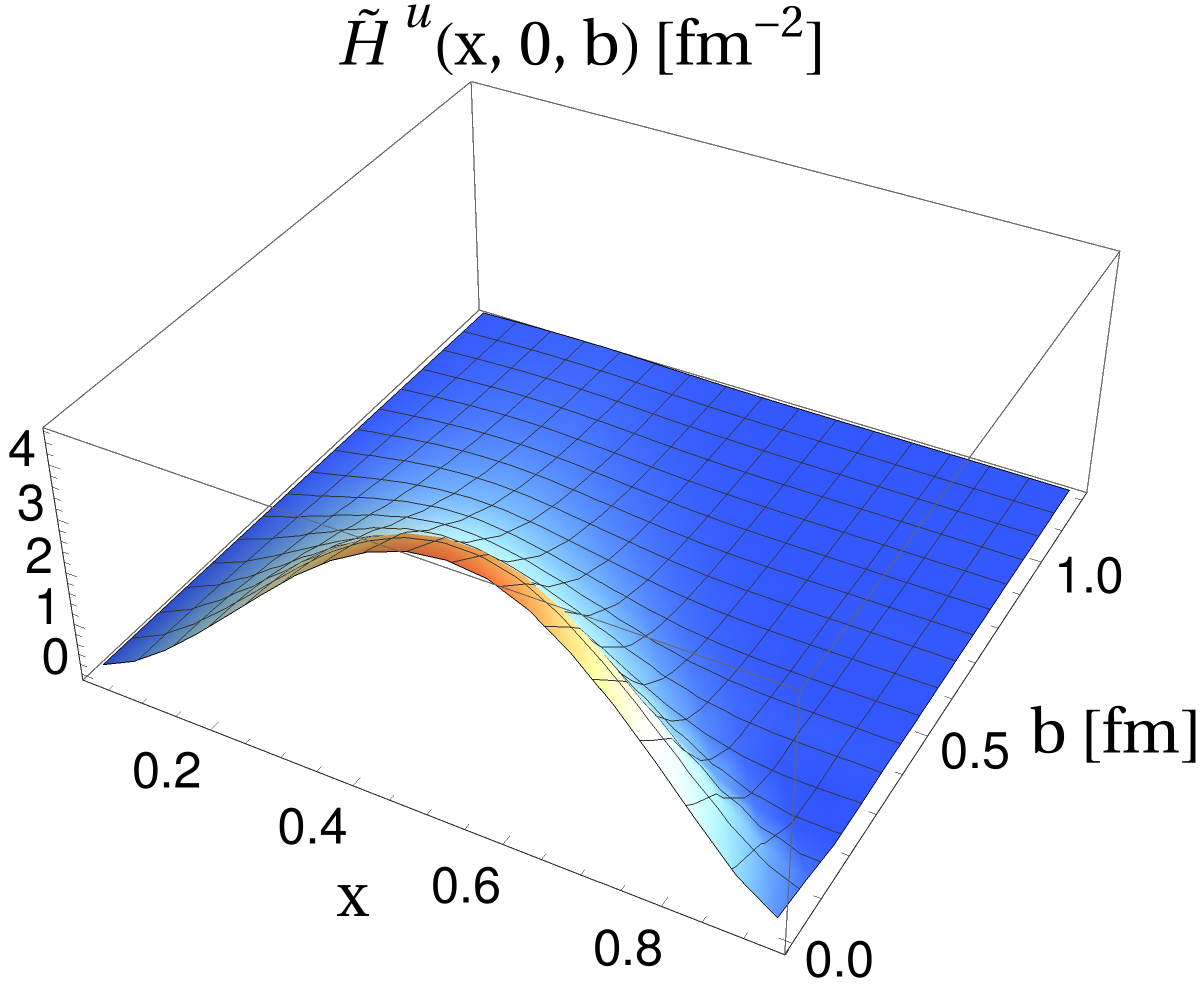}
      \caption{}
         \label{fig:Htilde_u_b}
     \end{subfigure}
     \begin{subfigure}[b]{0.41\textwidth}
         \centering
         \includegraphics[width=\textwidth]{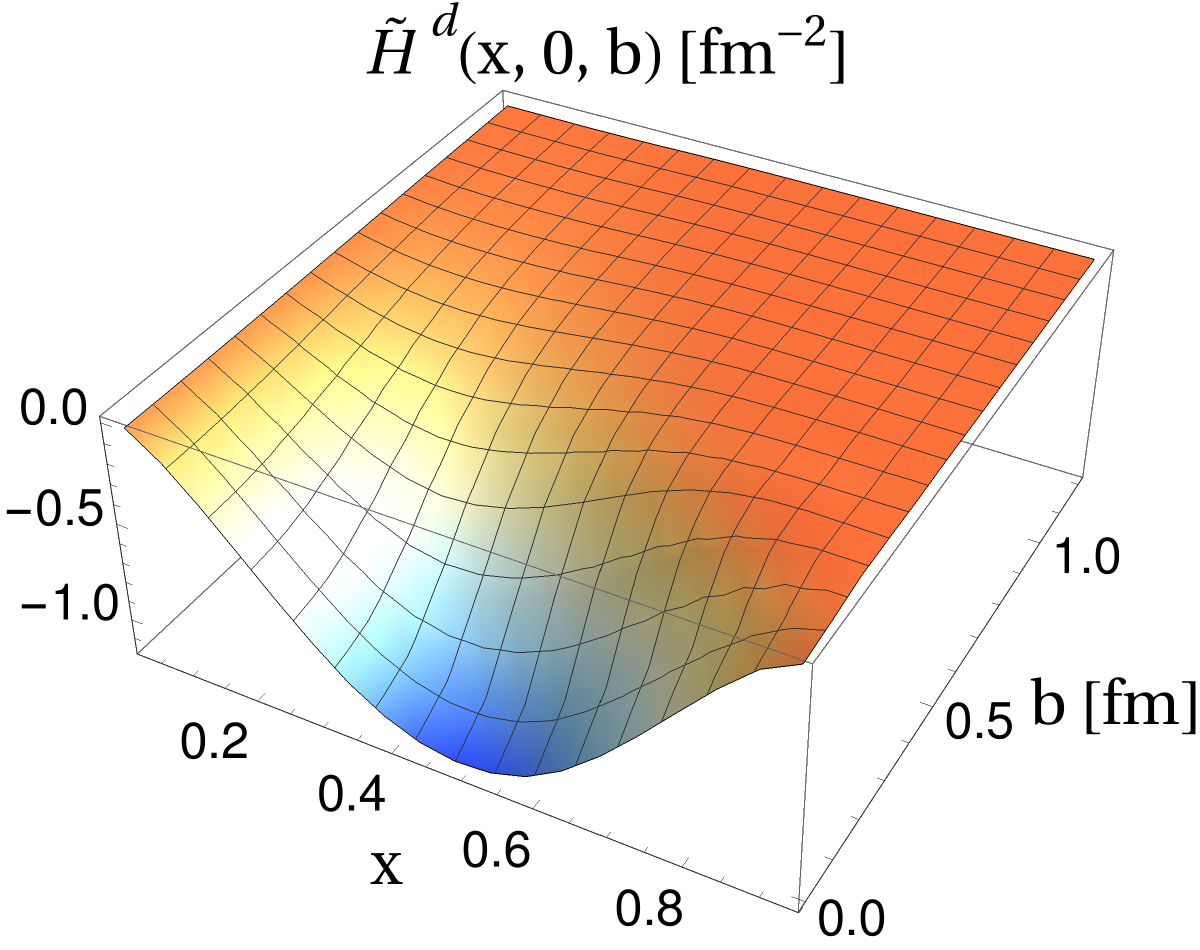}
        \caption{}
         \label{fig:Htilde_d_b}
     \end{subfigure}   
        \caption{The chiral-even GPDs: (a) $H(x,0,b)$, (c) $E(x,0,b)$ and (e) $\tilde{H}(x,0,b)$ for the $u$-quark, where the respective GPDs for the $d$-quark are shown in (b), (d) and (f). The GPDs are presented with respect to $x$ and $b$ (in fm).}
        \label{fig:chiral_even_GPDs_b}
\end{figure}
     
     \begin{figure}[hbt!]
     \centering
     \begin{subfigure}[b]{0.41\textwidth}
         \centering
         \includegraphics[width=\textwidth]{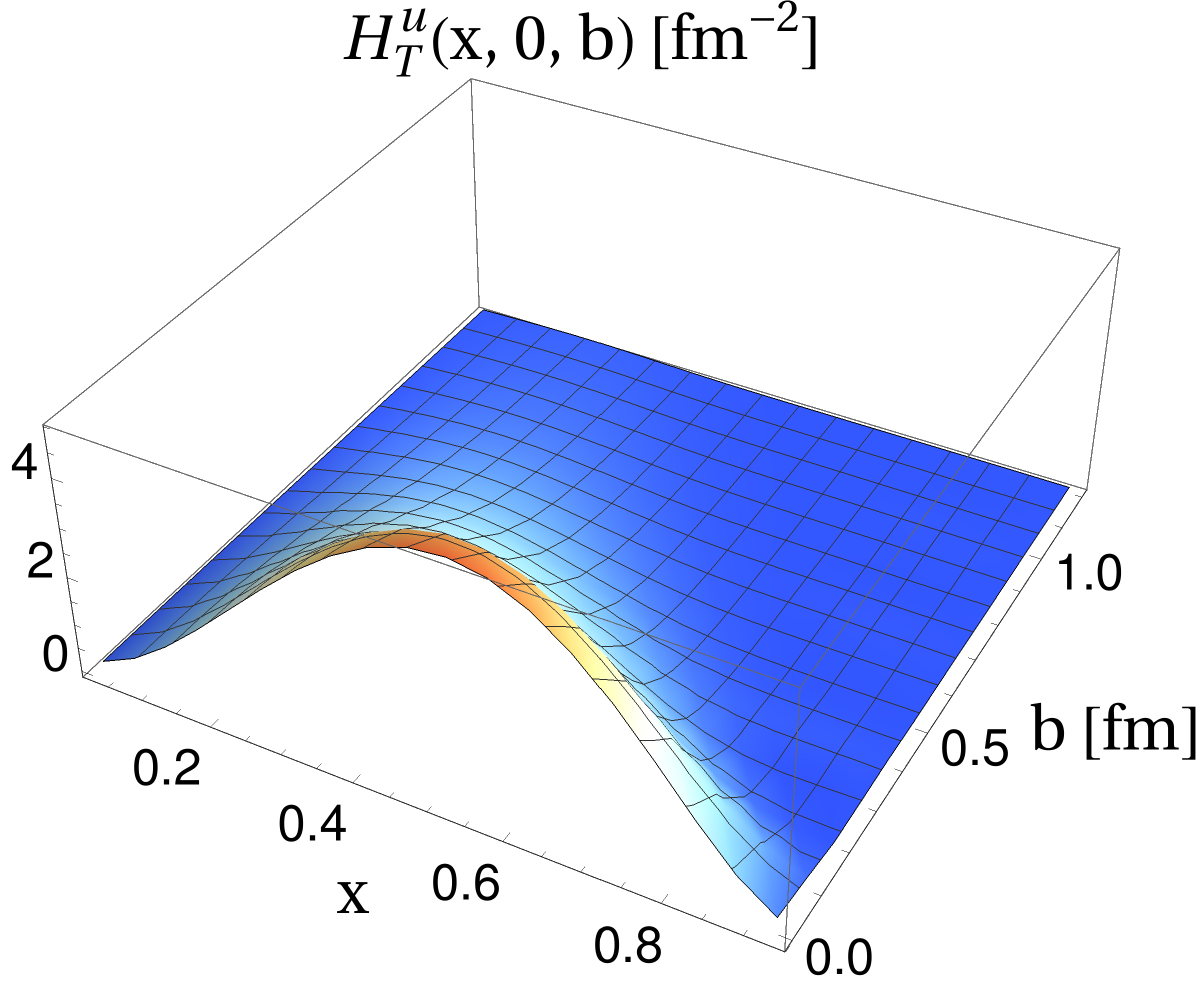}
         \caption{}
         \label{fig:HT_u_b}
     \end{subfigure}
     \begin{subfigure}[b]{0.41\textwidth}
         \centering
         \includegraphics[width=\textwidth]{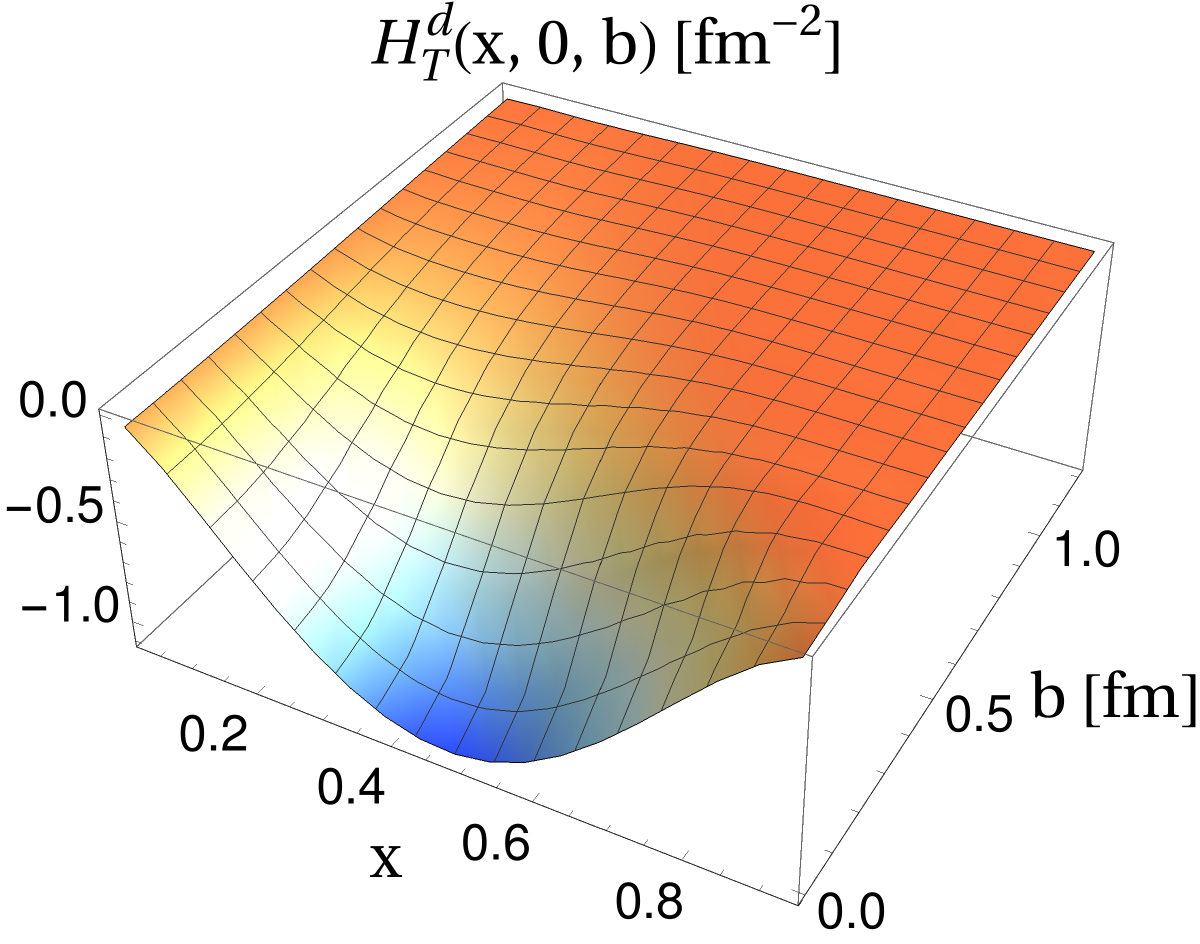}
        \caption{}
         \label{fig:HT_d_b}
     \end{subfigure}
     \begin{subfigure}[b]{0.41\textwidth}
         \centering
         \includegraphics[width=\textwidth]{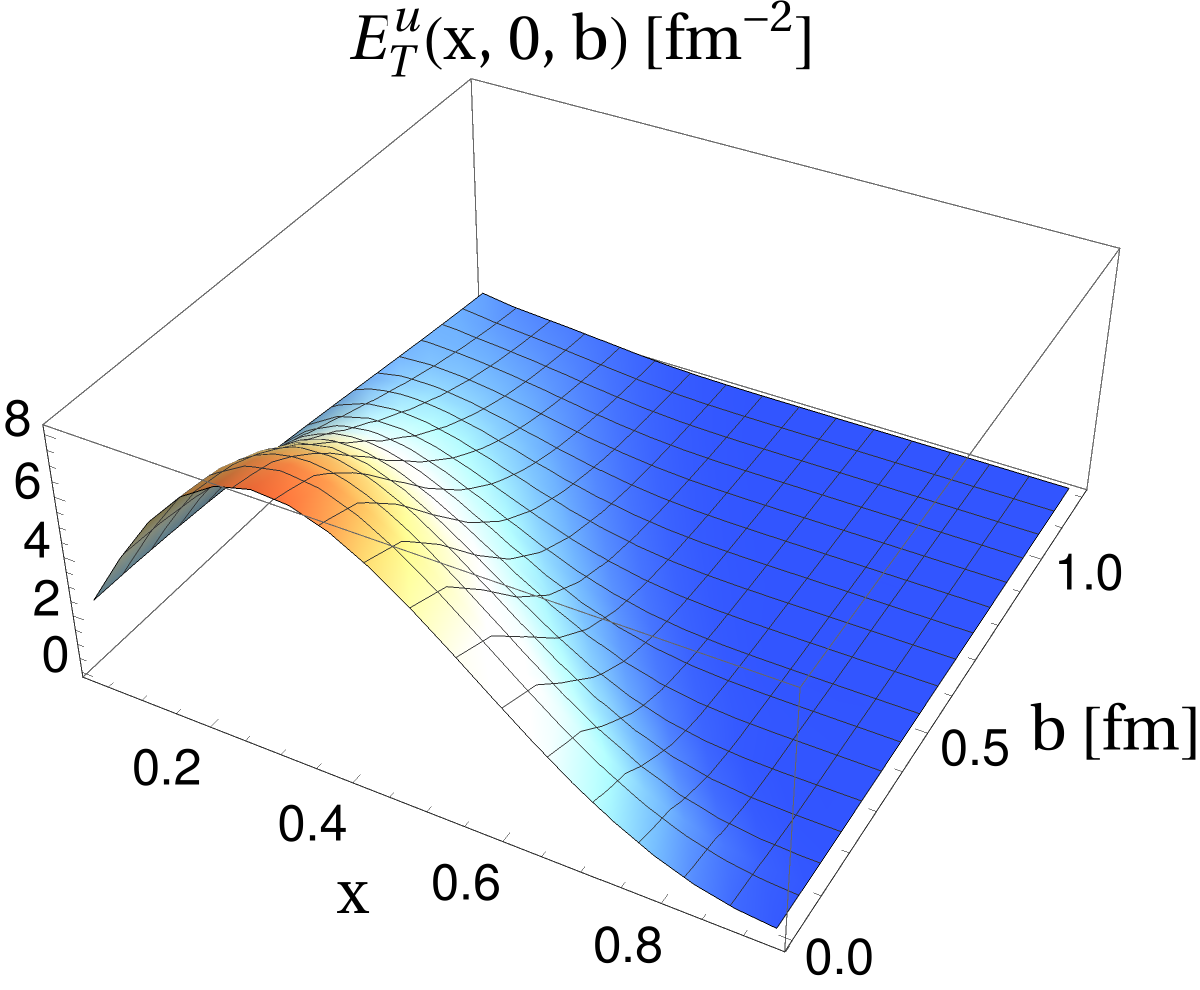}
       \caption{}
         \label{fig:ET_u_b}
     \end{subfigure}
          \begin{subfigure}[b]{0.41\textwidth}
         \centering
         \includegraphics[width=\textwidth]{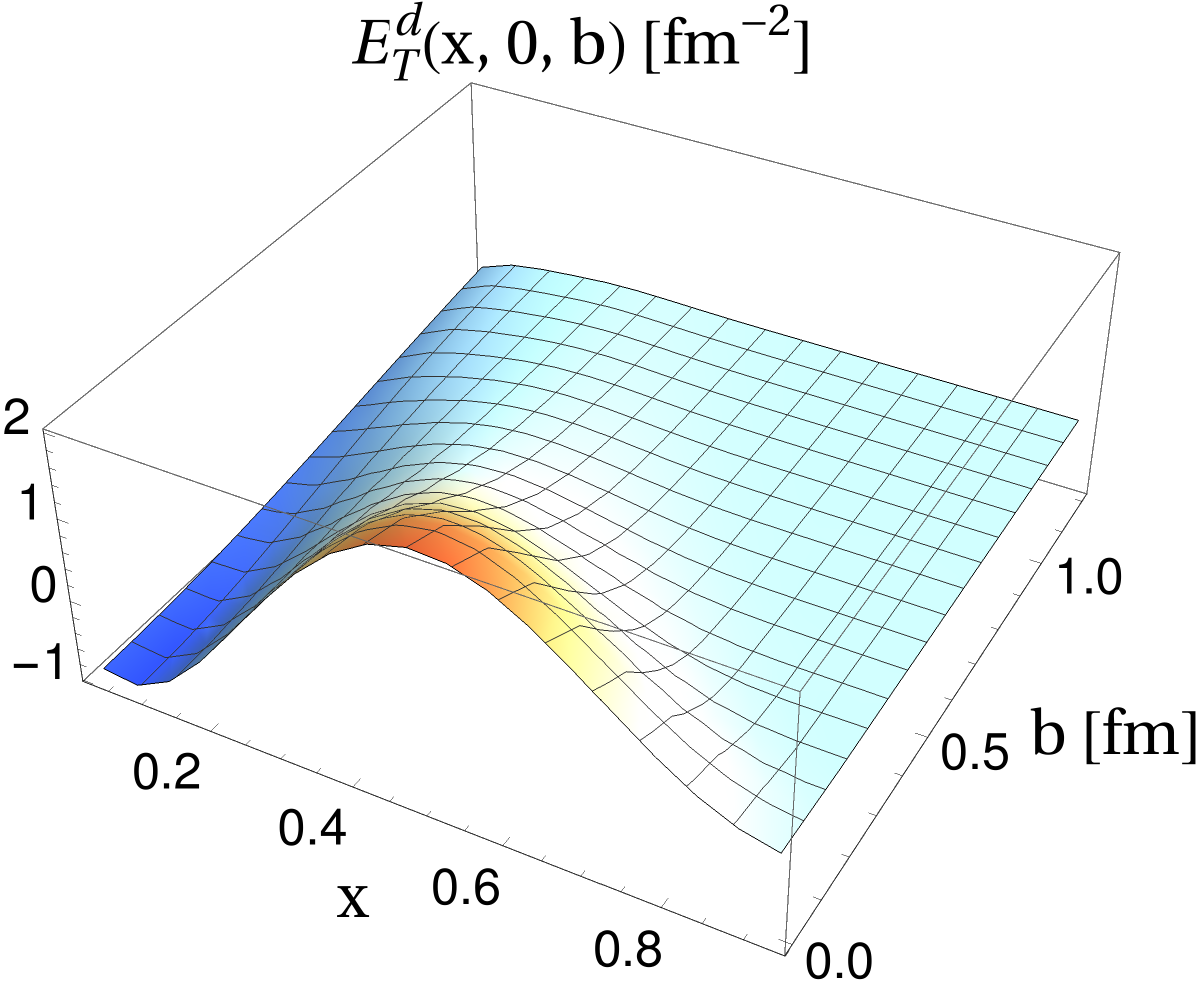}
         \caption{}
         \label{fig:ET_d_b}
     \end{subfigure}
     \begin{subfigure}[b]{0.41\textwidth}
         \centering
         \includegraphics[width=\textwidth]{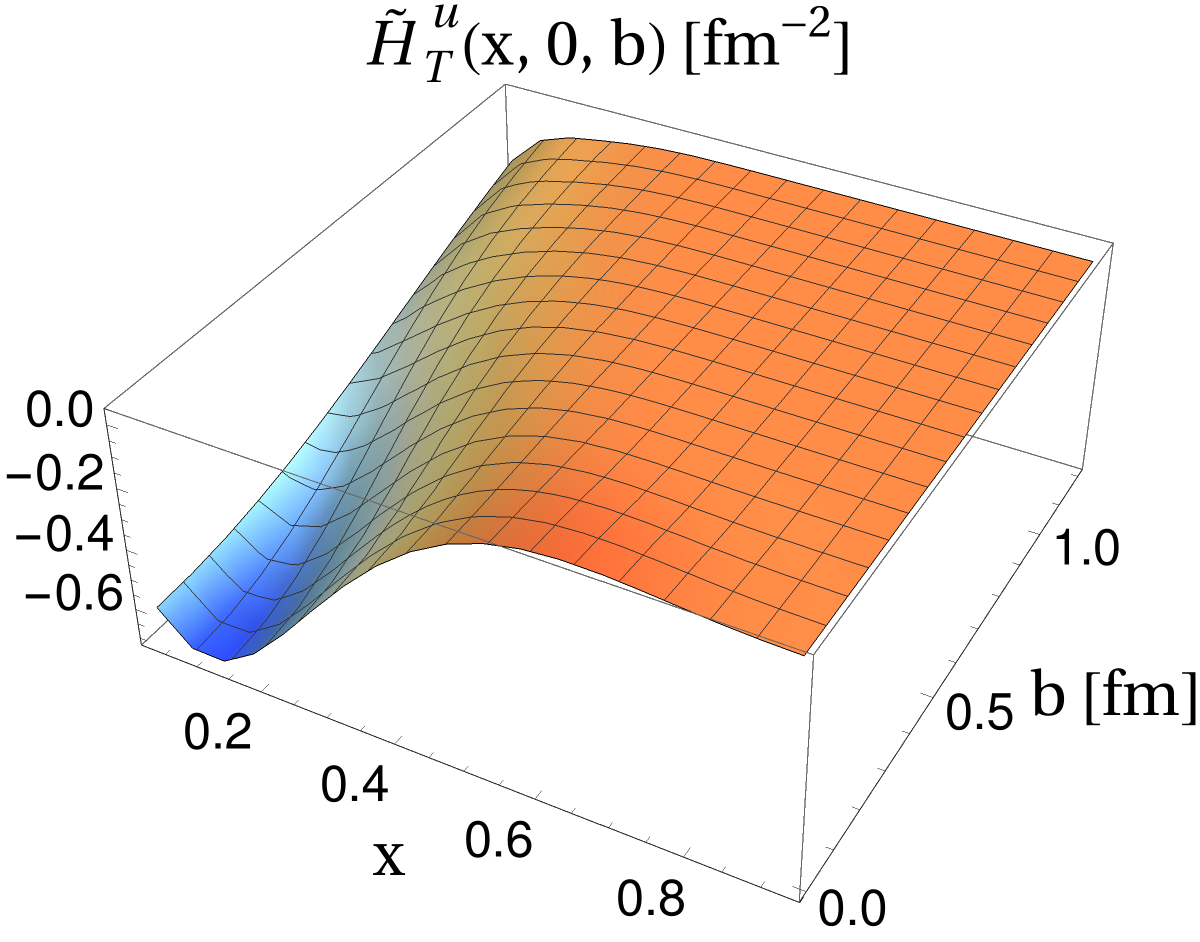}
      \caption{}
         \label{fig:HTtilde_u_b}
     \end{subfigure}
     \begin{subfigure}[b]{0.41\textwidth}
         \centering
         \includegraphics[width=\textwidth]{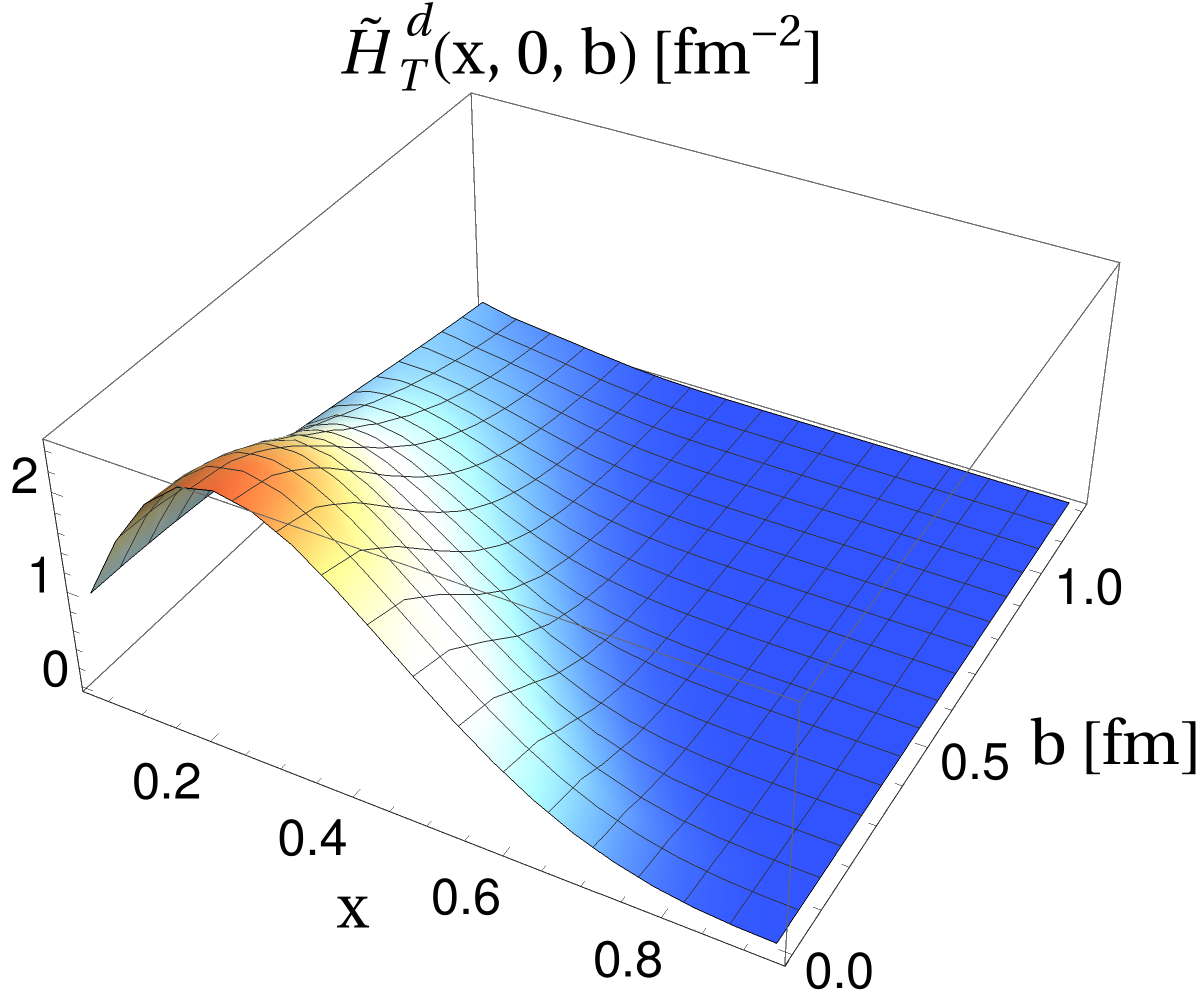}
        \caption{}
         \label{fig:HTtilde_d_b}
     \end{subfigure}
        \caption{The chiral-odd GPDs: (a) $H_T(x,0,b)$, (c) $E_T(x,0,b)$ and (e) $\tilde{H}_T(x,0,b)$ for the $u$-quark, where the respective GPDs for the $d$-quark are shown in (b), (d) and (f). The GPDs are presented with respect to $x$ and $b$ (in fm).}
        \label{fig:chiral_odd_GPDs_b}
\end{figure}

Taking the 2-d Fourier transform of GPDs w.r.t. $\mathbf{\Delta}_\perp$ leads to the GPDs in transverse impact-parameter $({\bf b}_\perp)$ plane~\cite{Burkardt:2000za, Burkardt:2002hr}. We have
\begin{align}
{\rm [GPD]}(x,0,{\bf b}_\perp) &= \frac{1}{(2\pi)^2} \int {\rm d}^2 \mathbf{\Delta}_\perp \, e^{-\iota \mathbf{\Delta}_\perp \cdot {\bf b}_\perp} \, {\rm [GPD]}(x,0,t) \nonumber\\
&= \frac{1}{2\pi} \int \Delta {\rm d} \Delta J_0(\Delta b) {\rm [GPD]}(x,0,t)\;,
\end{align}
where $\Delta=\vert \mathbf{\Delta}_\perp \vert$.
The parameter $b=\vert {\bf b}_\perp \vert$ describes the transverse distance between the active quark and the center of momentum of the proton and satisfies the condition that $\sum_i x_i b_i=0$, where the sum runs over the partons. 

We show the unpolarized and helicity GPDs for the valence quarks as a function of $x$ and $b$ in Fig.~\ref{fig:chiral_even_GPDs_b}. Furthermore, we present the 3-d graphical representation of the transversity GPDs with respect to $x$ and $b$ in Fig.~\ref{fig:chiral_odd_GPDs_b}. We observe that all the GPDs, regardless of their signs, show a decrease in the width of the valence quark distributions in transverse impact-parameter plane, as $x$ increases. For example, in Fig.~\ref{fig:H_u_b}, we find that the width decreases from $1.00$ fm to $0.58$ fm as $x$ increases from $0.5$ to $0.7$. This implies that when the quarks take a larger longitudinal momentum fraction, they locate near the center of transverse position $(b=0)$. On the other hand, the peak of distributions shift towards the lower values of $x$ accompanied by decreasing magnitude, and as we go away $b=0$. Eventually, the distribution vanishes with increasing transverse distance. The rate of the dropoff varies for different GPDs, depending upon the helicities of both target proton and the active quark. All the GPDs have maximum distributions at $b=0$. When the valence quarks carry more than $50\%$ of the longitudinal momentum, the GPDs $H^q, \tilde{H}^q$ and  $H^q_T$ are observed to have this maxima, while the other GPDs have peaks at $x<0.5$. Further, the flavor distributions $E^d,\tilde{H}^d, H^d_T, \tilde{H}_T^u$ are found negative. These signs in transverse $b$-space are directly traced to the GPDs in momentum space $(x, t)$. Our impact-parameter dependent GPDs show similarities with those of various studies in the literature~\cite{Burkardt:2002hr, Dahiya:2007mt, Mondal:2017wbf, Chakrabarti:2013gra, Vega:2010ns, Mondal:2015uha, Chakrabarti:2015ama} which leads us to suggest that there is an emerging trend towards model-independent characteristics.

\section{Conclusions}
In this work, we have presented the leading-twist generalized parton distributions (GPDs) for the proton at zero skewness using the basis light-front quantization (BLFQ) approach, where the effective light-front Hamiltonian includes the transverse and longitudinal confinement as well as the one gluon exchange interaction between the valence quarks. The proton light-front wave functions (LFWFs) have been obtained by treating it as a relativistic three-body system and by diagonalizing the effective Hamiltonian matrix numerically. The resulting LFWFs are utilized to study the various static and dynamic properties, relevant to the low-energy regime. The parameters used in this work were previously fixed to reproduce the proton mass and electromagnetic form factors (FFs)~\cite{Mondal:2019jdg, Xu:2021wwj}. 

The multi-dimensional GPDs, when taken to specific limits, represent unified versions of form factors (FFs) and parton distribution functions (PDFs). We have computed the chiral-even and chiral-odd non-skewed GPDs of the proton to observe the $u$ and $d$ quarks in momentum space, where we take into account the different configurations of helicities of both the active quark and target proton. We have found qualitatively similar behavior for the distributions at zero skewness to those in related studies~\cite{Ji:1997gm, Boffi:2002yy, Boffi:2003yj, Mondal:2015uha, Mondal:2017wbf, Freese:2020mcx, Pasquini:2005dk, Chakrabarti:2015ama, Hashamipour:2021kes, Hashamipour:2020kip}. Note that, we have not computed $\tilde{E}$ and $\tilde{E}_T$ as they require $\zeta \neq 0$ at the initial stage of evaluation in order to obtain the expressions which is beyond the scope of the present work.
Further, these functions are capable of producing the different FFs based on the helicity dependence of both quark and the proton, such that the unpolarized, helicity and transversity GPDs provide Dirac, Pauli, axial, and tensor FFs and are also known as the first Mellin moments of the GPDs. 
We observed that these FFs qualitatively match available predictions from other approaches~\cite{Gockeler:2005cj,QCDSF:2006tkx, Ledwig:2010tu, Pasquini:2005dk, Ledwig:2011qw, Chakrabarti:2015ama}. We have also computed the second moments of the GPDs, which provide precise information on the gravitational form factors and are linked with the total angular momentum contributions of partons inside the proton.
In addition, we have computed the GPDs in transverse position space. Again, we have found that our results show similar qualitative descriptions as obtained in other models. 

Our approach can be systematically improved by incorporating Fock sectors beyond the valence quark component ($\ket{qqq}$). Our future efforts will include higher Fock components of the proton, for example, $\ket{qqqq\bar{q}}$, $\ket{qqqg}$ and so on. It will also be of great interest to see the distributions of sea quarks and gluons describing the proton structure with input from light-front QCD to the model at its initial scale.


\section*{Acknowledgements}
We thank Dimitra Anastasia Pefkou for providing us with
the lattice data. SK is supported by Research Fund for International Young Scientists, Grant No. 12250410251, from the National Natural Science Foundation of China (NSFC), and China Postdoctoral Science Foundation (CPSF), Grant No. E339951SR0. CM thanks the Chinese Academy of Sciences President's International Fellowship Initiative for the support via Grants No. 2021PM0023. CM is also supported by new faculty start up funding by the Institute of Modern Physics, Chinese Academy of Sciences, Grant No. E129952YR0. XZ is supported by new faculty startup
funding by the Institute of Modern Physics, Chinese
Academy of Sciences, by Key Research Program of
Frontier Sciences, Chinese Academy of Sciences, Grant
No. ZDBS-LY-7020, by the Natural Science Foundation of
Gansu Province, China, Grant No. 20JR10RA067, by the
Foundation for Key Talents of Gansu Province, by the
Central Funds Guiding the Local Science and Technology
Development of Gansu Province, Grant No. 22ZY1QA006,
by international partnership program of the Chinese Academy
of Sciences, Grant No. 016GJHZ2022103FN, by National
Natural Science Foundation of China, Grant No. 12375143,
by National Key R\&D Program of China, Grant
No. 2023YFA1606903 and by the Strategic Priority
Research Program of the Chinese Academy of Sciences,
Grant No. XDB34000000. JPV acknowledges partial support from the Department of Energy under Grant Nos. DE-FG02-87ER40371 and DE-SC0023692. This research used resources of the National Energy Research Scientific Computing Center (NERSC), a U.S. Department of Energy Office of Science User Facility located at Lawrence Berkeley National Laboratory, operated under Contract No. DE-AC02-05CH11231 using NERSC award NP-ERCAP0020944. A portion of the computational resources were also provided by Gansu Computing Center. This research is supported by Gansu International Collaboration and Talents Recruitment Base of Particle Physics, and the International Partnership Program of Chinese Academy of Sciences, Grant No.016GJHZ2022103FN.

\bibliographystyle{apsrev}
\bibliography{ref}

 \end{document}